\documentclass[a4paper,12pt]{article}
\usepackage[]{epsfig}
\usepackage[]{rotating}
\usepackage[]{cite}
\newcommand{\ee}    {\mbox{${\mathrm{e}}^+ {\mathrm{e}}^-$}}
\newcommand{\tautau}{\mbox{$\tau^+\tau^-$}}
\newcommand{\mm}    {\mbox{$\mu^+\mu^-$}}

\newcommand{\evisn} {\mbox{$\protect E_{\mathrm {vis}}$/$\protect \sqrt{s}$}}
\newcommand{\rvis}  {\mbox{$R_{\mrm{vis}}$}}
\newcommand{\omclb} {\mbox{${\mathrm {1-CL_b}}$}}
\newcommand{\clb}   {\mbox{${\mathrm {CL_b}}$}}
\newcommand{\cls}   {\mbox{${\mathrm {CL_s}}$}}
\newcommand{\clsb}  {\mbox{${\mathrm {CL_{s+b}}}$}}

\newcommand{\qq}    {\mbox{$\mathrm{q}\bar{\mathrm{q}}$}}

\newcommand{\bb}    {\mbox{$\mathrm{b}\bar{\mathrm{b}}$}}
\newcommand{\cc}    {\mbox{$\mathrm{c}\bar{\mathrm{c}}$}}

\newcommand{\nunu}  {\mbox{$\nu\bar{\nu}$}}
\newcommand{\nuenue}{\mbox{$\nu_{\mathrm{e}}\bar{\nu_{\mathrm{e}}}$}}
\newcommand{\mZ}    {\mbox{$m_{\mathrm{Z}}$}}
\newcommand{\mH}    {\mbox{$m_{\mathrm{H}}$}}

\newcommand{\mHrec} {\mbox{$m_{\mathrm{H}}^{\mathrm{rec}}$}}
\newcommand{\mHtest} {\mbox{$m_{\mathrm{H}}$}}

\newcommand{\gevcs} {\mbox{${\mathrm{GeV}}/c^2$}}
\newcommand{\gevcm} {\mbox{${\mathrm{GeV}}/c$}}

\newcommand {\Ho}   {\mbox{$\mathrm{H}$}}

\newcommand {\Zo}   {\mbox{$\mathrm{Z}$}}

\newcommand{\qqp}   {\mbox{$\mathrm{q\overline{q}^\prime}$}}

\newcommand{\lnu}{\mbox{$\ell\nu$}}

\newcommand{\Zgs}        {\mbox{$\mathrm{(Z/\gamma)}^{*}$}}
\newcommand {\Wpm}       {\mbox{$\mathrm{W}^{\pm}$}}

\newcommand{\WW}         {\mbox{$\mathrm{W}^+\mathrm{W}^-$}}
\newcommand{\ZZ}         {\mbox{$\mathrm{Z}\mathrm{Z}$}}

\def\mrm       {\mathrm}

\newcommand{\sqrts}     {\mbox{$\sqrt{s}$}}
\newcommand{\sprime}     {\mbox{$\sqrt{s^\prime}$}}
\newcommand{\PhysLettB}[1] {Phys. Lett. {\bf B#1}}

\def\etal{\mbox{{\it et al.}}}
\newcommand{\ra}        {\mbox{$\rightarrow$}}   
\begin{document}
\begin{titlepage}
\centerline{\Large EUROPEAN ORGANIZATION FOR NUCLEAR RESEARCH}

\begin{flushright}
       CERN-EP/2002-059\\ 23rd July 2002\\ 
\end{flushright}

\bigskip\bigskip\bigskip\bigskip\bigskip
\begin{boldmath}
\begin{center}{\LARGE\bf Search for the Standard Model Higgs 
 Boson with the OPAL Detector at LEP}
\end{center}
\end{boldmath}
\bigskip

\vspace{0.1cm}  
\begin{center}
 {\Large {The OPAL Collaboration}} \\
\end{center}

\bigskip
\begin{center}{\large  Abstract}\end{center}

This paper summarises the search for the Standard Model Higgs boson in
\ee\ collisions at centre-of-mass energies up to 209~GeV performed
by the OPAL Collaboration at LEP.  The consistency of the data with
the background hypothesis and various Higgs boson mass hypotheses is
examined.  No indication of a signal is found in the data and a lower
bound of 112.7~\gevcs\ is obtained on the mass of the Standard Model
Higgs boson at the 95\% CL.

\bigskip\bigskip\bigskip\bigskip\bigskip\bigskip\bigskip\bigskip
\begin{center}
  {\LARGE
    To be submitted to the European Physical Journal {\bf C}}
\end{center}

\end{titlepage}

%
%
\newpage

\begin{center}{\Large        The OPAL Collaboration
}\end{center}\bigskip
\begin{center}{
G.\thinspace Abbiendi$^{  2}$,
C.\thinspace Ainsley$^{  5}$,
P.F.\thinspace {\AA}kesson$^{  3}$,
G.\thinspace Alexander$^{ 22}$,
J.\thinspace Allison$^{ 16}$,
P.\thinspace Amaral$^{  9}$, 
G.\thinspace Anagnostou$^{  1}$,
K.J.\thinspace Anderson$^{  9}$,
S.\thinspace Arcelli$^{  2}$,
S.\thinspace Asai$^{ 23}$
% ...
}\end{center}\bigskip
\bigskip
$^{  1}$School of Physics and Astronomy, University of Birmingham,
Birmingham B15 2TT, UK
\newline
$^{  2}$Dipartimento di Fisica dell' Universit\`a di Bologna and INFN,
I-40126 Bologna, Italy
\newline
$^{  3}$Physikalisches Institut, Universit\"at Bonn,
D-53115 Bonn, Germany
\newline
% ...
$^{  5}$Cavendish Laboratory, Cambridge CB3 0HE, UK
\newline
% ...
$^{  9}$Enrico Fermi Institute and Department of Physics,
University of Chicago, Chicago IL 60637, USA
\newline
% ...
$^{ 16}$Department of Physics, Schuster Laboratory, The University,
Manchester M13 9PL, UK
\newline
% ...
$^{ 22}$Department of Physics and Astronomy, Tel Aviv University,
Tel Aviv 69978, Israel
\newline
$^{ 23}$International Centre for Elementary Particle Physics and
Department of Physics, University of Tokyo, Tokyo 113-0033, and
Kobe University, Kobe 657-8501, Japan
\newline
% ...

\section{Introduction}
\label{intro}
The $SU(2)_L \times U(1)_Y$ gauge theory of the electroweak
interaction~\cite{gws} accurately predicts all electroweak phenomena
observed so far.  It describes, with a minimum of assumptions, the
structure of the electroweak gauge boson sector and its interactions
with the fermions, and constrains the structure of the fermion
multiplets.  The gauge theory, $SU(2)_L \times U(1)_Y$, is referred to
as the Standard Model (SM) of electroweak interactions.  An important
ingredient of the SM is the Higgs sector which provides masses for the
gauge bosons \Zo\ and \Wpm\ and for the charged fermions without
violating the principle of gauge invariance.  This is done by
introducing a doublet of scalar ``Higgs'' fields and their
interaction~\cite{higgsmech}.  One consequence of this extension is
the prediction of a neutral scalar Higgs boson; its mass is a free
parameter of the theory.

\begin{figure}[p]
  \centerline{
    \epsfig{file=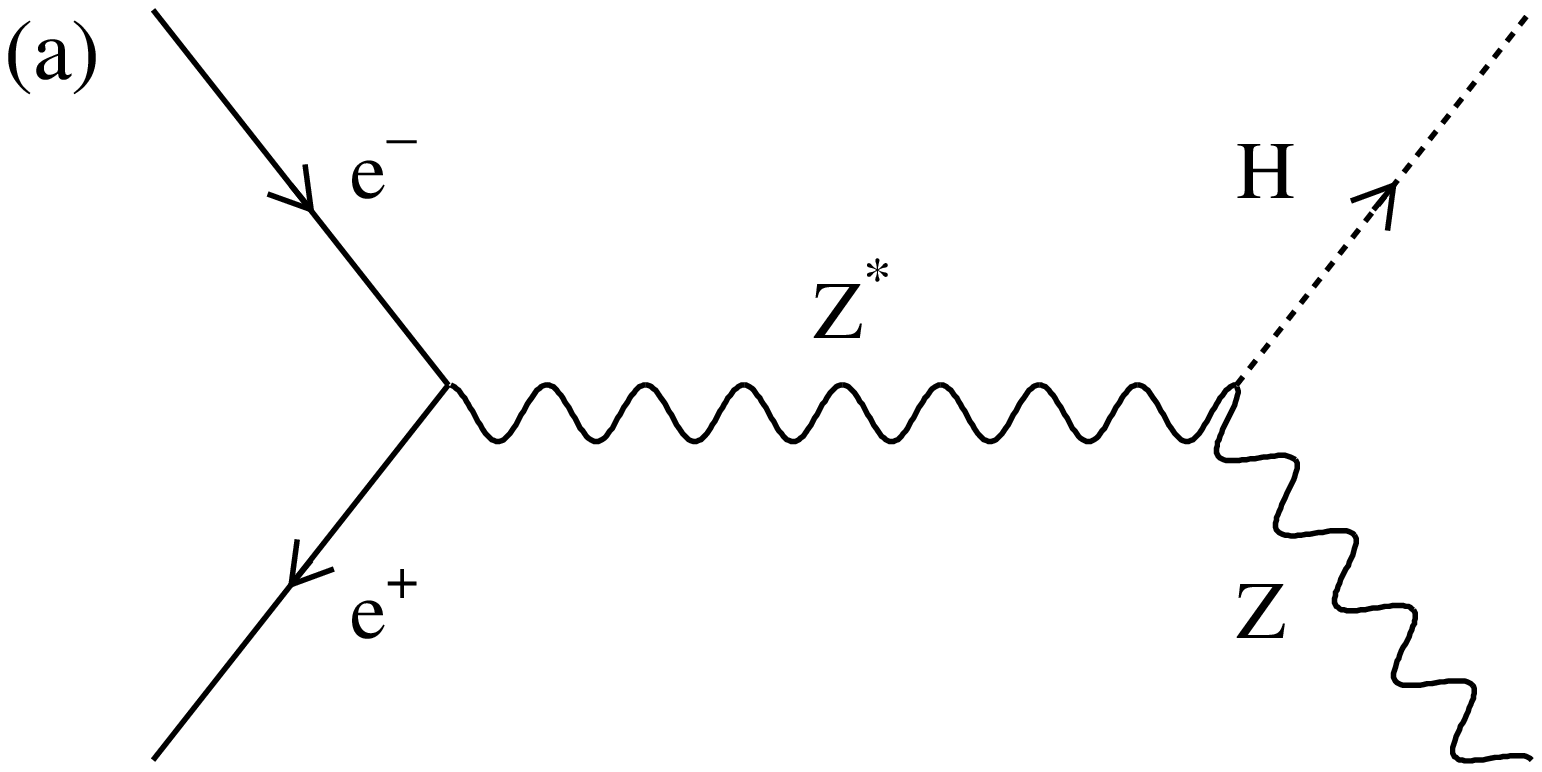,width=0.34\textwidth}
    \epsfig{file=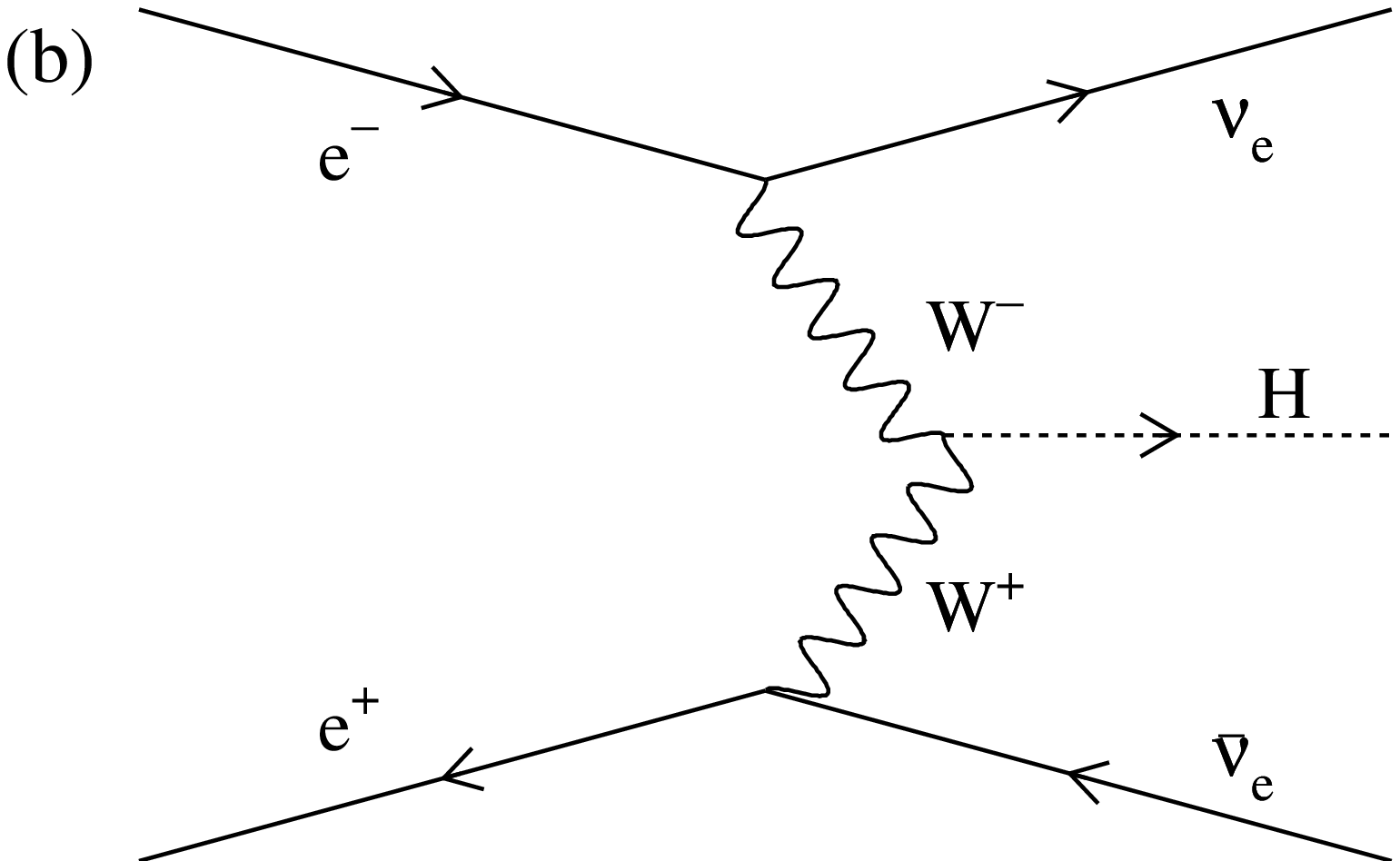,width=0.3\textwidth}
    \epsfig{file=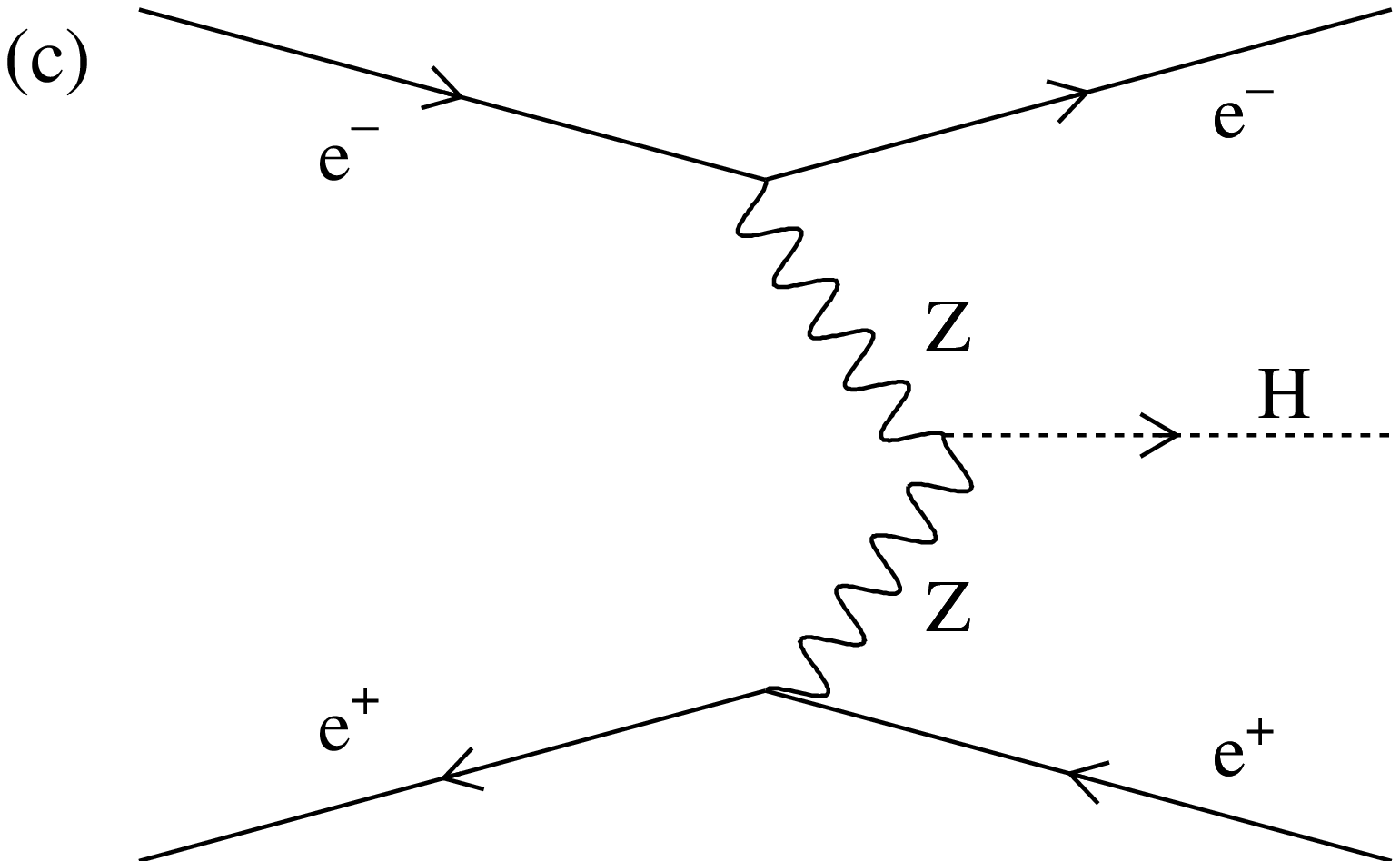,width=0.3\textwidth}}
  \caption[]{\label{fig:diagrams}{\sl Feynman diagrams for (a) the
      Higgs-strahlung process, (b) the \WW fusion process and (c) the
      \ZZ\ fusion process for the production of Higgs bosons in \ee\
      collisions.}}
\end{figure}

The SM predicts that Higgs bosons may be produced at \ee\ colliders
via the Higgs-strahlung process, \ee\ra\Ho\Zo.  The cross-section
depends only on the centre-of-mass energy $\sqrt{s}$ and the Higgs
boson mass $\mH$.  Higgs bosons may also be produced via the \WW\ and
\ZZ\ fusion processes \ee\ra\Ho\nuenue\ and \ee\ra\Ho\ee.  The same
final states are also accessible via the Higgs-strahlung process and
the contributions interfere. The diagrams for the Higgs-strahlung and
fusion processes are shown in Figure~\ref{fig:diagrams}.  The fusion
processes extend beyond the kinematic range of the Higgs-strahlung
process. The fusion contributions are comparatively small at LEP2 energies, 
but taken into account in this analysis.

Previous searches performed by the OPAL Collaboration, at
centre-of-mass energies up to 189 GeV, have excluded a Higgs boson
with mass less than 91.0~\gevcs~\cite{pr285}. Using the data taken in
the years 1999 and 2000 at \sqrts\ between 192 and 209 GeV, and based
on a preliminary calibration of the detector and the beam energy,
OPAL has raised this limit to 109.7 \gevcs~\cite{pr329}.
 
This paper reports on the final results of the search for the Higgs
boson, after the complete calibration, and supersedes the previous
published results in~\cite{pr329}. Moreover, the analysis procedures
have been modified to increase the search sensitivity in the mass
region from 100~\gevcs\ up to about 115~\gevcs. In this range, the
Higgs boson is expected to decay predominantly to \bb\ (73\% to 80\%,
depending on \mH ), and \tautau (7\%), with the remaining branching
fraction shared between \cc, gg , and off shell \WW\ decays.  The
final states arising from the Higgs-strahlung process are determined
by these decay properties of the Higgs boson and by those of the
associated \Zo\ boson. The searches, therefore, encompass the
``four-jet'' channel (\Ho\Zo\ra\bb\qq ), the ``missing-energy''
channel (\Ho\Zo\ra\bb\nunu ), the ``tau'' channels
(\Ho\Zo\ra\bb\tautau, \tautau\qq) and the ``electron'' and ``muon''
channels (\Ho\Zo\ra\bb\ee, \bb\mm).  The dominant backgrounds arise
from quark-pair production $\ee\ra\qq(\gamma)$ and four-fermion
production including \WW\ and \ZZ\ production.  In particular, the
\ZZ\ background constitutes an irreducible background for
\mH~$\sim$~\mZ\ if one of the \Zo\ bosons decays to \bb.

Section~\ref{sect:data} summarises the properties of the OPAL
detector, the data samples, and Monte Carlo simulation.  Event
reconstruction and b-tagging tools, which are common to the searches
in all channels, are described in Section~\ref{sect:recon}. The event
selections addressing separately the four final states are discussed in
detail in Section~\ref{sect:selections}.  The statistical procedures
to investigate the compatibility of the observation with the
background hypothesis and various Higgs signal hypotheses are
described in Section~\ref{limitcalcappendix}. Finally, the results
are summarised in Section~\ref{sect:results}.

\section{Detector, Data Samples, and Monte Carlo Simulations}
\label{sect:data}
The tracking and calorimetry systems of the OPAL detector have nearly
complete solid angle coverage.  The central tracking detector is
placed in a uniform 0.435~Tesla axial magnetic field.  The innermost
part is occupied by a high-resolution silicon microstrip
vertex (``microvertex'') detector~\cite{simvtx} which surrounds the
beam pipe and covers the polar angle\footnote{ OPAL uses a
right-handed coordinate system where the $+z$ direction is along the
electron beam and where $+x$ points to the centre of the LEP ring.
The polar angle $\theta$ is defined with respect to the $+z$ direction
and the azimuthal angle $\phi$ with respect to the $+x$ direction.
The centre of the \ee\ collision region defines the origin of the
coordinate system.} range $|\cos\theta|<0.93$.  This detector is
followed by a high-precision vertex drift chamber, a large-volume jet
chamber, and chambers to measure the $z$ coordinates along the
particle trajectories.  A lead-glass electromagnetic calorimeter with
a presampler is located outside the magnet coil.  In combination
with the forward calorimeters, a forward ring of lead-scintillator
modules (the ``gamma catcher''), a forward scintillating tile 
counter~\cite{detector,mip}, and the silicon-tungsten
luminometer~\cite{sw}, the calorimeters provide 
a geometrical acceptance down to 25~mrad from the beam direction. The
silicon-tungsten luminometer measures the integrated luminosity using
Bhabha scattering at small angles~\cite{lumino}.  The iron return-yoke
of the magnet is instrumented with streamer tubes and thin-gap
chambers for hadron calorimetry.  Finally, the detector is completed
by several layers of muon chambers.

\begin{figure}[p]
  \centerline{\epsfig{file=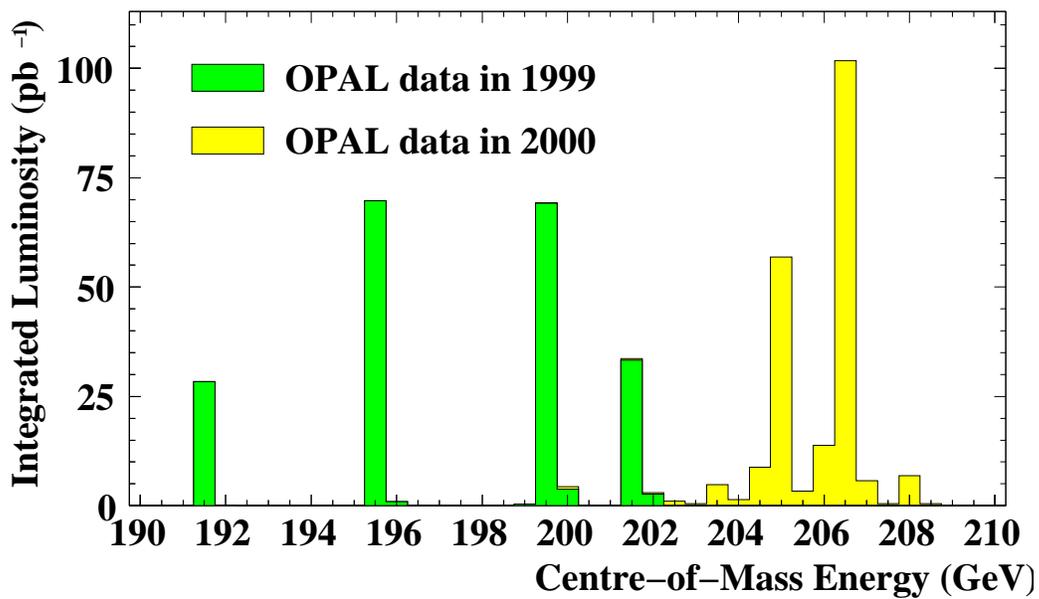,width=0.85\textwidth}}
  \caption[]{\label{fig:lumi}\sl Integrated luminosity collected in the
    years 1999 and 2000, as a function of the centre-of-mass energy.}
\end{figure}

The data used for the present analysis were collected during the years
1999 and 2000 in \ee\ collisions at centre-of-mass energies between
192 and 209 GeV.  During the year 2000 the LEP collider ran in a mode
optimised for the highest possible luminosity at the highest available
energies.  The distribution of the integrated luminosity collected by
OPAL at various LEP energies is shown in Figure~\ref{fig:lumi} and
summarised in Table~\ref{tab:lumi_st}.

A variety of Monte Carlo samples have been generated to estimate the
detection efficiencies for the Higgs boson signal and to optimise the
rejection of SM background processes.  The cross-section and kinematic
properties of the Higgs boson signal depend very strongly on
$\sqrt{s}$ for Higgs boson masses near the limit of sensitivity.  For
an accurate modelling, the signal and background samples were
generated at several centre-of-mass energies, from 192~GeV to 210~GeV,
and in steps of 1~\gevcs\ from \mH=80~\gevcs\ to \mH=120~\gevcs.  For
the Higgs boson signal the HZHA generator is used~\cite{hzha}.  For
the background processes the following event generators are used:
KK2f~\cite{kk2f} for \Zgs\ra\qq($\gamma$), \mm$(\gamma)$ and
\tautau$(\gamma)$, grc4f~\cite{grc4f} for four-fermion processes (4f),
BHWIDE~\cite{bhwide} for \ee$(\gamma)$, and PHOJET~\cite{phojet},
HERWIG~\cite{herwig}, and Vermaseren~\cite{vermaseren} for hadronic
and leptonic two-photon processes ($\gamma\gamma$).
JETSET~\cite{pythia} is used as the principal model for fragmentation.
The detector response to the generated particles is simulated in full
detail~\cite{gopal}.

\section{Event Reconstruction and B-Tagging}
\label{sect:recon}
Some analysis procedures, described in this section, are common to all
search channels.  These include the reconstruction of particles and
basic quality requirements, the assignment of particles to jets and
the identification of jets which contain hadrons with b-flavour.

\subsection{Particle Reconstruction Requirements}
\label{sect:recon2}
Events are reconstructed from charged-particle tracks and energy
deposits (``clusters'') in the electromagnetic and hadronic calorimeters 
(the forward calorimeters are not used in the cluster reconstruction).
Tracks have to satisfy the following quality requirements.

\begin{itemize}
\item The number of hits in the central jet chamber must exceed 20,
  and must be more than half the number of possible hits along the
  track, given its polar angle and origin.
\item The polar angle of the track must satisfy $|\cos\theta|<0.962$.
\item The distance of closest approach of the track to the beam axis
  must not exceed 2.5~cm.
\item The $|z|$ coordinate of the track at the point of closest
  approach to the beam axis must not exceed 30 cm.
\item The component of the momentum transverse to the beam axis must
  exceed 120~MeV$/c$. 
\end{itemize}
Clusters have to satisfy the following quality requirements.
\begin{itemize}
\item At least one lead-glass block must contribute to an
  electromagnetic cluster in the barrel part of the electromagnetic
  calorimeter and at least two in the endcap part.
\item These clusters must have an energy of at least 100 MeV in the
  barrel part and 250~MeV in the endcap part.
\item Hadronic calorimeter clusters in the barrel or endcap parts
 must have an energy of at least 600 MeV;  in the poletip hadron
  calorimeter the energy must exceed 2.0~GeV.
\end{itemize}

Charged particle tracks and energy clusters satisfying these quality
requests are associated to form ``energy flow objects''.
A matching algorithm is employed to
reduce double counting of energy in cases where 
charged tracks point towards electromagnetic clusters.
Specifically, the expected calorimeter energy of the associated
tracks is subtracted from the cluster energy.
If the energy of a cluster is smaller than that expected for the associated 
tracks, the cluster is not used.
Each accepted track and cluster is considered to be a particle.
Tracks are assigned the pion mass. Clusters are assigned zero mass since they 
originate mostly from photons.
The resulting energy flow objects are then grouped into jets and
contribute to the total energy and momentum, $E_{\mathrm{
vis}}$ and $P_{\mathrm{vis}}$, of the event.  Unassociated tracks and
clusters are left as single energy-flow objects.  Corrections are
applied to prevent multiple counting of energies in the case of
particles which produce signals in several subdetectors.  The
association of energy flow objects (particles) into jets is
provided primarily by the Durham jet finder algorithm~\cite{durham}
where the number of reconstructed jets in the event is controlled by
the ``jet resolution parameter'' $y_{\mathrm {cut}}$. The value of
$y_{\mathrm {cut}}$ is chosen in the various search channels according
to the desired signal topology.

\subsection{B-Tagging}
\label{sect:btag}

We combine three nearly independent properties to identify jets
containing b-hadron decays~\cite{pr285}: the high-$p_t$ lepton from
the semi-leptonic decay $\mathrm{b \ra c \ell^{\pm} \nu_ {\ell}} $,
the detectable lifetime of b-hadrons, and kinematic differences
between b-decays and light-quark decays.

For each jet, using the jet-definition given by the channels, the outputs 
of the lifetime artificial neural network (ANN), 
the kinematic ANN, and the lepton tag are combined into a single likelihood 
variable ${\cal B}$.

\begin{itemize}
\item For the high-$p_t$ lepton tag, semileptonic b-decays are
  identified using electron and muon identification and rejection of
  $\mathrm{\gamma}$ conversions~\cite{pr33}.  The high
  transverse momentum of the lepton with respect to the jet axis 
  is used as a b-tag variable~\cite{pr183}.
  
\item The lifetime-sensitive tagging variables are combined to
  reconstruct secondary vertices.  Tracks in the jet are ranked by an
  ANN (track-ANN) using the impact parameter of tracks with respect to
  the primary vertex.  The first six tracks (or all tracks if their
  number is less than six) are used to form a ``seed''
  vertex~\cite{btag1} by a technique where first a vertex is formed
  from all input tracks, and then the track with the highest
  contribution to the vertex $\chi^2$ is removed.  This procedure is
  repeated until no track contributes more than 5 to the $\chi^2$, and
  the seed vertex is obtained by a fit to the selected tracks. After
  the seed vertex is formed, the remaining tracks in the jet are then
  tested, in the order of distance to the seed vertex, and added to
  this vertex if their contributions to the vertex $\chi^2$ are
  smaller than 5.
  
  In addition to identifying displaced vertices, we also use track
  impact parameters to gain further separation. The impact parameter
  significances $S^{r\phi}$ and $S^{rz}$, in the $r\phi$ and $rz$
  projections, respectively, are formed by dividing the track impact
  parameters by their estimated errors.  The distributions of
  $S^{r\phi}$ and $S^{rz}$ for each quark flavour, obtained from
  Z-peak Monte Carlo simulation, are used as the probability density
  functions (PDF's) $f^{r\phi}_{q}$ and $f^{rz}_{q}$ ($q$=uds, c and
  b).  The combined estimator $F_{q}$ for each quark flavour $q$ is
  computed by multiplying the $f^{r\phi}_{q}$ and $f^{rz}_{q}$ for all
  tracks.  The final estimator ${\cal L}_{\mathrm{IP}}$ is obtained as
  the ratio of $F_b$ and the sum of $F_{uds}$, $F_c$ and $F_b$.

  The following four variables are used as inputs to the lifetime ANN.
\begin{itemize}
\item The combined impact parameter likelihood
  ($\cal{L}_{\mathrm{IP}}$) described above.
 \item The vertex significance likelihood ($\cal{L}_{\mathrm V}$): The
   likelihood for the vertex significance is computed analogously to
   the ${\cal L}_{\mathrm{IP}}$ above, using the decay length
   significance of the secondary vertex rather than the impact
   parameter significance of the tracks. 
 \item The reduced impact parameter likelihood
   ($\cal{R}_{\mathrm{IP}}$): To reduce sensitivity to single
   mismeasured tracks, the track with the largest impact parameter
   significance with respect to the primary vertex is removed from the
   secondary vertex candidate and the remaining tracks are used to
   recompute the likelihood $\cal{L}_{\mathrm{IP}}$.  If the original
   vertex has only two tracks, the function is calculated from the
   impact parameter significance of the remaining track.
 \item The reduced vertex significance likelihood ($\cal{R}_{\mathrm
     V}$).  The track having the largest impact parameter significance
   is removed in the calculation of $\cal{L}_{\mathrm V}$.
\end{itemize}

\item Three kinematic variables are combined in the jet-kinematics
  part of the tag, using an ANN: The number of energy-flow objects
  around the central part of the jet, the angle between the jet
  axis and its boosted sphericity axis, i.e. the
  sphericity~\cite{spher} in the jet rest frame obtained by
  boosting back the system using the measured mass and momenta of
  the jet, and the $C$-parameter~\cite{cpar} for the jet boosted
  back to its rest frame.

\end{itemize}

\begin{figure}[p]
  \centerline{\epsfig{file=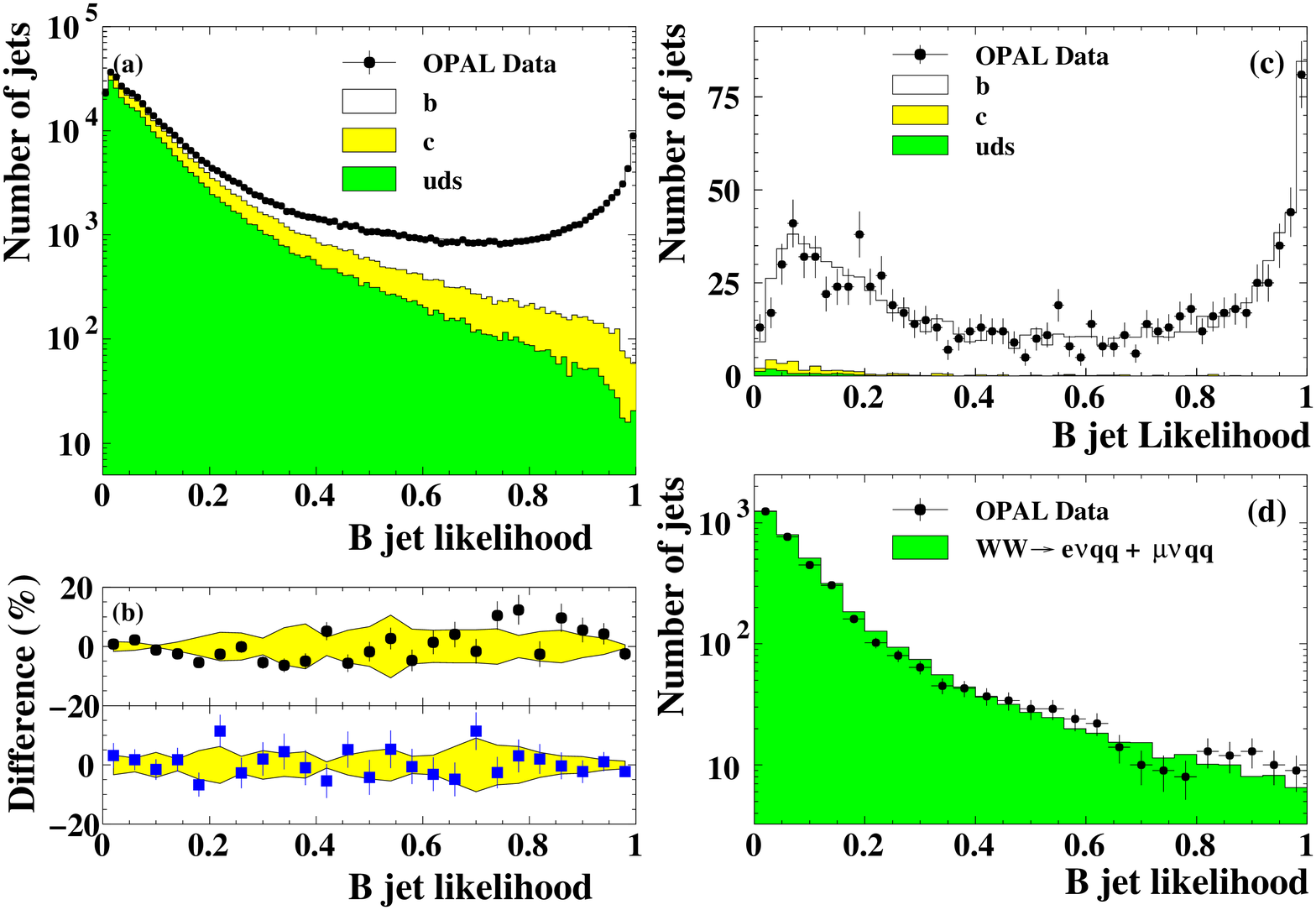,width=0.95\textwidth}}
  \vspace*{-0.8cm}
  \caption[]{\label{fig:btag}\sl (a) The b-tagging algorithm output
    $\cal{B}$, for all jets in the \Zo\ calibration data.  (b) The
    comparison between calibration data and simulation of the b-tag
    values opposite to an anti-tagged(upper part, dots) and a
    tagged(lower part, rectangles) b-jet. The shaded bands indicate
    the systematic uncertainty on the difference. (c) The distribution of
    $\cal{B}$ for high-energy \ee\ra\qq$(\gamma)$ events. One of the
    jets has been tagged as a b-jet and the $\cal{B}$ of the other jet
    is shown.  (d) Distribution of $\cal{B}$ for jets in events
    identified as $\ee\ra\WW\ra\qq\mathrm{e}\nu$ or $\qq\mu\nu$. }
\end{figure}

The outputs from the lifetime ANN, the high-$p_t$ lepton tag and the
jet-kinematics ANN are combined with an unbinned likelihood
calculation~\cite{pr183}, and a final likelihood output ${\cal B}$ is
computed for each jet.  Figure~\ref{fig:btag} shows distributions of
${\cal B}$ obtained from various data sets.  Figure~\ref{fig:btag}(a)
shows the distribution for data taken at $\sqrt{s}=m_{\mathrm{Z}}$ for
calibration purposes~\cite{pr329}, using the same detector
configuration and operating conditions as for the high energy data.
In this data set, the simulation reproduces the observed b-tagging
efficiency within a precision of 1\%, see Figure~\ref{fig:btag}(b).
The shaded band indicates the systematic uncertainties on the
efficiency. In Figure~\ref{fig:btag}(c), the distribution of ${\cal
  B}$ is shown for high energy data enriched in
\Zgs$\rightarrow{\mathrm {q\bar{q}}}$ processes with and without hard
initial-state photon radiation. We find agreement with the SM Monte
Carlo, within a relative statistical uncertainty of 5\%.  For the
lighter flavours, the efficiency has been examined by vetoing
b-flavoured jets in the opposite hemisphere, and the resulting
estimate of the background is found to be described by Monte Carlo
within a precision of 5--10\%.  In Figure~\ref{fig:btag}(d) the
distribution ${\cal B}$ for a high-purity sample of light-flavour jets
in \WW\ra\qq\lnu\ decays is shown.

\section{Event Selection }
\label{sect:selections}

For each of the four topological searches which may characterise
\Ho\Zo\ signal events (see Section~\ref{intro}) the events are first
required to pass a loose preselection.  A finer selection is then
applied using a likelihood function~\cite{barlow}, or an
ANN~\cite{jetnet} with b-tagging and kinematic discriminating
information as inputs.  For the candidates passing the preselection, a
reconstructed mass \mHrec\ is obtained.  For each channel, two
distributions are built for the expected signal, for the backgrounds and
for the data. One is formed by mass dependent and the other by mass
independent variables.  For each Higgs mass to be tested, a
discriminator, constructed as the product of these two distributions,
is then fed into the statistical evaluation described in
Section~\ref{sect:clcalc}.  Although the discriminator distributions
are used in the confidence level calculations, a cut in the selection
likelihood or ANN output is applied to evaluate the systematic
uncertainties on the accepted rates and to illustrate the level of
agreement between the data distributions and the Monte Carlo
simulations (as shown in the figures and tables).

\subsection{Four-Jet Channel}
\label{sect:fourjet}

The four-jet channel selection is sensitive to the process
$\ee\ra\Ho\Zo\ra\bb\qq$.  The branching fraction of the \Zo\ to
hadrons is approximately 70\%, giving the four-jet channel a
potentially large signal rate.  The main backgrounds are hadronic \WW\ 
events, \ZZ\ events, and $\qq (\gamma)$ events in which two or more
energetic gluons are radiated from the quarks.  The b-tag removes
nearly all of the \WW\ background, as well as much of the other
backgrounds.  The \ZZ\ background, with one \Zo\ decaying into \bb, is
irreducible when searching for Higgs bosons of mass near \mZ, but for
higher masses the separation provided by the reconstructed mass,
\mHrec, becomes effective.

The Durham algorithm~\cite{durham} is used to reconstruct four jets in
each event; {\it i.e.}, the resolution\ parameter $y_{\mathrm {cut}}$
is chosen to be between $y_{34}$ and $y_{45}$, where $y_{34}$ is the
transition point from three to four jets, and $y_{45}$ is the
transition point from four to five jets.  These jets are used as
reference jets in the following procedure.  Each particle (energy flow
object) is reassociated to the jet with the smallest ``distance'' to
the particle.  The ``distance'' is defined to be $E_{\mathrm {jet}}
\cdot E_{\mathrm {particle}} \cdot (1-\cos{\theta})$, where
$E_{\mathrm {jet}}$ is the energy of the reference jet, $E_{\mathrm
  {particle}}$ is the energy of the particle (energy-flow object) and
$\theta$ is the angle between them. As a result of this procedure the
di-jet mass resolution before kinematic fitting improves by about
10\%.

\subsubsection{Preselection}
\label{sect:4jpresel}

The preselection is designed to enrich the selected sample in four-jet events.
\begin{itemize}
\item Events must satisfy the standard hadronic final-state 
  requirement~\cite{l2mh}.
\item The effective centre-of-mass energy, $\sqrt{s^\prime}$, obtained
  by kinematic fits assuming that initial state radiation photons are
  lost in the beampipe or seen in the detector~\cite{l2mh}, must be at
  least 80\% of $\sqrt{s}$.
\item The value of $y_{34}$ must exceed 0.003.
\item The event shape parameter $C$~\cite{cpar} must be larger than 0.25.
\item Each of the four jets must have at least two tracks.
\item The $\chi^2$ probabilities must be larger than 10$^{-5}$ both
  for a four-constraint (4C) kinematic fit, which requires energy and
  momentum conservation, and for a five-constraint (5C) kinematic fit,
  $P_{\mathrm{HZ}}$, additionally constraining the invariant mass of
  one pair of jets to the mass of the \Zo, $m_{{\mathrm Z}}$.  The
  algorithm used to select the jet pair which is constrained to the
  \Zo\ mass is described below.
\end{itemize}
The numbers of events remaining after each preselection requirement are
given in Table~\ref{tab:smflow1999} for the data taken in 1999 and 
Table~\ref{tab:smflow2000} for the data taken in 2000.

\subsubsection{Jet-Pairing and Reconstructed Mass}
\label{sect:4jpairing}
There are six possible ways of grouping the four jets into pairs
associated to two bosons.  A likelihood method is used to choose the
jet-pairing, where the combination with the highest likelihood value is
retained for the final likelihood selection and the mass
reconstruction.  The correct assignment of particles to jets plays an
essential role in accurately reconstructing the masses of the initial
bosons.

The likelihood function is constructed from reference histograms for
right and wrong pairing in signal and four-fermion background events.
The signal samples used have 105~\gevcs$<\mH <$115~\gevcs.
All events in the \qq\ background sample are classified as having
wrong combinations.

The following six variables are used. 
\begin{itemize}
\item The logarithm of the probability of a 5C kinematic fit
  constraining two of the jets to have an invariant mass \mZ,
  $\log_{10}(P_{\mathrm{HZ}})$.
\item ${\cal B}_1\cdot{\cal B}_2$, the product of the b-tagging
  discriminant variables for the two jets assumed to come from
  the Higgs boson decay.
\item The quantity $(1-{\cal B}_3)(1-{\cal B}_4)$, formed from the
  b-tagging discriminant variables of the other two jets in the
  event.
\item $|\cos\theta_{\mathrm Z}|$, where $\theta_{\mathrm Z}$ is the 
 polar angle of the momentum sum of the two jets assigned to the
  Z.
\item The 4C fit mass of the \Zo\ boson candidate, $m_{\mathrm
  Z}^{4C}$.
\item The logarithm of the probability from a 6C fit, requiring energy
  and momentum conservation and constraining the masses of the pairs
  of jets to the WW hypothesis, $\log_{10}(P_{\mathrm{WW}})$.

\end{itemize}

For a signal of mass \mH=115~\gevcs, the fraction of swapped events is
7\% (the two jets of the \Ho\ incorrectly identified as originating
from the \Zo), and the fraction of correct-pairing is 58\%, while for
a sample of \Zo\Zo\ (\WW ) events the correct-pairing fraction is 60\%
(75\%).

Once the jet pairing is established, a Higgs boson mass is assigned to
the event using
\mHrec$=m^{\mathrm{4C}}_{\mathrm{H}}+m^{\mathrm{4C}}_{\mathrm{Z}}-
\mZ$ where $m^{\mathrm{4C}}_{\mathrm{H}}$ and
$m^{\mathrm{4C}}_{\mathrm{Z}}$ are the di-jet invariant masses of the
assumed \Ho\ and \Zo\ respectively.  Hence the jet pairings which have
the correct jet assignment to \Zo/\Ho\ and the swapped assignment have
the same reconstructed mass.

\subsubsection{Likelihood Selection}
\label{sect:4jlikelihood}
Events passing the preselection are assigned a discriminator function
which is the product ${\cal D{\mathrm {( \mH )}}}={\cal L}_1 \cdot
{\cal L}_2$(\mH) of two separate likelihood functions. ${\cal L}_1$ is
a likelihood using input variables which are not explicitly mass
dependent, while ${\cal L}_2$(\mH) uses explicitly mass dependent
characteristics of the event.  To obtain ${\cal L}_1$ a set of signal
samples with 105~\gevcs$<$\mH$<$115~\gevcs\ are combined into one
sample, while the likelihood function ${\cal L}_2$(\mH) is obtained
for individual \mH\ values in the range of
80~\gevcs$<$\mH$<$120~\gevcs.

\begin{figure}[p]
  \centerline{
    \epsfig{file=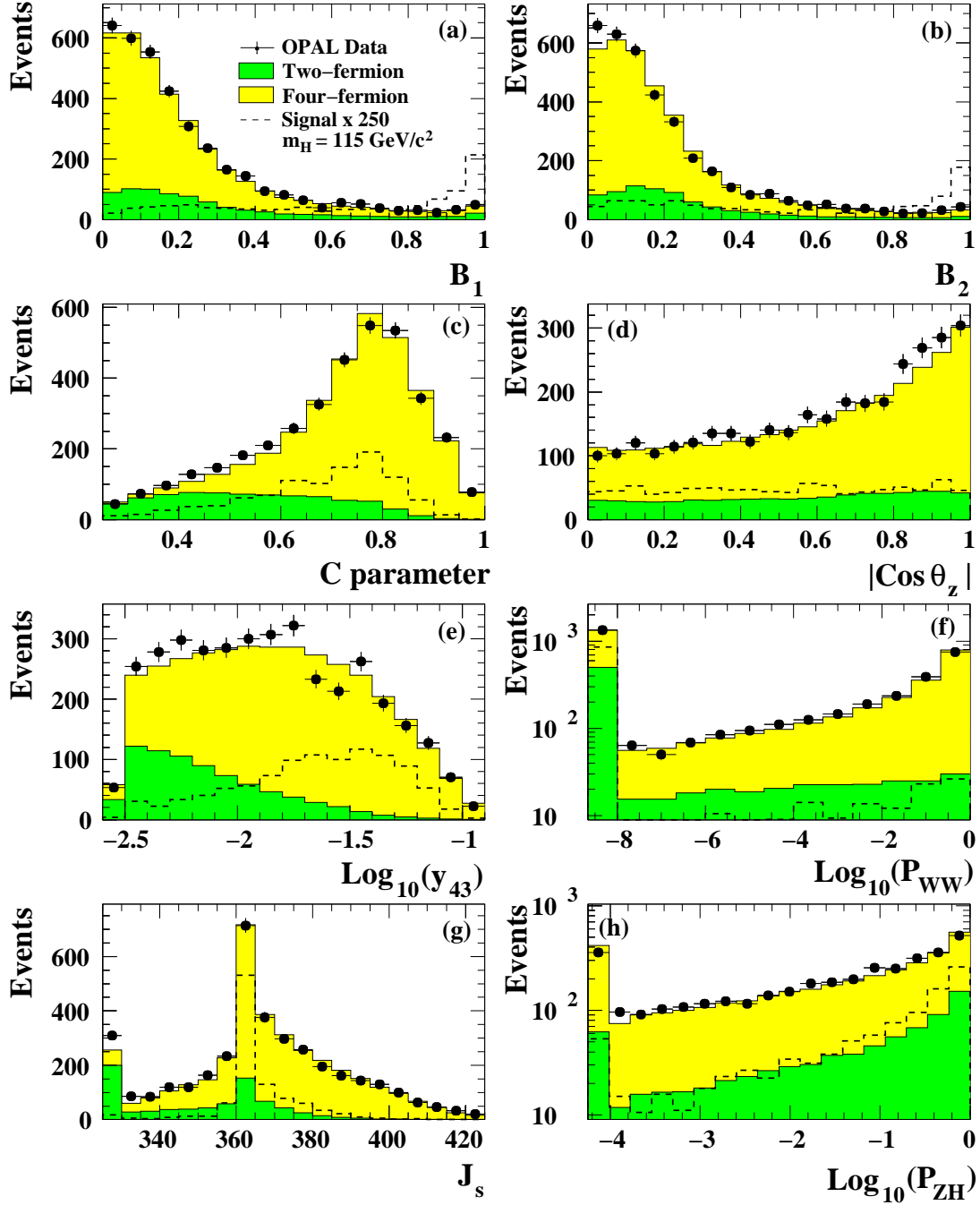,width=0.9\textwidth}}
  \caption[]{\label{fig:fourjetvars3999}\sl Four-Jet Channel:
    Distributions of the variables used as input to the likelihood
    function ${\cal{L}}_1$. Distributions of (a) ${\cal B}_1$ and (b)
    ${\cal B}_2$, (c) the $C$ parameter, (d) $|\cos\theta_{\mathrm
    Z}|$, (e) $\log_{10}(y_{34})$ using the Durham jet-finding
    algorithm, (f) $\log_{10}(P_{\mathrm{WW}})$, (g) jet-angle sum
    $J_s$, and (h) $\log_{10}(P_{\mathrm{HZ}})$. }
\end{figure}

\begin{figure}[p]
  \centerline{
    \epsfig{file=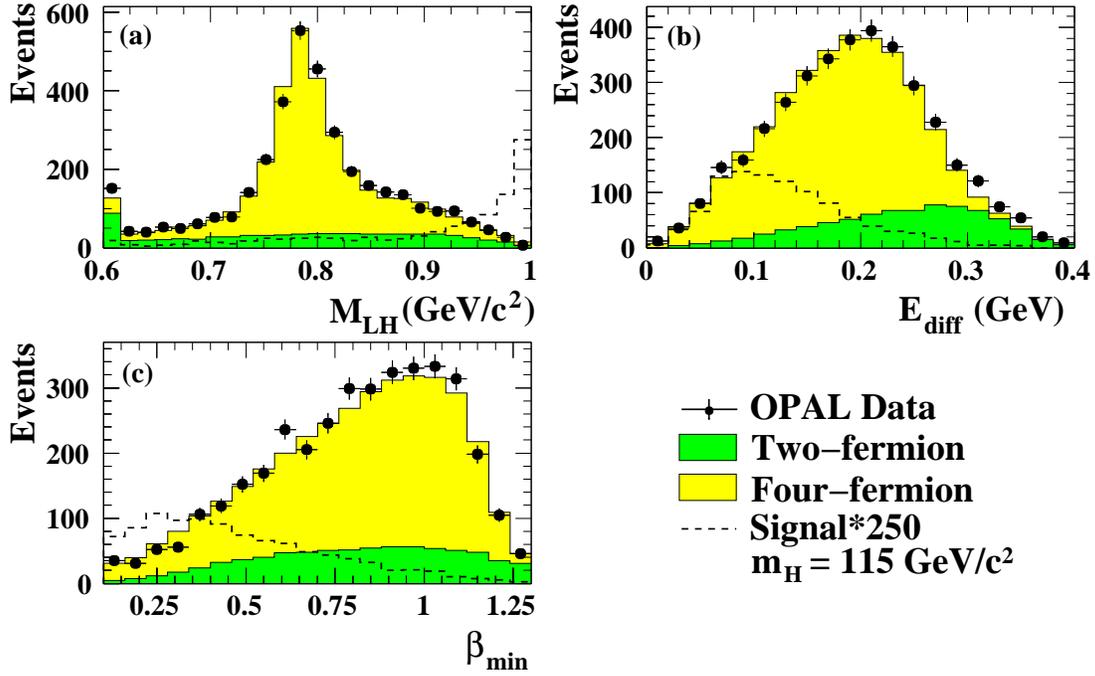,width=0.9\textwidth}}
  \caption[]{\label{fig:fourjetvars3999_2}\sl Four-Jet Channel:
    Distributions of the variables used as input to the likelihood
    function ${\cal L}_2$; (a) $\mathrm{M_{LH}}$, (b)
    $E_{\mathrm{diff}}$, and (c) $\beta_{\mathrm{min}}$. }
\end{figure}

\begin{figure}[p]
  \centerline{
    \epsfig{file=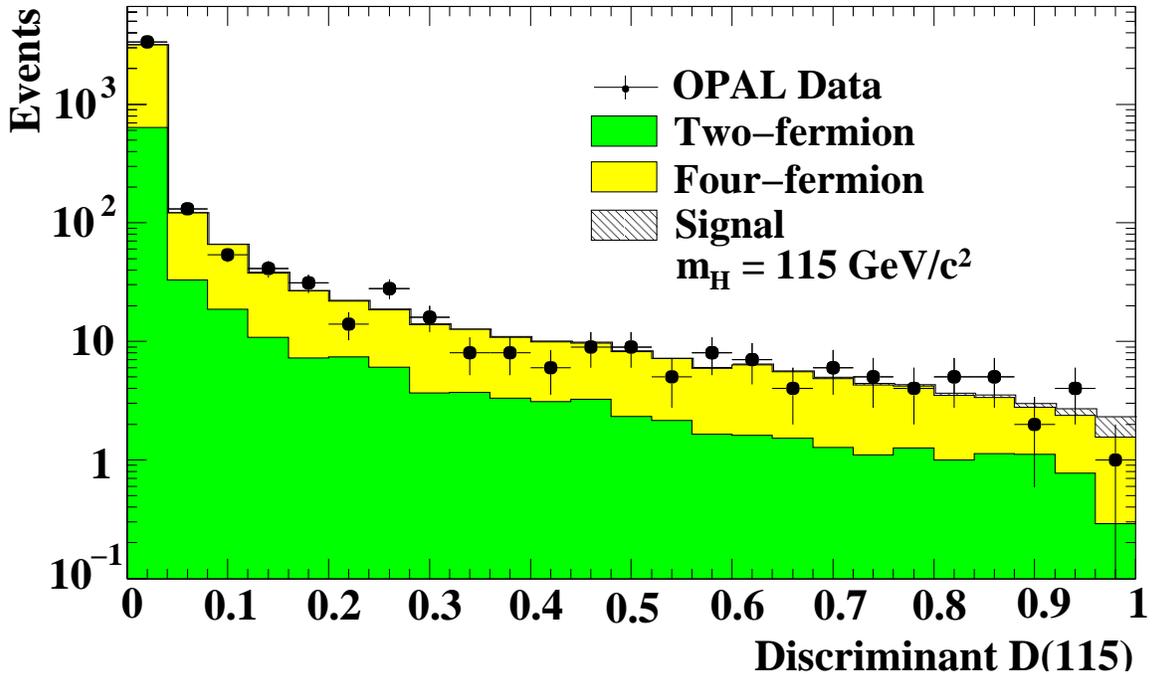,width=0.95\textwidth}}
  \caption[]{\label{fig:fourjlike3999}\sl Four-Jet Channel: Distribution
    of the final discriminating function for a hypothetical \mH=115~\gevcs\
    signal added to the expectations from SM backgrounds.}
\end{figure}

\begin{figure}[p]
\centerline{
  \epsfig{file=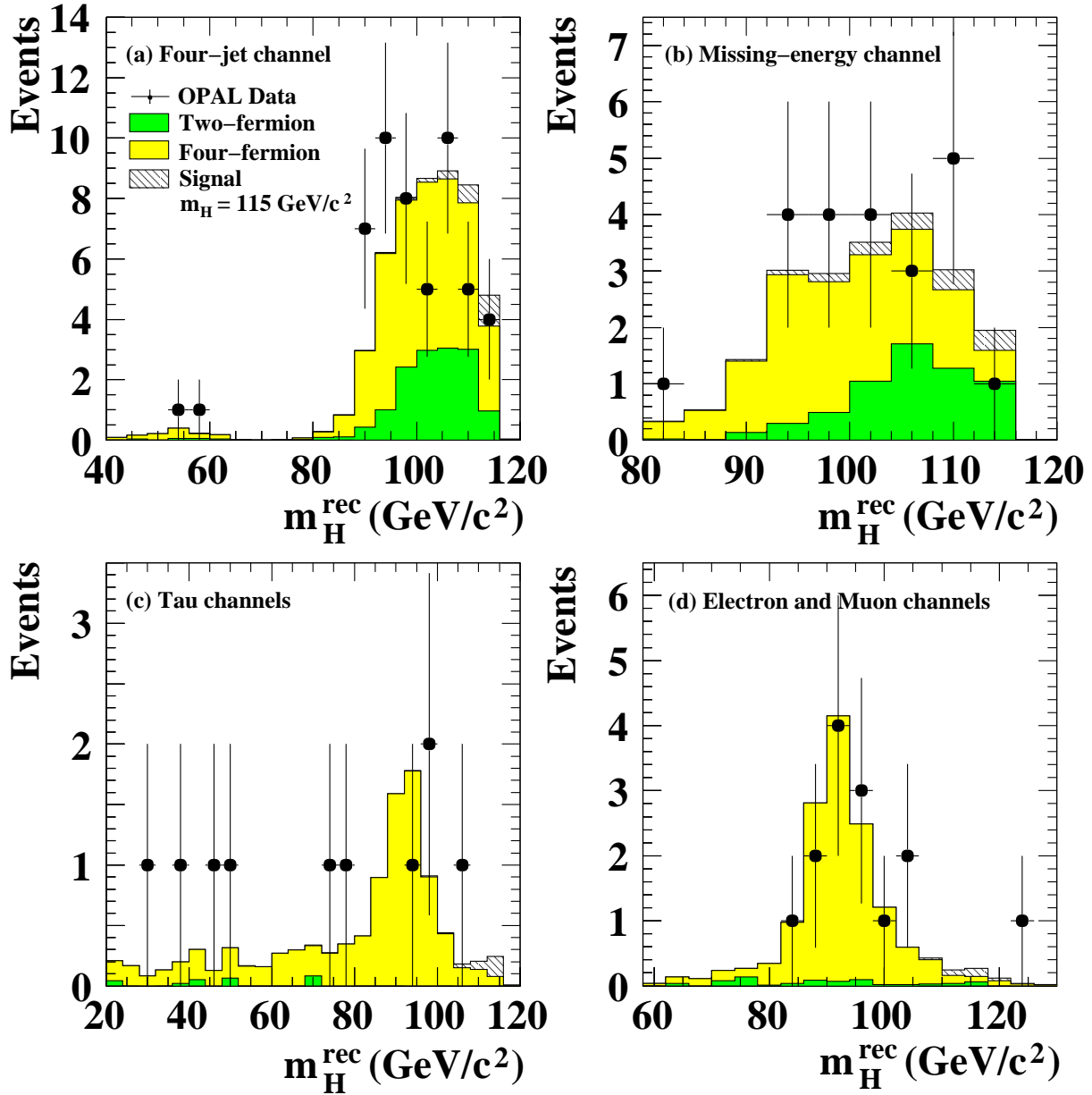,width=\textwidth}}
\vspace*{-0.3cm}
\caption[]{\label{fig:massplot3999}\sl
  Distributions of the reconstructed mass for the selected events
  after a cut on the likelihood/ANN for (a) the four-jet channel, (b)
  the missing-energy channel, (c) the tau channels, and (d) the
  electron and muon channels combined.  }
\end{figure}

The input variables used in the likelihood
${\cal L}_1$ are listed below.
\begin{itemize}
\item ${\cal B}_1$ and ${\cal B}_2$, the b-tagging discriminant variable of the
  jet with the higher and lower energy assigned to the Higgs boson.
\item The jet resolution parameter, $\log_{10}(y_{34})$.
\item The event shape parameter $C$. 
\item $|\cos\theta_{\mathrm Z}|$ defined above in
  Section~\ref{sect:4jpairing}.
\item $\log_{10}({P_{\mathrm{WW}}})$, for the dijet assignment chosen by
      the jet-pairing likelihood algorithm.
\item $J_s$, the sum of the four smallest dijet angles.
\item $\log_{10}(P_{\mathrm{HZ}})$, where $P_{\mathrm{HZ}}$ is
   the probability of a 5C kinematic fit
  constraining two of the jets to have invariant mass \mZ.  The two
  jets are those identified as the \Zo\ jets by the jet-pairing
  likelihood algorithm.
\end{itemize}
The input variables used in the likelihood
${\cal L}_2$ are:
\begin{itemize}
\item $M_{\mathrm {LH}}=(m^{\mathrm{rec}}_{\mathrm{H}}+91.2)/206$~\gevcs, the
  scaled mass of the reconstructed hypothetical Higgs boson.
\item $\beta_{\mathrm {min}}$, the minimum of $\beta_{\mathrm
  {dijet1}}+\beta_{\mathrm {dijet2}}$ for each of the possible dijet
  combinations, where $\beta_{\mathrm {dijet}}$ is the ratio of dijet
  momentum and energy after the 4C fit.
\item $E_{\mathrm{diff}}=(E_{\mathrm{max}}-E_{\mathrm{min}})/\sqrt{s}$,
  the difference of the largest and the smallest jet energies
  normalised to the centre-of-mass energy.
\end{itemize}
The distributions of these input variables are shown in
Figure~\ref{fig:fourjetvars3999} and
Figure~\ref{fig:fourjetvars3999_2} for the data, the SM backgrounds
and for a signal of mass \mH=115~\gevcs.  The distributions of the
resulting discriminator function are shown in
Figure~\ref{fig:fourjlike3999} and of the reconstructed Higgs boson
mass in Figure~\ref{fig:massplot3999}(a) for the data, the SM
background, and for a signal with \mH=115~\gevcs.  For the purpose of
this figure, ${\cal D{\mathrm {( \mH )}}}>0.54$ has been
required. This cut retains 50 events in the data while 45.5$\pm$4.7
are expected on the basis of the SM background simulation.

\subsection{Missing-Energy Channel}
\label{sect:emis}

The \ee\ra\Ho\Zo\ra\bb\nunu\ process is characterised by a substantial
amount of missing energy and a missing mass close to \mZ.  
After preselection, an ANN is used to separate the signal from the
background and forms the mass-independent part of the total
discriminating variable. The mass dependent part is simply the 
reconstructed Higgs boson mass.

\subsubsection{Preselection}
\label{sect:emispresel}

The preselection is designed to remove accelerator-related
backgrounds (such as beam-gas interactions and instrumental noise),
dilepton final states, two-photon processes and radiative \qq\ events,
and to select events with a significant amount of missing energy.
Events with large visible energy in the forward regions of the
detector are also rejected since they are less well modelled and more
likely to have mismeasured missing energies.
\begin{enumerate}
\item Dilepton final states and two-photon processes are reduced by
  the following requirements:
\begin{itemize}
\item The number of tracks passing the quality requirements of 
  Section~\ref{sect:recon} must be greater than six and must
  exceed 20\% of the total number of reconstructed tracks.
\item No track momentum and no energy cluster in the electromagnetic 
  calorimeter may exceed $\sqrt{s}/2$.
\item The visible energy must be less than 80\% of
  $\sqrt{s}$.
\item The energy deposited in either side of the forward calorimeter 
  must not exceed
  2~GeV, and the energy deposits in either side of the silicon-tungsten
  luminometer and the gamma catcher must not exceed 5~GeV.
\item The component of the total visible momentum vector transverse to
  the beam axis must exceed 3~\gevcm.
\item The visible mass must be greater than 4~\gevcs, to
  suppress unmodelled two-photon events.
\item The thrust value $T$ must exceed 0.6.
\item The tracks and clusters in the event are grouped into two jets
  using the Durham algorithm.  The $y_{\mathrm {cut}}$ at which
  the event changes its classification from a two-jet to a
  three-jet event, $y_{23}$, is required to be less than 0.3.
\item The chi-squared of the one-constraint (1C) HZ fit,
  $\chi^2_{\mathrm{HZ}}$, constraining the missing mass 
  to \mZ, is required to be less than 35.
\end{itemize}
\item The energy deposited in the forward region ($|\cos\theta| > 0.9$)
  must not exceed 20\% of the visible energy.
\item  The missing mass must be in the range
  50~\gevcs$<M_{\mathrm{miss}}<130$~\gevcs.
\item The effective centre-of-mass energy \sprime~\cite{l2mh} must
  exceed 60\% of $\sqrt{s}$, to reject events with large
  amounts of initial-state radiation.
\item The acoplanarity angle ($180^\circ$ minus the angle between the
  two jets when projected into the $xy$ plane) must be between  $3^\circ$
  and $100^\circ$, to reject \qq\ events, which often have nearly
  back-to-back jets.  
\item The event must not have any identified isolated
  lepton~\cite{smpaper172}, to reduce the background from
  \WW\ events.
\item To reduce background further, in particular \qq ($\gamma$) 
background, the
  following cuts are applied:
\begin{itemize}
\item The projection of the visible momentum along the beam
  axis, $|P^z_{\mathrm {vis}}|$, must not exceed 25\% of $\sqrt{s}$.
\item The polar angle of the missing momentum vector must lie within
  the region $|\cos\theta_{\mathrm{miss}}|<0.95$ to reject radiative
  events and also to ensure that the missing momentum is not a result
  of mismeasurement.
\item The jet closest to the beam axis is required to have
  $|\cos\theta_{\mathrm {jet}}|<0.95$ to ensure complete
  reconstruction.
\item The polar angle of the thrust axis is required to satisfy
  $|\cos\theta_{\mathrm {thr}}|<0.95$ to ensure good containment of the
  event.
\end{itemize}
\end{enumerate}
The numbers of events remaining after each selection requirement are
given in Table~\ref{tab:smflow1999} for the data taken in 1999 and in
Table~\ref{tab:smflow2000} for the data taken in 2000.  

\subsubsection{Neural Net Selection}
\label{sect:emisann}

The 13 variables used as inputs to the ANN are listed below.  All
variables are scaled to values between zero and one, and some of the
variables with peaking distributions are subject to logarithmic
transformations to give smoother distributions better suited for use
as ANN input variables.

\begin{itemize}
\item The effective centre-of-mass energy \sprime\, divided by $\sqrt{s}$,
\item The missing mass $M_{\mathrm{miss}}$,
\item The polar angle of the missing momentum
  vector $|\cos\theta_{\mathrm{miss}}|$,
\item The thrust value, $\ln{(1-T)}$,
\item The polar angle of the thrust axis $|\cos\theta_{\mathrm{thr}}|$,
\item The jet resolution parameter, $\ln(y_{32})$,
\item The acoplanarity angle of the jets, 
      $\ln{(\phi_{\mathrm{acop}})}$,
\item The polar angle of the jet closest to the beam pipe,
      $|\cos\theta_{\mathrm {jet}}|$,
\item The b-tag likelihood output ${\cal B}_1$ of the more energetic jet,
\item The b-tag likelihood output ${\cal B}_2$ of the less energetic jet,
\item The angle between the more energetic jet and the missing momentum
  vector, \\ $\ln(1-\cos\angle(j_1,p_{\mathrm{miss}}))$,
\item The angle between the less energetic jet and the
  missing momentum vector, \\ $\cos\angle(j_2,p_{\mathrm{miss}})$,
\item The logarithm of the $\chi^2$ of the 1C HZ fit,
  $\ln(\chi^2_{\mathrm{HZ}}$).
\end{itemize}
The jet b-tag variables ${\cal B}_1$ and ${\cal B}_2$ differ from the
ones described in Section~\ref{sect:btag} in that the high-$p_t$
lepton information is not used.  Removing it from the b-tag reduces the
rate at which \WW\ra\qqp\lnu\ events with leptons close to or inside
jets have spurious b-tags. 

\begin{figure}[p]                                
  \centerline{\epsfig{file=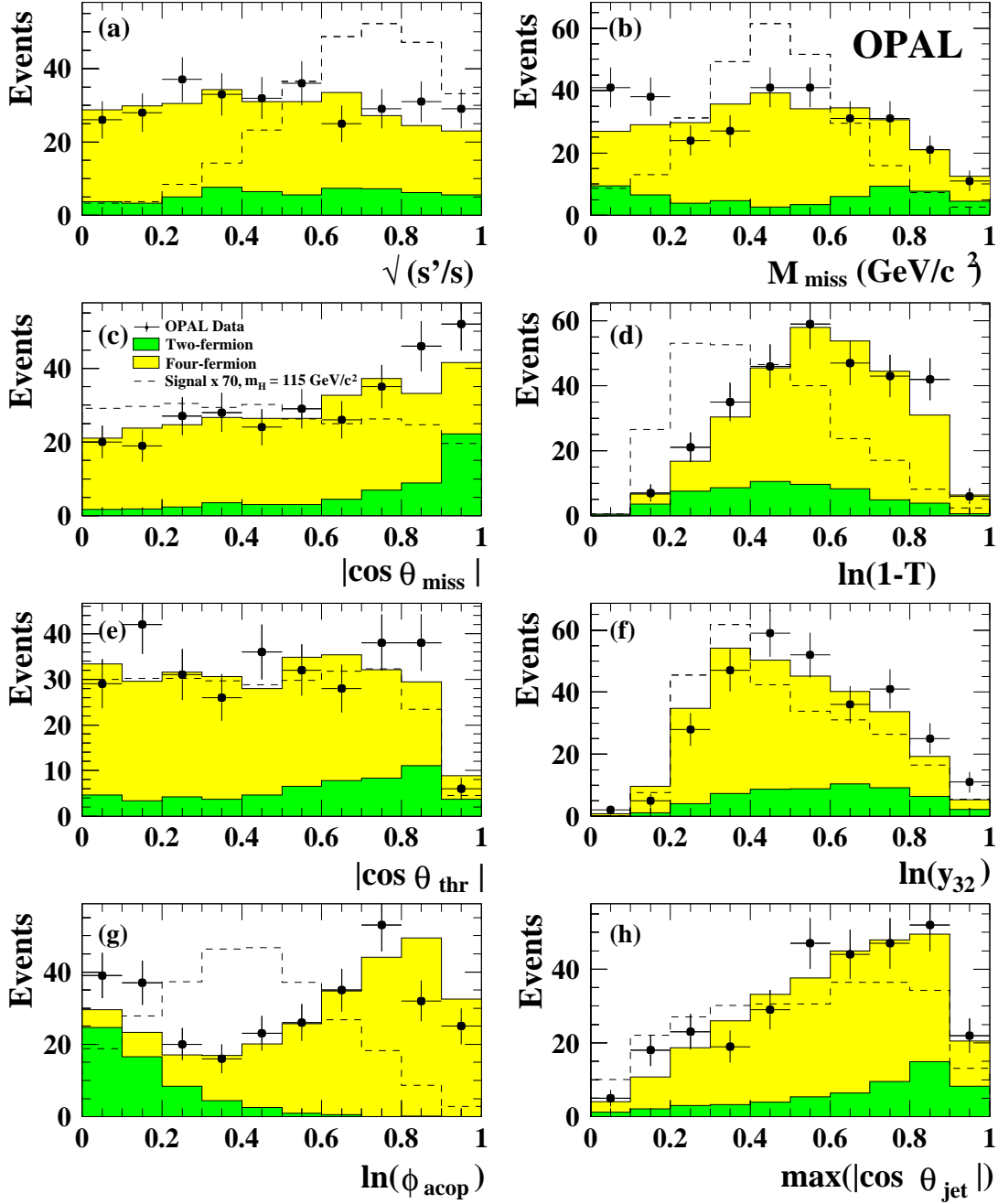,width=0.9\textwidth}}
  \caption[]{\label{fig:emisanninputvars1}\sl Missing-Energy Channel:
    Distributions of the variables used as inputs to the ANN.  (a)
    $\sqrt{s^{\prime}/s}$, (b) $M_{\mathrm{miss}}$, (c)
    $|\cos\theta_{\mathrm{miss}}|$, (d) the logarithm of (1-$T$), (e)
    the cosine of the thrust polar angle, (f) the logarithm of
    $y_{32}$, (g) the logarithm of the acoplanarity, (h) the cosine of
    the polar angle of the jet closest to the beam pipe.}
\end{figure}

\begin{figure}[p]
  \centerline{\epsfig{file=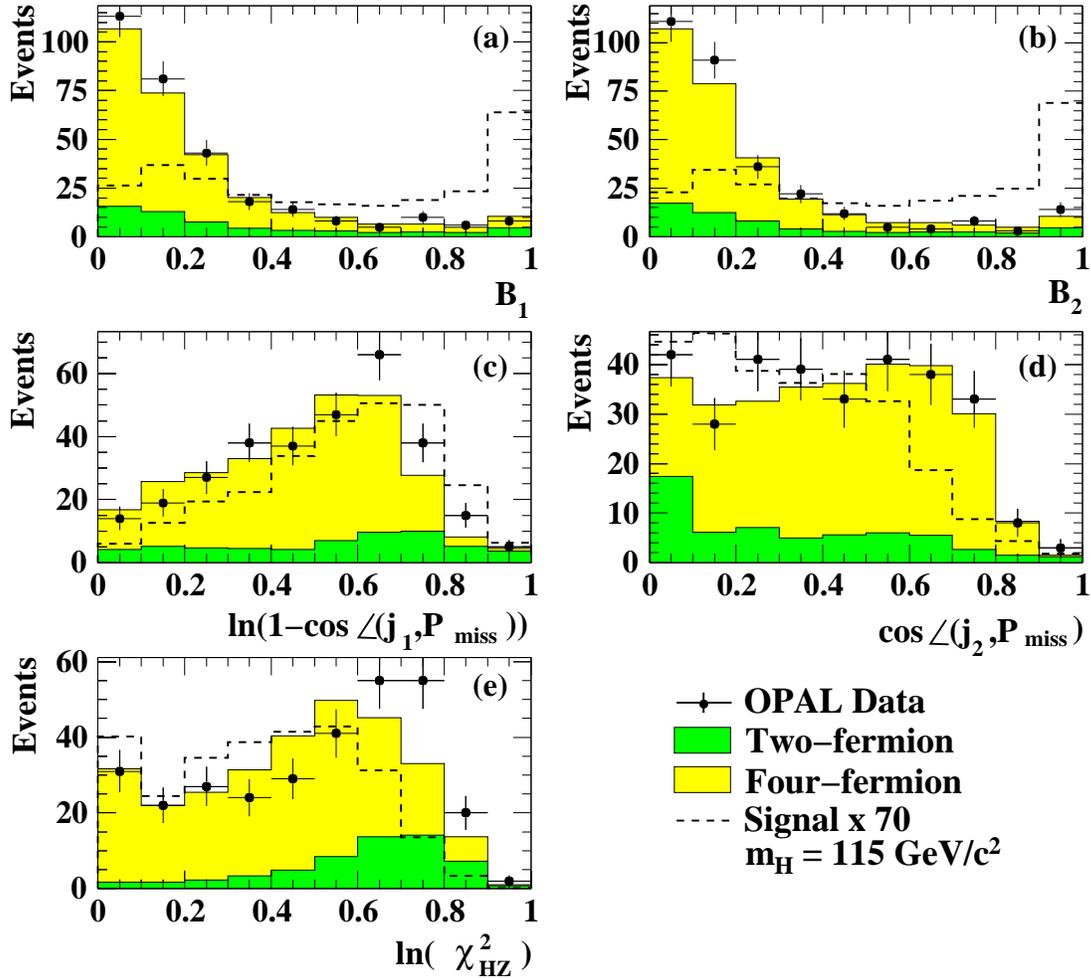,width=0.9\textwidth}}
  \caption[]{\label{fig:emisanninputvars2}\sl Missing-Energy Channel:
    Continuation of Figure~\ref{fig:emisanninputvars1}. (a) and (b)
    the b-tag likelihood output of the more and less energetic jets,
    (c) the logarithm of $(1-\cos\angle(j_1,p_{\mathrm{miss}}))$, (d)
    $\cos\angle(j_2,P_{\mathrm{miss}})$, (e) the logarithm of 
    $\chi^2_{\mathrm{ZH}}$. }
\end{figure}

The distributions of the missing energy channel ANN
input variables for the data, SM background and for a Higgs boson
signal with \mH=115~\gevcs\ are shown in
Figures~\ref{fig:emisanninputvars1} and~\ref{fig:emisanninputvars2}.
The input values of each preselected event are passed on to an ANN,
which is trained to give zero for background and one for signal events.
The ANN's were trained at three centre-of-mass energies, which
cover the range of the data.

\begin{figure}[p]
  \centerline{
    \epsfig{file=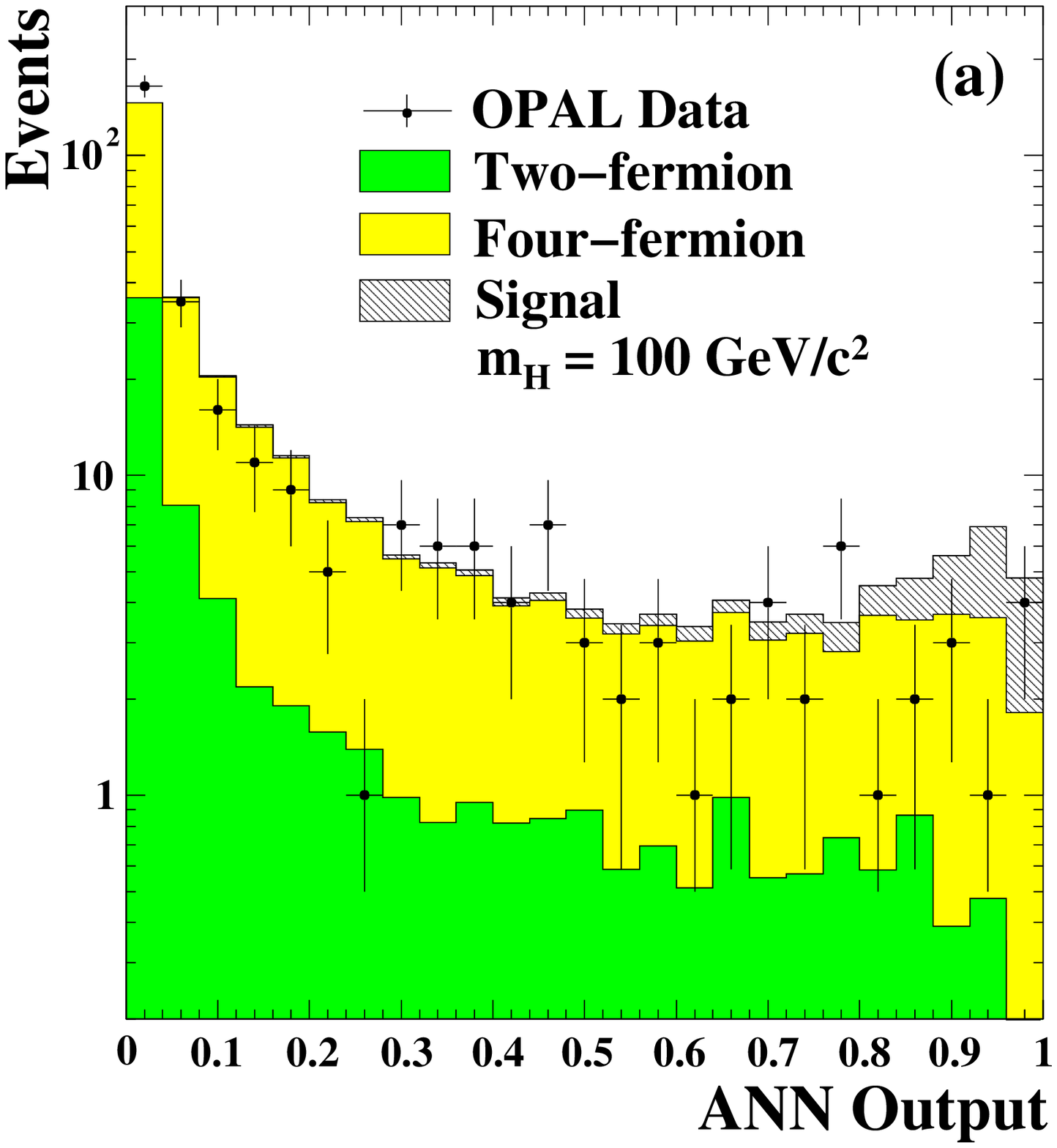,width=0.50\textwidth}
    \epsfig{file=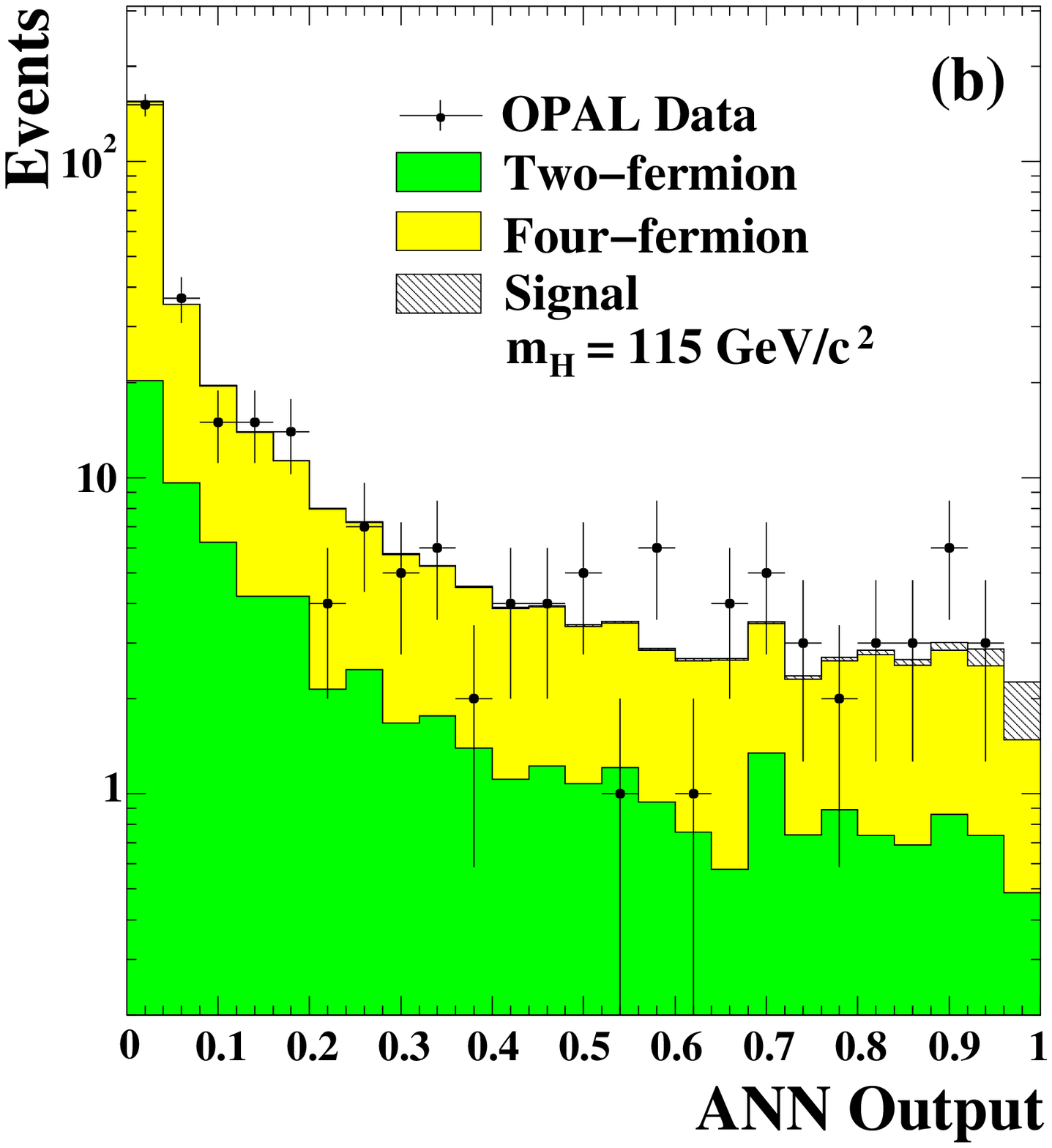,width=0.50\textwidth}}
  \caption[]{\label{fig:emisannoutput3999}\sl Missing-energy channel:
   Distribution of the outputs of the two different ANN's.  Plots (a)
   and (b) shows the output of the ANN trained with \mH=100~\gevcs\
   signal and \mH=110~\gevcs\ signal, respectively.}
\end{figure}

\begin{itemize}
\item At 196 GeV for year 1999 data with $\sqrt{s}=192$ and 196 GeV,
\item At 200 GeV for year 1999 data with $\sqrt{s}=200$ and 202 GeV,
\item At 207 GeV for all year 2000 data. 
\end{itemize}
At each energy, two separate ANNs were trained, one for low Higgs
boson masses ($\mH<107~\gevcs$), and one for high masses
($\mH>107~\gevcs$). The Monte Carlo samples for training contained a
full SM background simulation as well as Higgs signal with
\mH=100~\gevcs\ for the low-mass and \mH=110~\gevcs\ for the high-mass
ANN. Monte Carlo studies have shown this training strategy to be optimal.
The output distributions of the two ANNs in data and Monte Carlo
are shown in
Figure~\ref{fig:emisannoutput3999}(a) and
~\ref{fig:emisannoutput3999}(b) for the years 1999 and 2000.

A cut at ANN$>$0.7  has been used to calculate
all systematic and statistical errors and is used 
in the tables and figures.
This cut retains 21 events in the data while 22.8$\pm$2.7 are expected on the
basis of the SM background simulation.
Figure~\ref{fig:massplot3999}(b)
shows the distributions of the reconstructed mass for the SM background and 
for an \mH=115~\gevcs\ signal.

\subsection{Tau Channels}
\label{sect:tau}

The tau channel selection is sensitive to the two processes
\ee\ra\Ho\Zo\ra\bb\tautau\ and \ee\ra\Ho\Zo\ra\tautau\qq.  The
preselection is designed to identify events with two jets and two tau
leptons. Efficient tau identification is important to separate the
signal from the background which arises mainly from $\qq (\gamma)$,
\ZZ\ and \WW\ processes, with either real tau leptons or with hadronic
jets or isolated tracks with high transverse momentum misidentified as
tau leptons.  The two signal processes have different kinematic and
b-tagging characteristics; therefore two likelihood functions are
created, one optimised for each signal process.  If the event passes
the requirements for either of them, it is counted as a candidate
event.

\subsubsection{Tau Identification}
\label{sect:tauid}

The tau identification procedure~\cite{pr183} makes use of an ANN and
combines the ANN outputs for two oppositely-charged tau candidates
with a likelihood.  The ANN is a track-based algorithm trained to
discriminate real tau decay tracks from tracks arising from the
hadronic system.  
Tracks were not considered if they had a ratio of the momentum
resolution over the momentum, $\sigma_p/p$, larger than 0.5,
if they passed within $1.5^\circ$ of a jet chamber anode plane or
if they were identified as being consistent with coming from a
photon conversion.

To construct the reference histograms a signal sample is used in
which two bosons, both in the mass range of 20 to 170~\gevcs, decay into
\tautau\bb\ so that the networks are trained on a wide variety of
tau momenta. The background sample consists of $\ee\ra\qq$ events.


Any track with momentum greater than $2$~GeV$/c$ and with no other
good track within a narrow cone of half-angle $10^{\circ}$ is
considered as a one-prong tau candidate.  Any family of exactly three
charged tracks, having a total charge of $\pm 1$ and a total momentum
greater than $2$~GeV$/c$, within a $10^{\circ}$ cone centred on the
track with the highest momentum is considered as a three-prong tau
candidate.  Separate ANN's were trained to identify one-prong and
three-prong tau decays.

Around each candidate, an isolation
cone of half-angle $30^{\circ}$ is constructed, concentric with and
excluding the narrow $10^{\circ}$ cone.  Both the one-prong and the
three-prong ANN's use as inputs the invariant mass of all tracks and
neutral clusters (i.e. with no associated tracks) in the $10^{\circ}$ cone, 
the ratio of the total
energy contained in the  $30^{\circ}$ isolation cone to that in the 
$10^{\circ}$ cone, and the total number of tracks and neutral clusters with 
energy greater than $750$~MeV in the isolation cone.  The one-prong ANN
additionally takes as input the total energy in the $10^{\circ}$ cone,
and the track energy in the isolation cone.  The three-prong ANN
additionally uses the largest angle between the most energetic track
and any other track in the $10^{\circ}$ cone.


To select signal candidates, we use the two-tau
likelihood~\cite{pr183},
\begin{equation}
  \mathcal{L}_{\tau\tau}={{\frac{P_1P_2}{P_1P_2+(1-P_1)(1-P_2)}}},
\end{equation}
  where $P_i$ is the probability that the $i^{\mathrm {th}}$ tau
candidate originates from a real tau lepton.  This probability is
calculated from the shapes of the ANN output for signal and fake taus.
The distribution of the ANN output for signal events was computed from
Monte Carlo simulations.  
The distribution of the ANN output in hadronic \Zo\ decay data
collected in the calibration run at $\sqrt{s}\approx 91$~GeV, which
has a low fraction of events with real taus, was used as the reference
distribution for fake taus. This estimation of the fake tau rate from
data reduces the corresponding systematic uncertainty.  To pass the
selection, an event must have $\mathcal{L}_{\tau\tau}$ of at least
0.10.


\subsubsection{Preselection}
\label{sect:taupresel}

The preselection requirements are as follows.
\begin{enumerate}
\item A basic selection is made to ensure well-measured
  events and reject accelerator backgrounds, dilepton events,
  two-photon events, and two-fermion events with ISR.  
\begin{itemize}
\item The event must satisfy the standard hadronic final state requirement~\cite{l2mh}.
\item To ensure that the selected events are well measured, the
  magnitude of the missing momentum $p_{\mathrm{miss}}$ has to be less
  than $0.3 \sqrts$.
\item The direction of the missing momentum must be well within the
  detector acceptance, $\left| \cos\theta_{\mathrm{miss}} \right| \leq
  0.95$, in order to reject events with substantial ISR energy
  escaping undetected along the beam axis.
\item The scalar sum of the transverse momenta of the particles in the
  event must exceed 45~GeV$/c$.
\end{itemize}
\item At least one pair of oppositely charged tau candidates must be
  identified by the two-tau likelihood.

\item The particles in the events are subdivided into two tau
  candidates and two jets.  A two-constraint kinematic fit is applied using 
  total energy and momentum conservation constraints, where the tau momentum
  directions are taken from their visible decay products while leaving
  the energies free.  The $\chi^2$-probability of the fit is
  required to be larger than $10^{-5}$.
\item In events where both taus are classified as one-prong decays,
  the sum of the momenta of the charged particles assigned to the tau
  decays must be less than 80~GeV$/c$, in order to reduce backgrounds
  from $\ZZ\ra\mu^+\mu^-\qq$ and $\ZZ\ra{\mathrm{ e^+e^-}}\qq$.
\end{enumerate}

\subsubsection{Likelihood Selection}
\label{sect:taulikelihood}

Two final likelihoods are then constructed from reference histograms.
One of them, $\mathcal{L}( \bb\tautau )$, is optimised for the
$\Ho\Zo\ra\bb\tautau$ final state and makes use of the b-tag of
Section~\ref{sect:btag}.  The other, $\mathcal{L}( \tautau\qq )$, is
optimised for the $\Ho\Zo\ra\tautau\qq$ process and does not use
b-tagging.  The following variables serve as inputs to both likelihoods:
\begin{itemize}
\item The ratio of the visible energy and the beam energy, \rvis = \evisn,
\item The polar angle of the missing momentum
      $\left| \cos \theta_{\mrm{miss}} \right|$,
\item $\log_{10}{(y_{34})}$, using the Durham algorithm, and
  allowing the tau candidates to be identified as low-multiplicity
  jets,
\item The angles between each tau candidate and the nearest jet,
\item The logarithm of the larger of two 3C kinematic fit
  probabilities, in which the additional constraint comes from fixing
  either the tau pair invariant mass or the jet pair invariant mass to
  the \Zo\ mass,
\item The tau likelihood, $1-\sqrt{1-\mathcal{L}_{\tau\tau}}$,
\item The energy of the most energetic electron or muon identified in
  the event (if any),
\item $\sqrt{P_{\mathrm {joint}}}$, where  $P_{\mathrm {joint}}$ is the probability that the $N$ tracks identified as the products of the tau decay
 come from the primary vertex. This is used to 
    separate real taus from prompt electrons, muons and hadronic 
    fake taus~\cite{aleph_btag}. 
%
\end{itemize}
In addition to the above variables, the $\mathcal{L}(
\bb\tautau )$ likelihood uses as an input $1-\sqrt{1-{\cal B_{\tau}}}$, where
\begin{equation}
{\cal B_{\tau}} = \frac{{\cal B}_1{\cal B}_2}
                 {{\cal B}_1{\cal B}_2+(1-{\cal B}_1)(1-{\cal B}_2)},
\end{equation}
and ${\cal B}_1$ and ${\cal B}_2$ are the b-tag discriminating variables
described in Section~\ref{sect:btag} for the two hadronic jets in the
event.  

\begin{figure}[p]
\centerline{\epsfig{file=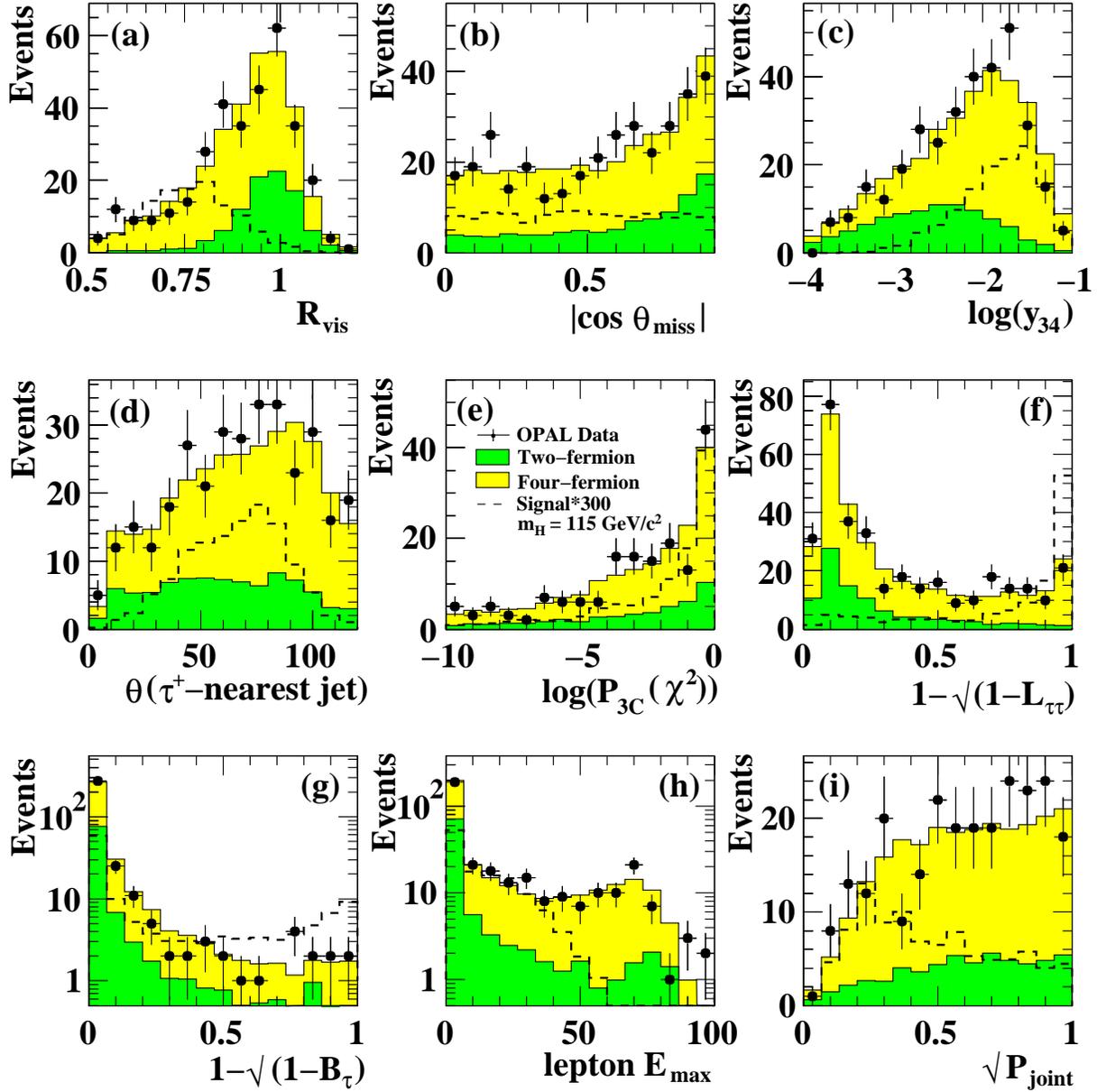,width=1.0\textwidth}}
\caption[]{\label{fig:tau1}\sl Tau channels: Distributions of the
  likelihood input variables, (a) $R_{\mathrm{vis}}$, (b)
  $|\cos\theta_{\mathrm{miss}}|$, (c) $\log_{10}(y_{34})$, (d) the
  angle between the $\tau^+$ candidate and the nearest jet in degrees,
  (e) the logarithm of the larger of two 3C fit probabilities, fixing
  either the dijet mass or the tau pair mass to \mZ, (f) the combined
  tau likelihood, (g) the combined b-tag, (h) the lepton
  $E_{\mathrm{max}}$, and (i) $\sqrt{P_{\mathrm {joint}}}$.}
\end{figure}

\begin{figure}[p]
\centerline{
  \epsfig{file=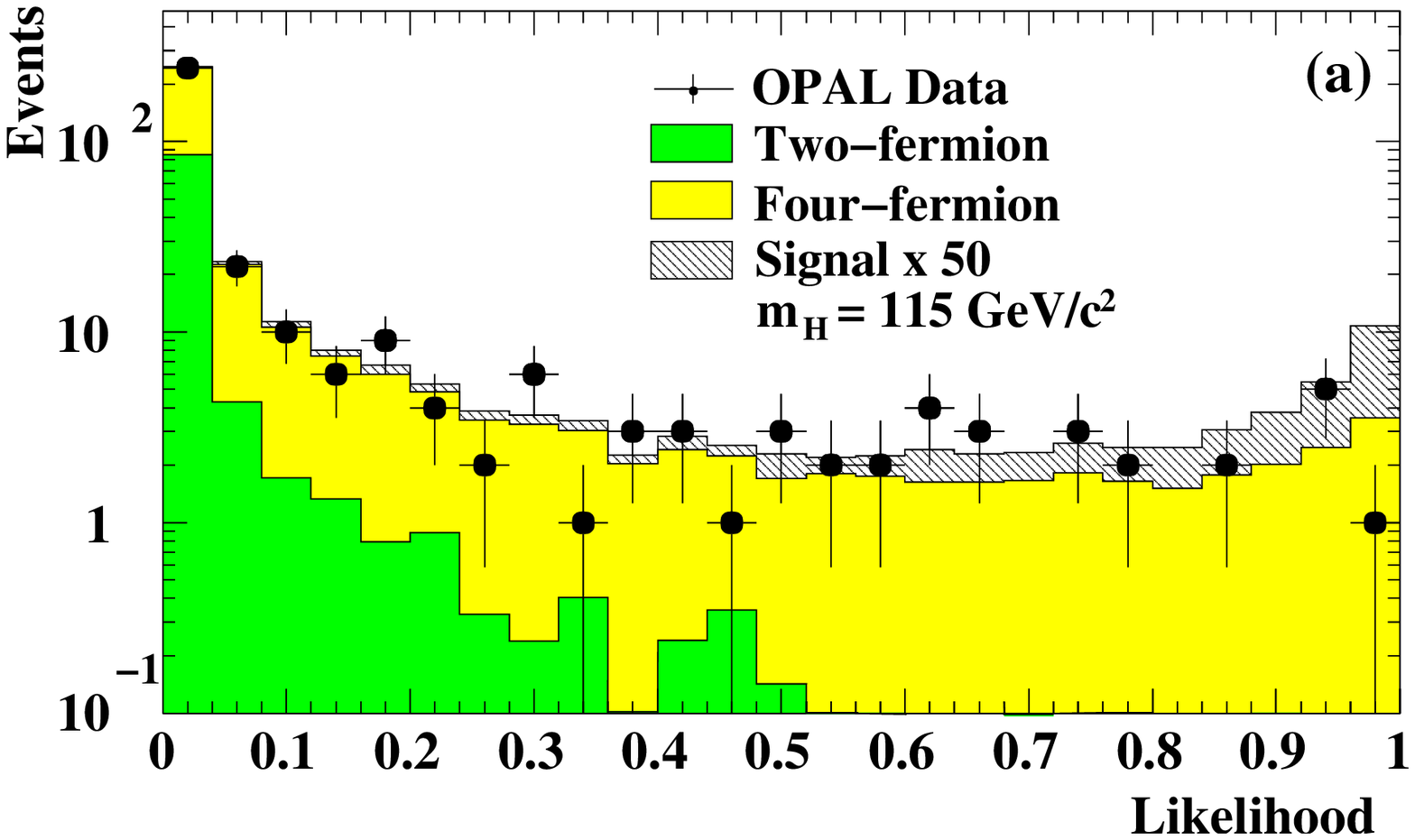,width=0.85\textwidth}}
\vspace{0.5cm}
\centerline{
  \epsfig{file=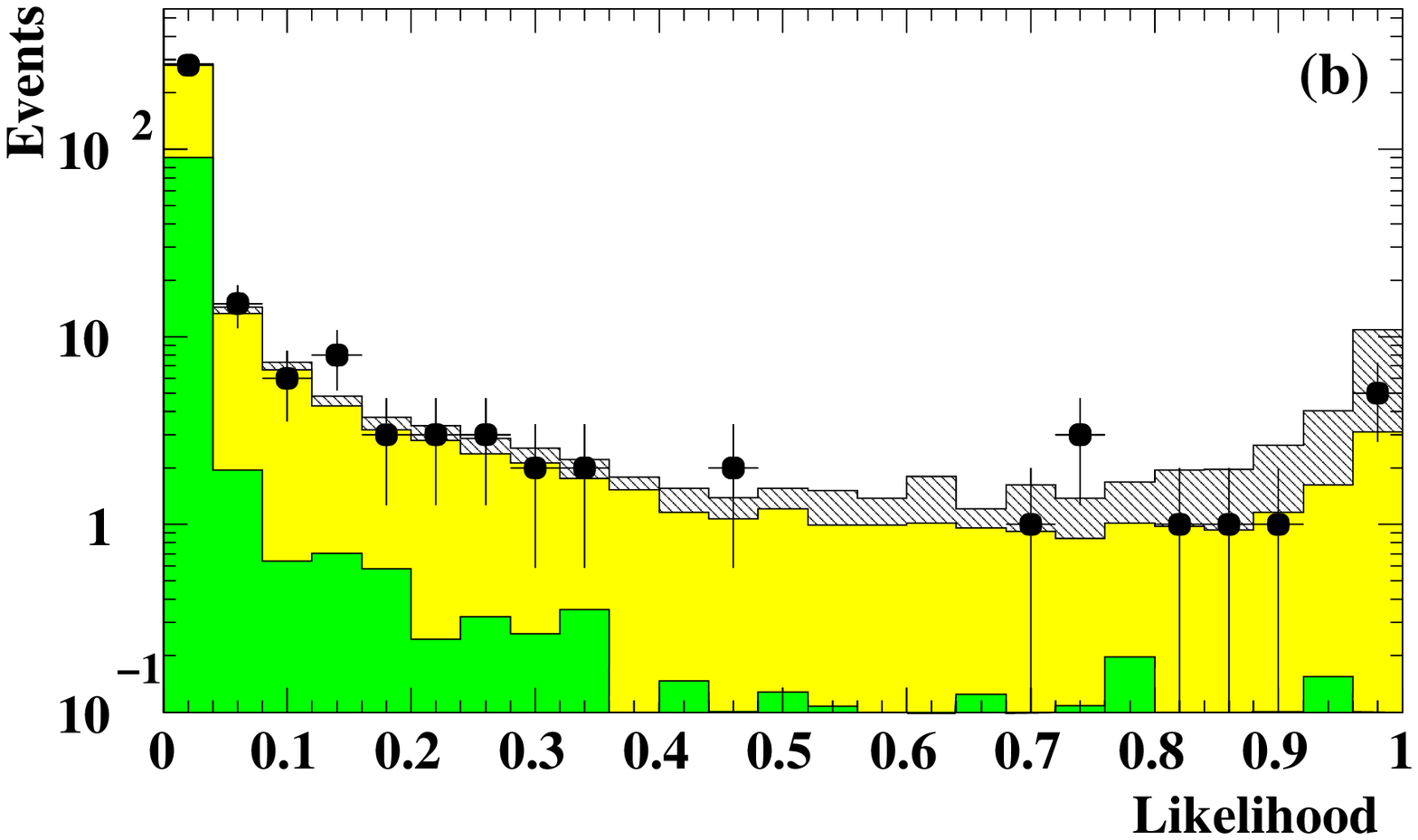,width=0.85\textwidth}}
\caption[]{\label{fig:tau3}\sl Tau channels: Distributions of the two
  likelihood output variables (a) ${\cal L}(\tau\tau\qq)$ and (b)
  ${\cal L}(\bb\tau\tau)$. }
\end{figure}

The reference histograms for $\mathcal{L}( \bb\tautau )$
include contributions only from signal and background events
containing B hadrons, while the reference histograms for $\mathcal{L}(
\tautau\qq )$ include contributions only from events without B
hadrons.  Distributions of the likelihood input variables for the
data, SM backgrounds,
and a signal with \mH=115~GeV, are shown in
Figure~\ref{fig:tau1} for the year 2000 data sample.  The distributions of
the likelihood discriminants $\mathcal{L}( \bb\tautau )$ and
$\mathcal{L}( \tautau\qq )$ are shown in Figure~\ref{fig:tau3}.  All
events passing the preselection are retained for the 
calculation of the confidence levels.
 
As discussed in Section 4, the discriminator used for the confidence
level calculation is composed of mass-dependent and mass-independent
parts. In this channel, the reconstructed Higgs boson mass is used for
the mass-dependent part, and either the \bb\tautau\ or the \tautau\qq\ 
likelihood is used for the mass-independent part. The choice of the
likelihood is based on a study performed on $\Ho\Zo\ra\tautau\qq$ and
$\Ho\Zo\ra\bb\tautau$ signal events.  The distribution of the signal
events in the $\mathcal{L}( \tautau\qq )$ versus $\mathcal{L}(
\bb\tautau)$ plane shows two well separated regions.  The likelihood
to be used for each candidate is chosen according to the region 
of the likelihood plane in which it falls.


For the calculation of systematic errors and for illustration purposes
in tables and figures, an event is selected as a candidate if either  
$\mathcal{L}(\bb\tautau)>0.92$ 
or  $\mathcal{L}(\tautau\qq )>0.88$. 
The distributions of the reconstructed mass for the data, the Standard
Model backgrounds, and a signal of mass \mH=115~\gevcs\  
are shown in Figure~\ref{fig:massplot3999}(c).

The numbers of observed and expected events after each stage of the
selection are given in Table~\ref{tab:smflow1999}, together with the
detection efficiency for a 100~\gevcs\ SM Higgs boson, for the data
taken in 1999, and in Table~\ref{tab:smflow2000}, with the detection
efficiency for a 115~\gevcs\ SM Higgs boson, for the data taken in
2000.  Ten events survive the likelihood cut, to be compared to
the expected background of $9.6  \pm 1.1$.

\subsection{Electron and Muon Channels}
\label{sect:lepton}
The signature for the Higgs signal in the light-lepton channels,
$\ee\ra\Ho\Zo\ra\bb\ee$ and $\bb\mm$,
is two hadronic jets from
the decay of the \Ho\ and two leptons, either electrons or muons, 
from the decay of the \Zo.  The
hadronic jets are expected to contain B hadrons, and the invariant
mass of the leptons is expected to be close to the \Zo\ mass.  The
main background to these channels arises from \ZZ\ production, in which
one \Zo\ boson decays leptonically; the background from
$\qq (\gamma)$ plays only a minor role.

\subsubsection{Preselection}
\label{sect:leptonpresel}

The preselection is intended to enhance the \ee\qq\ and \mm\qq\ topologies.
\begin{enumerate}
\item The event must have at least six charged tracks, a jet resolution
  parameter  $y_{34} >
  10^{-4}$ (Durham scheme), $|P^z_{\mathrm{vis}}| < (E_{\mathrm{vis}}-0.5\sqrt{s})$ and
  $E_{\mathrm {vis}} > 0.6\sqrt{s}$.  These requirements select
  multihadronic events, and reduce the backgrounds from two-photon and
  $\qq (\gamma)$ events with large amounts of ISR along the beam axis.
\item At least one pair of oppositely charged leptons of the same
  flavour (e or $\mu$) must be identified. The identification of muons
  is described in~\cite{oldline} and electrons are identified using
  the information on the association between tracks and the
  electromagnetic calorimeter clusters, as described
  in~\cite{smpaper172}.
\item The events are reconstructed as two leptons and two jets.  In
  the case of the muon channel, a 4C kinematic fit, requiring total
  energy and momentum conservation, is applied to improve the mass
  resolution of the muon pair; the $\chi^2$ probability of the fit  
  is required to be larger than 10$^{-5}$. 
  The energy of an electron candidate is obtained from the energy of the
  electromagnetic cluster associated with the track, while for a muon
  candidate it is approximated by the track momentum.
  The invariant mass of the lepton pair is
  required to be larger than 40~\gevcs.  
\end{enumerate}

\subsubsection{Likelihood Selection}
\label{sect:leptonlikelihood}

\begin{figure}[p]
  \centerline{\epsfig{file=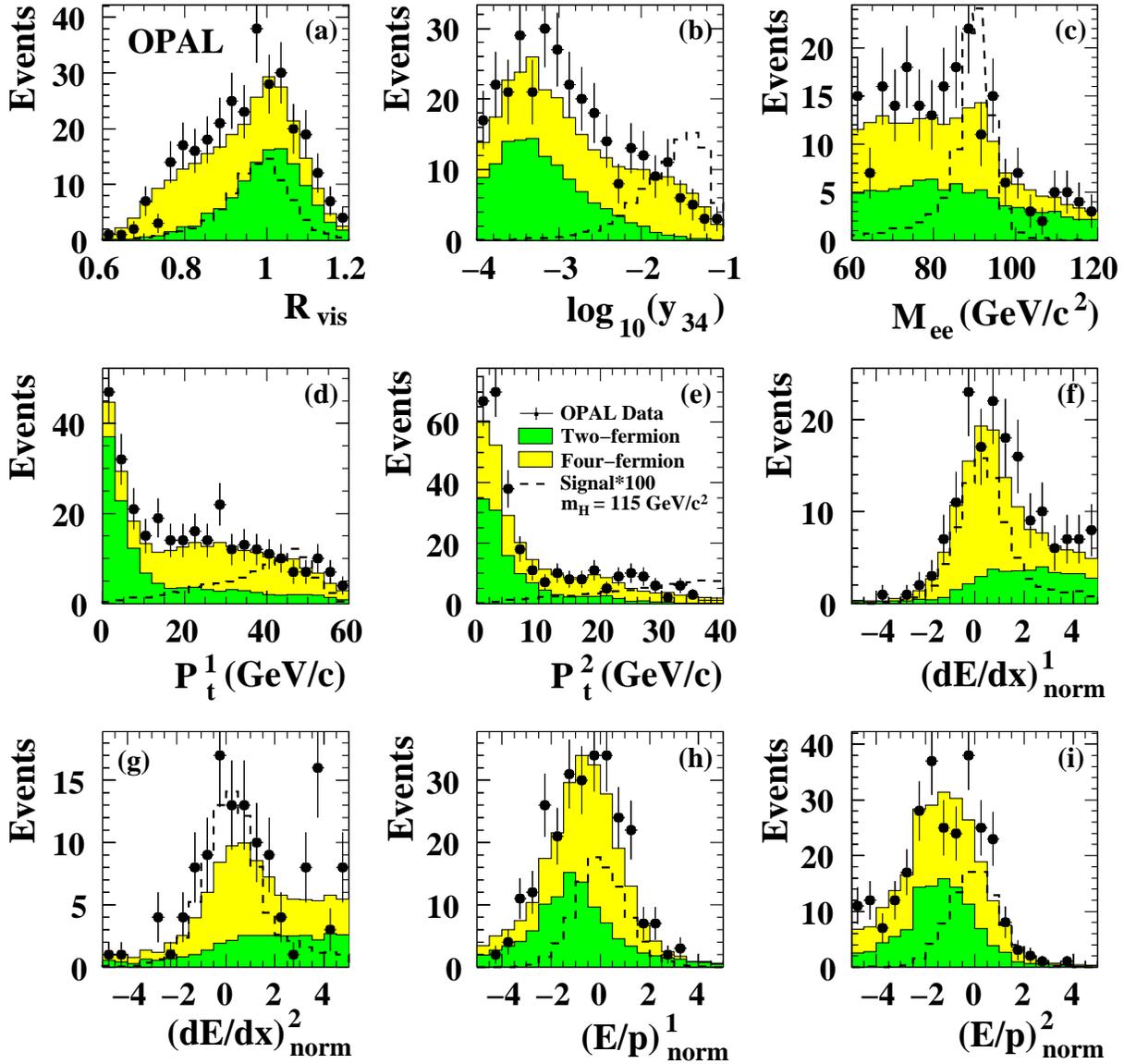,width=1.0\textwidth}}
  \caption[]{\label{fig:electron1}\sl Electron channel: Distributions of
    the variables used as input to the selection likelihood function.
    (a) The distribution of the ratio of the visible energy and the
    energy in the centre-of-mass, (b) $\log_{10}(y_{34})$, (c) the
    invariant mass of the two electrons, (d) and (e) the transverse
    momentum of the more and less energetic electron, respectively,
    (f) and (g) the normalised ionisation energy loss of the more and
    less energetic electron, (h) and (i) $(E/p)_{\mathrm{norm}}$, of the
    more and less energetic electron. }
\end{figure}

\begin{figure}[p]
  \centerline{\epsfig{file=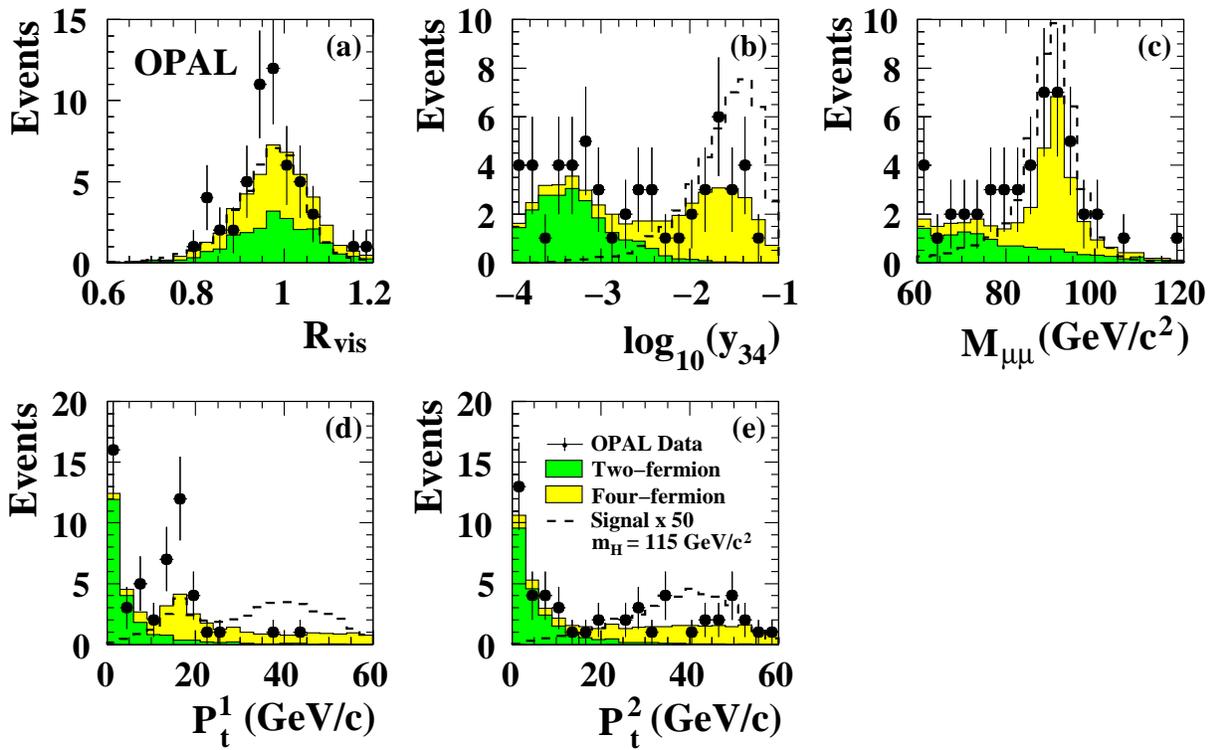,width=1.0\textwidth}}
  \caption[]{\label{fig:muon1}\sl Muon channel: Distributions of the
    variables used as input to the selection likelihood function. (a)
    The distribution of the ratio of the visible energy and the energy
    in the centre-of-mass, (b) $\log_{10}(y_{34})$, (c) the invariant
    mass of the two muon candidates, (d) and (e) the transverse
    momentum of the more and less energetic muon, respectively. }
\end{figure}

Two likelihoods are calculated, one based on kinematic variables,
$\cal K$, and one based on b-tagging in the two hadron jets, ${\cal
B}_{\mathrm{2jet}}$.  The kinematic likelihood is formed using the
variables:
\begin{itemize}
\item The ratio of the visible energy and the beam energy,
     $R_{\mathrm{vis}}=E_{\mathrm{vis}}/\sqrts$,
\item $\log_{10}(y_{34})$, using the Durham algorithm, 
\item The transverse momenta of the two leptons calculated with respect to 
      the nearest jet axis,
\item The invariant mass of the two leptons.  
\end{itemize}
For the electron channel, the following
electron identification variables are used for both electron
candidates:
\begin{itemize}
\item The normalised energy and momentum ratio, $(E/p)_{\mathrm{norm}}\equiv
[(E/p)-1]/\sigma$, where $E$ and $p$ are cluster energies and track
momenta, and $\sigma$ is the error in $E/p$ obtained from
the measurement errors of $E$ and $p$,
\item The normalised ionisation energy
loss, $(dE/dx)_{\rm norm}=\left[(dE/dx)-(dE/dx)_{\rm
nominal}\right]/\sigma$, where $(dE/dx)$ is the ionisation
energy loss in the jet chamber, $(dE/dx)_{\rm nominal}$ is the nominal
ionisation energy loss for an electron, and $\sigma$ is the
error of $(dE/dx)$.
\end{itemize}
The distributions of the input variables to the ${\cal K}$ likelihood are 
shown in Figure~\ref{fig:electron1} and in Figure~\ref{fig:muon1} for the
electron and muon channels, respectively.

The b-flavour requirement, ${\cal B}_{\rm 2jet}$, is the likelihood
obtained by combining in one single likelihood the b-probabilities of
the two jets.  The background is weighted according to the branching
fractions for \Zo\ decay since the dominant background arises from
\ZZ\ production.

The signal likelihood is given by combining ${\cal K}$ and 
${\cal B}_{\rm 2jet}$:

\begin{equation}
 {\cal L} = \frac{{\cal K} \cal B_{\mathrm{2jet}}}
{{\cal K} {\cal B}_{\rm 2jet} + (1 - {\cal K})(1 - {\cal B}_{\rm
    2jet})} 
\end{equation}

\begin{figure}[p]
  \centerline{\epsfig{file=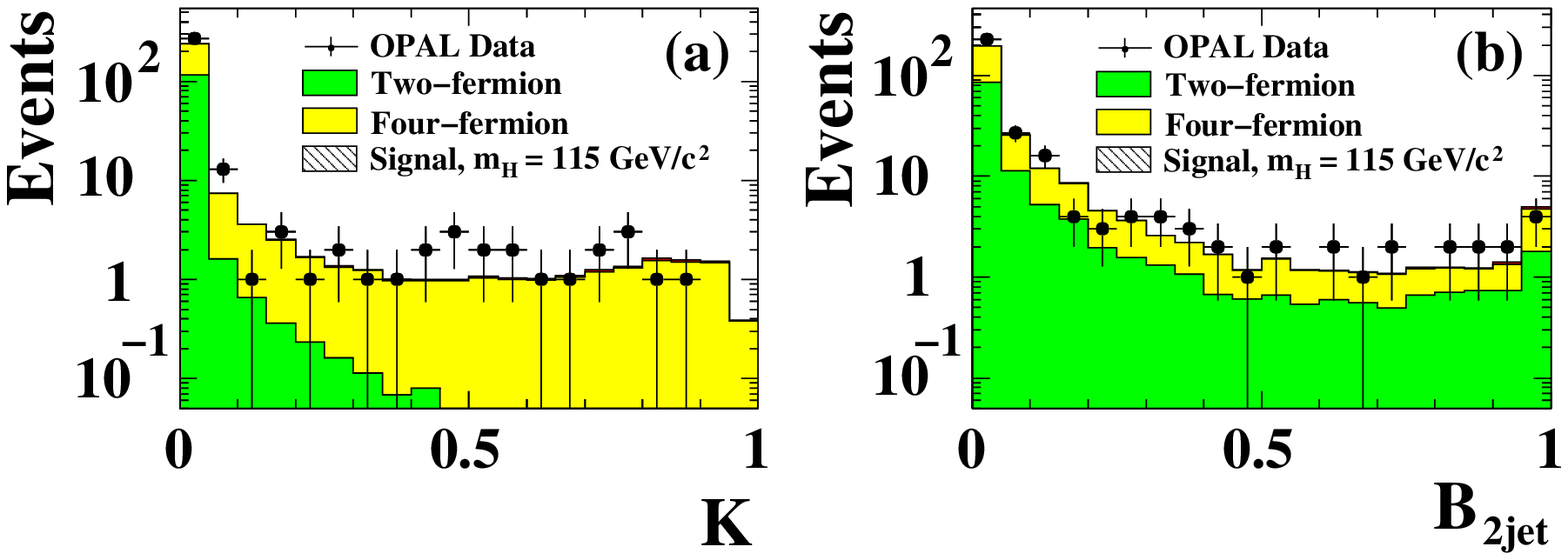,width=0.95\textwidth}}
  \centerline{
    \epsfig{file=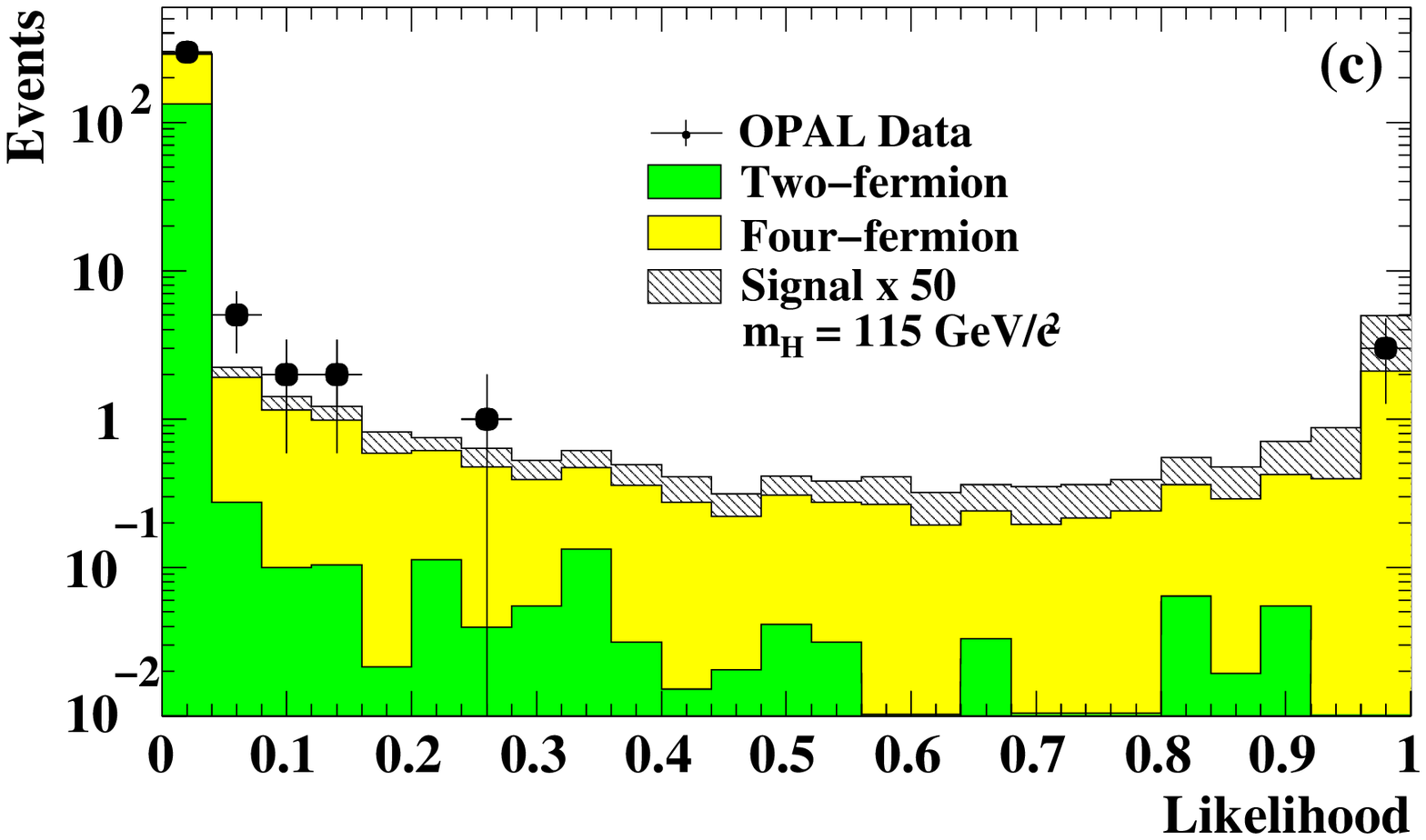,width=0.9\textwidth}}
  \caption[]{\label{fig:electron2}\sl Electron channel: Plots (a), (b)
    and (c) show the distributions of the ${\cal K}$, ${\cal
    B}_{\mathrm{2jet}}$ and ${\cal L}$ likelihoods, respectively.}
\end{figure}

\begin{figure}[p]
\centerline{\epsfig{file= 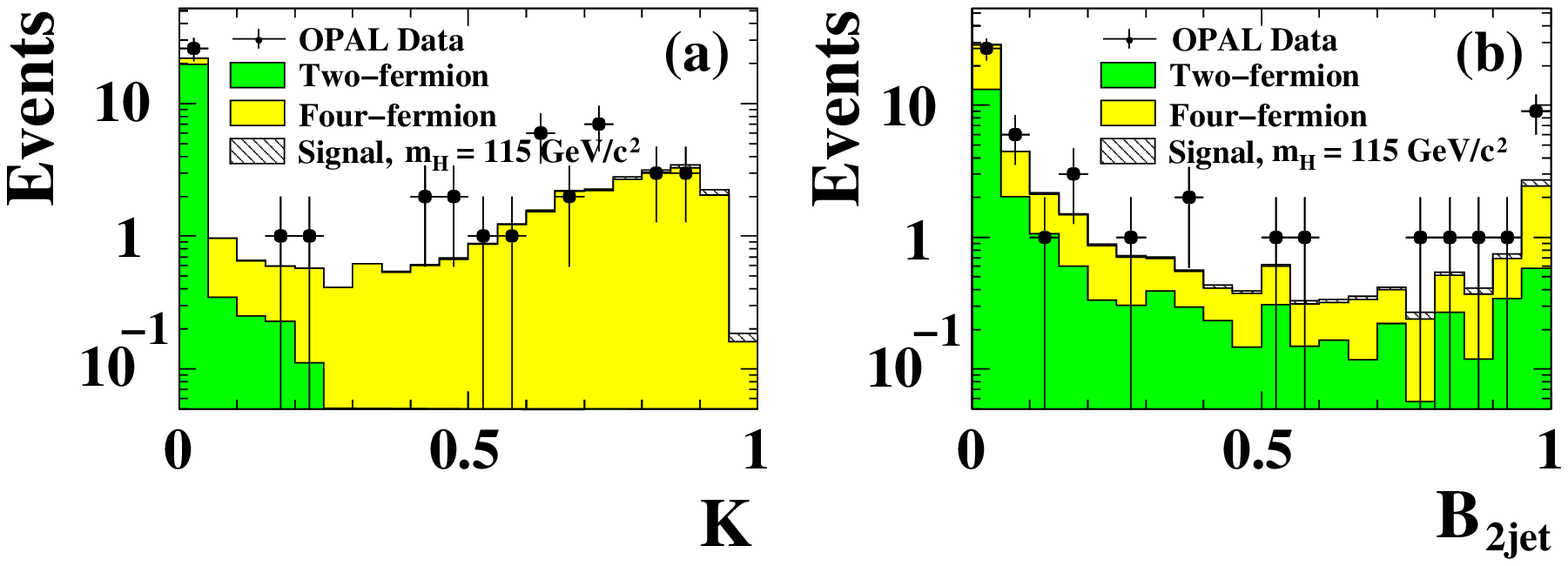,width=1.\textwidth}}
\centerline{
\epsfig{file=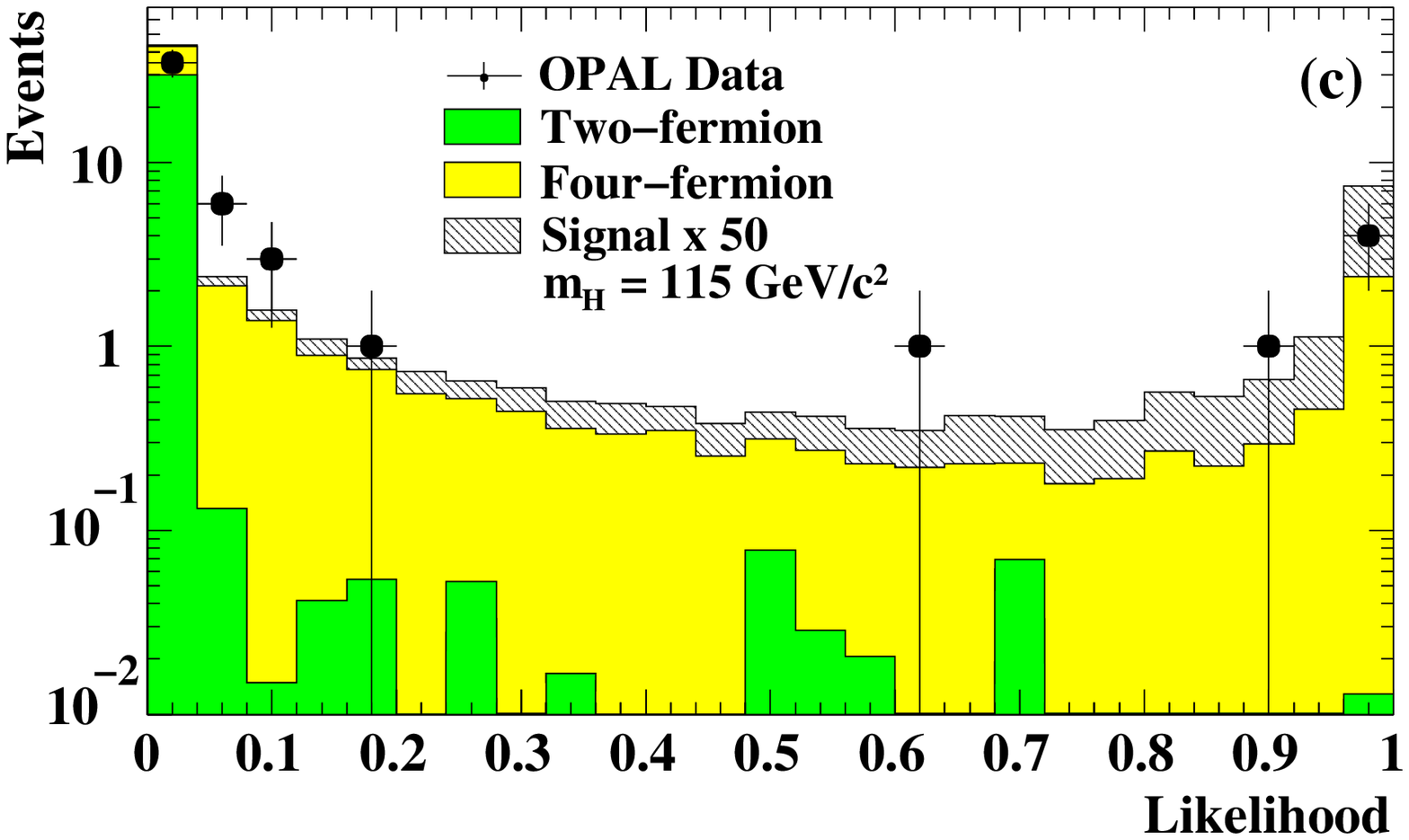,width=0.9\textwidth}}
\caption[]{\label{fig:lepton_lhout2}\sl Muon Channel: Plots (a), (b) and
  (c) show the distribution of the ${\cal K}$, ${\cal
  B}_{\mathrm{2jet}}$ and ${\cal L}$ likelihoods, respectively. }
\end{figure}

The distributions of ${\cal L}$ are shown in Figure~\ref{fig:electron2}(c) 
and Figure~\ref{fig:lepton_lhout2}(c) for the electron and muon channels,
respectively. 


In this channel, the reconstructed Higgs boson mass is used for the 
mass-dependent part of the discriminator calculation, and the 
likelihood output ${\cal L}$ is used for the mass-independent part.

Likelihood cuts at 0.2 for the electron channel and 0.3 for the muon
channel are introduced for the calculation of systematic errors and for
illustration of event rates and distributions in tables and figures.
The reconstructed Higgs boson mass \mHrec\ is given by the recoil mass
to the lepton pair and is shown in Figure~\ref{fig:massplot3999}(d).

The numbers of observed and expected events after each stage of the
selection are given in Tables~\ref{tab:smflow1999}
and~\ref{tab:smflow2000} for the data taken in 1999 and 2000,
respectively.  The selection retains 4 events in the
electron channel, while one expects a background rate of
7.7$\pm$1.4 events. In the muon channel 10 events remain, with
an expectation from background of 6.7$\pm$1.0 events.
\subsection{Systematic Uncertainties}
\label{sect:syst}

A large variety of systematic uncertainties have been
investigated. Many of them affect both the estimates of the signal
efficiencies and those of the background rates. Some of them are
shared between decay channels, or between data sets taken at different
centre-of-mass energies (for example the modelling of the variables
with the same Monte Carlo generators), while others are not. For this
reason, the relative signs of the errors as well as their absolute size
are relevant in the statistical limit
setting procedure discussed in Section~\ref{limitcalcappendix} below.

The different sources of systematic errors which have been considered
are enumerated below and listed, with their relative signs, in
Table~\ref{tab:systable}.  Note that in the statistical procedure
errors from different sources are considered to be uncorrelated.
\begin{itemize}
\item {\bf Monte Carlo statistics}: These uncertainties affect the
  signal and background rates and are uncorrelated between channels,
  energies, and signal and background.
\end{itemize}
The following uncertainties are correlated between all channels and
centre-of-mass energies:
\begin{itemize}
\item {\bf Tracking resolution in
    \begin{boldmath}$r\phi$\end{boldmath}}: This uncertainty is
  evaluated with the Monte Carlo simulation by multiplying the
  discrepancy between the true and reconstructed values of the track's
  impact parameter in the $r\phi$ plane, azimuthal angle $\phi$ and
  curvature by smearing factors of 1.05 and comparing efficiencies to
  the simulation without extra smearing. The smearing factor 1.05
  adequately covers the discrepancies seen in Figure~\ref{fig:btag}(b)
  and (d).
 
  
\item {\bf Tracking resolution in \begin{boldmath}$z$\end{boldmath}}:
  This uncertainty is evaluated by treating the track impact parameter
  in $z$ and $\tan\lambda=\cot\theta$ in the same way as described
  above, again using smearing factors of 1.05.
  
  
\item {\bf Hit-matching efficiency for \begin{boldmath}$r\phi$-hits in
  the silicon microvertex detector\end{boldmath}}: One percent of
  the hits on the $r\phi$ strips of the silicon microvertex detector,
  which are associated to tracks, are randomly dropped and the tracks
  are refitted. This uncertainty is chosen positive if the selection
  rate increases as hits are dropped. The hit dropping fractions
  were obtained from studies of the \Zo\ calibration data.

\item {\bf Hit-matching efficiency for \begin{boldmath}$z$-hits in the
  silicon microvertex\end{boldmath} detector}: This uncertainty
  is evaluated in the same way as for the $r\phi$ hits, except
  that 3\% of the $z$-hits are dropped.
  
\item {\bf B hadron charged decay multiplicity}: The average
  number of charged tracks in B hadron decay is varied within the
  range recommended by the LEP Electroweak Heavy Flavour Working
  Group~\cite{LEPHFEWWG}, $n_{\mathrm{B}}=4.955\pm 0.062$.  The
  uncertainty is given a positive sign if the selection efficiency
  increases with the average decay multiplicity.  
  
\item {\bf B hadron momentum spectrum}: The b fragmentation function
  has been varied so that the mean fraction of the beam energy
  carried by B hadrons, $\langle x_E(\mathrm{b}) \rangle$, 
  is varied in the range
  $0.702\pm 0.008$~\cite{LEPHFEWWG} using a reweighting technique.
  The uncertainty is given a positive sign if the selection efficiency
  rises with increasing average momentum.
  
\item{\bf Charm hadron production fractions}: The branching ratios of
  charm hadrons in charm quark decays have been varied within the ranges
  BR$(\mathrm{c}\ra \mathrm{D}^+)=0.237\pm0.016$, BR$(\mathrm{c}\ra
  \mathrm{D}_{\mathrm{s}})=0.130\pm0.027$ and BR $ (\mathrm{c}\ra
  \mathrm{c}_{\mathrm{baryon}})=~0.096\pm 0.023$~\cite{lepelch}.
  The uncertainty is given a positive sign if the
  selection efficiency increases with the charm hadron multiplicity.
  Each uncertainty is considered individually in the limit
  calculation, but they are summed in quadrature in
  Table~\ref{tab:systable}.
  
\item {\bf Charm hadron decay multiplicity}: The average number of
  charged and neutral particles in charm hadron decay is varied within
  the measured ranges~\cite{mark3}.  The number of charged particles in
  $\mathrm{D}^0, \mathrm{D}^{+}$ and $\mathrm{D}_\mathrm{s}$ decays
  has been varied within $n_{\mathrm{D}^0}=2.56\pm 0.05$,
  $n_{\mathrm{D}^{+}}=2.38\pm 0.06$,
  $n_{\mathrm{D}_{\mathrm{s}}}=2.69\pm 0.32$, respectively.  The
  number of $\pi^0$s in $\mathrm{D}^0$ and $\mathrm{D}^{+}$
  decays has been varied within the ranges $n_{\mathrm{D}^0}=1.31\pm
  0.27$, $n_{\mathrm{D}^{+}}=1.18\pm 0.33$, respectively.  The
  uncertainty is given a positive sign if the selection efficiency
  increases with the average decay multiplicity.  Each uncertainty is
  considered individually in the limit calculation, but they are
  summed in quadrature in Table~\ref{tab:systable}.
  
\item {\bf Charm hadron momentum spectrum}: As for the B hadron
  momentum spectrum, $\langle x_E(\mathrm{c}) \rangle$ has been varied
  in the range $0.484\pm 0.008$~\cite{LEPHFEWWG}.
  
\item {\bf Comparison of different SM background Monte Carlo
    generators}: Besides the main generators used (see
  Section~\ref{sect:data}), the background simulations are
  cross-checked with alternative generators and fragmentation models
  such as KORALW~\cite{koralw} and HERWIG~\cite{herwig}. 
  
\item {\bf Four-Fermion production cross-section}: This is taken to
  have a 2\% relative uncertainty, arising from the uncertainty
  in the \ZZ\ and \WW\ cross-sections~\cite{ref:zzxsuncert}.  
\end{itemize}
The remaining uncertainties are channel dependent and assumed
uncorrelated between the channels, but correlated between
centre-of-mass energies for the same channel:
\begin{itemize}
\item {\bf Modelling of likelihood variables}: These uncertainties are
  evaluated by rescaling each input variable in the Monte Carlo
  individually so as to reproduce the mean and variance of the data.
  This scaling is done at the level of the preselection cuts and the
  contributions evaluated for each of the variables are then summed in
  quadrature. 
\item {\bf Double-ISR rate}: For the missing-energy channel, the
  systematic uncertainty on the background rate includes, in addition
  to the shared systematic uncertainties, a relative error of 1.24\%
  on the rate of \ee\ra\qq$\gamma\gamma$ events (``double-ISR'').  This
  systematic uncertainty was evaluated by comparing the selections for
  $\qq$ background events generated with two different settings of the
  KK2F generator~\cite{kk2f}: one using the Coherent Exclusive
  EXponentiation (CEEX) matrix elements up to first order and one
  using up to second order QED corrections, both for initial and final
  state radiation (FSR). Interference effects between ISR and FSR were
  neglected in this study.
\item {\bf Tau identification}: In the tau channels the modelling of the
  fake rates is studied using high-statistics $\ee\ra\qq$ data sets
  taken at $\sqrt{s}\approx\mZ$.  The modelling of the signal inputs
  is studied using mixed events which are constructed by overlaying
  $\ee\ra\qq$ events with single hemispheres of $\ee\ra\tautau$ events
  at $\sqrt{s}\approx\mZ$. The systematic error estimated from these
  studies is $\pm 10\%$ for the single tau fake rate.  To be
  conservative, the error on the single tau efficiency is chosen as
  the largest of the errors obtained when applying a cut at the tau
  ANN$>$0.5 and ANN$>$0.75 resulting in a $\pm 3\%$ contribution.
\item {\bf Electron and muon identification}: The estimated
  contribution accounting for the observed differences between data
  and Monte Carlo simulation in the lepton identification are
  estimated to be, together with the contributions from the modelling
  of the likelihood variables, 1.2\% and 8.5\% in the electron
  channel, and 0.5\% and 3.5\% in the muon channel for signal and
  background, respectively.

\end{itemize}

\section{Statistical Procedures}
\label{limitcalcappendix}
The statistical evaluation of the data events which remain after the
selection criteria is done in two steps. Firstly all candidate events
are classified according to their signal-likeness, and secondly the
calculation of confidence levels is made, assuming background-only
and signal+background hypotheses. This section describes the
procedures used.

\subsection{Event Classification}
\label{sect:2d}
In the four-jet channel only one test-mass dependent discriminating
variable is used for the limit calculation, the discriminator function
${\cal{D}}(m_{\mathrm{H}})$, which already incorporates both
mass-dependent and mass-independent information.  For all the other
channels an event classification function is calculated, using the
likelihood or ANN value, ${\cal L}$, and the reconstructed Higgs boson
mass.
Using the two binned distributions directly in the limit calculation
is inconvenient due to the large amount of Monte Carlo statistics
needed to estimate accurately the expected signal and background in
each bin.  Furthermore, the histograms must be interpolated between
$\sqrt{s}$ and test-mass values where they are evaluated.
  
A simplification is to form histograms of \mHrec\ and ${\cal L}$ in
Monte Carlo events, and to smooth them into separate functions for
signal and background, $f_s(\mHrec )$, $f_b(\mHrec )$, $g_s({\cal L})$
and $g_b({\cal L})$.  These functions are normalised to the expected
rates of signal and background, respectively.  The event
classification function, $W$, is then defined as:

\begin{equation}
W = f_s(\mHrec )g_s({\cal L})/[f_s(\mHrec )g_s({\cal L})+2f_b(\mHrec )g_b({\cal L})].
\end{equation}
The values of this function are, by construction, between zero and
one.  It is chosen so that the separation of the mean values for a
background-only and a signal+background scenario, divided by the
variance, is maximised.

The Monte Carlo signal and background samples are then used to form
histograms of $W$, separately for the signal and background, which are
then smoothed.  There are signal and background distributions of $W$
for each channel at each centre-of-mass energy at which Monte Carlo
was generated, and at each test-mass.  The distributions of $W$ are
interpolated between neighbouring test-masses and centre-of-mass
energies to obtain the distributions used in the limit calculation.
The value of $W$ for a candidate event at a particular test-mass is
also interpolated between $W$ functions evaluated at nearby
centre-of-mass energies and test-masses.  The smoothed distributions
of $W$ for the signal and background are then binned, and the number of
candidates in the data in each bin of $W$ are counted.
The estimated signal, $s_i$, background, $b_i$, and number of candidates,
$n_i$, in each bin are used in the calculation of confidence levels.

\subsection{Confidence Level Calculation}
\label{sect:clcalc}
Confidence levels are computed by comparing the observed data
configuration to the expectations, for two hypotheses. In the
background hypothesis, only the SM background processes contribute to
the accepted event rate, while in the signal+background hypothesis the
signal from SM Higgs boson production adds to the
background. Each assumed Higgs boson mass (test-mass \mH) corresponds
to a separate signal+background hypothesis.

In order to test the signal+background and background hypotheses
optimally with the data, a {\it test statistic} is defined which
summarises the results of the experiment with expectations of the
signal+background and background hypotheses maximally different.  An
optimal choice~\cite{barlow} is the likelihood ratio of
Poisson probabilities.
\begin{equation}
  Q = P_{\mathrm{poiss}}(data|signal+background)/
  P_{\mathrm{poiss}}(data|background),
\end{equation}
where 
\begin{equation}
  P_{\mathrm{poiss}}(data|signal+background) = 
  \prod_{i=1}^{n_{\mathrm{bins}}}\frac{(s_i+b_i)^{n_i}e^{-(s_i+b_i)}}{n_i!},
\end{equation}
and
\begin{equation}
  P_{\mathrm{poiss}}(data|background) = 
  \prod_{i=1}^{n_{\mathrm{bins}}}\frac{(b_i)^{n_i}e^{-b_i}}{n_i!}.
\end{equation}

The products runs over all bins of all distributions to be combined.
The signal estimation, $s_i$, depends on the expected signal
cross-section, the decay branching ratios of the Higgs boson, the
integrated luminosity and the detection efficiency for the signal.  
The background estimation, $b_i$, depends on the SM background 
cross-sections, the integrated luminosity, and selection
efficiencies.  The number of observed events in bin $i$ is $n_i$.  The
test statistic is more conveniently expressed in the logarithmic form
\begin{equation}
  -2\ln Q = 2\sum_{i=1}^{n_{\mathrm{bins}}} 
  s_i - 2\sum_{i=1}^{n_{\mathrm{bins}}}n_i\ln(1+s_i/b_i),
\end{equation}
which reduces to a sum of event weights, $w=\ln(1+s_i/b_i)$, depending
on the local $s_i/b_i$ for each candidate event observed and on the
test-mass.

\begin{figure}[p]
  \vspace*{-0.5cm}
  \centerline{
    \epsfig{file=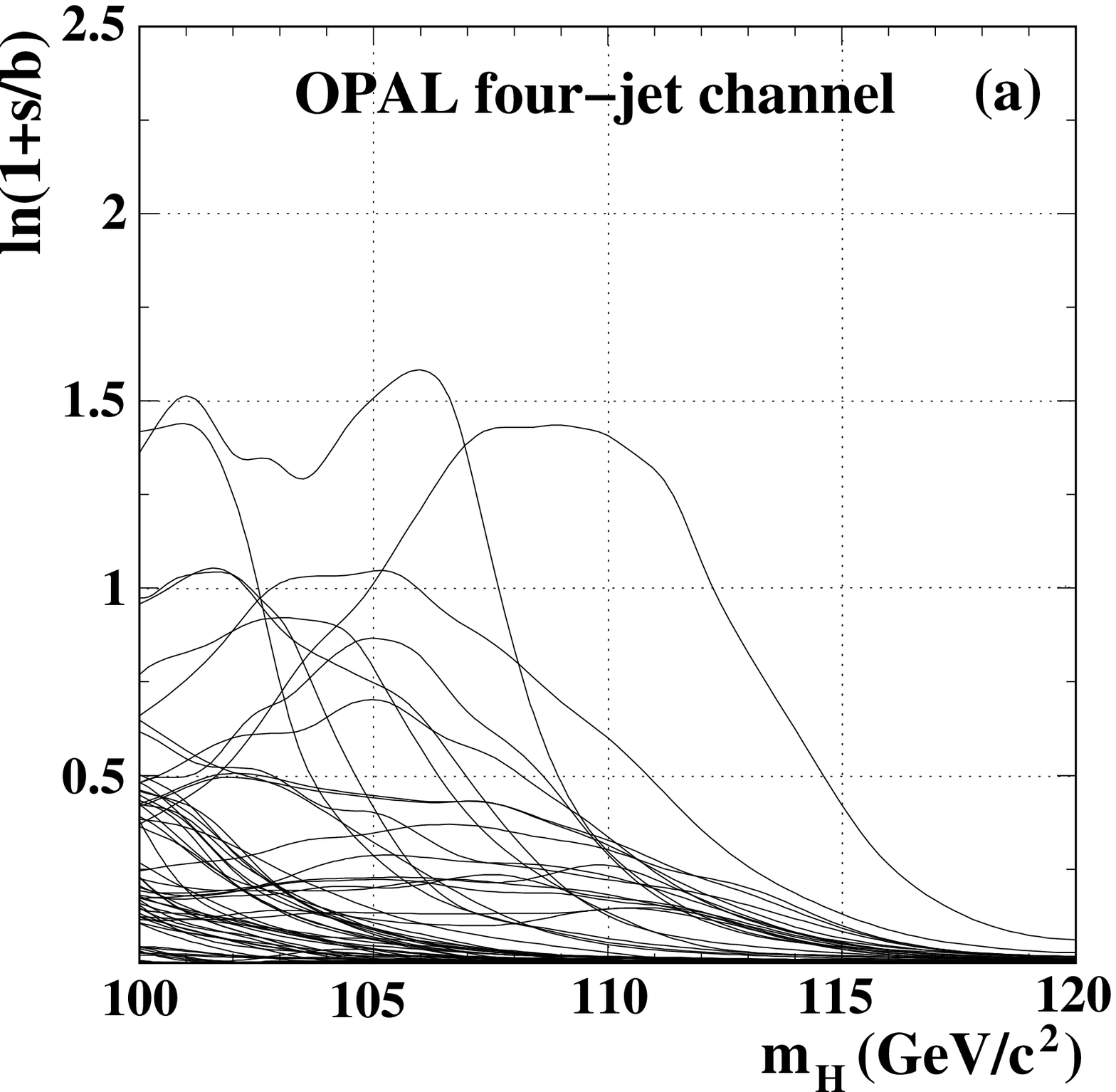,width=0.55\textwidth}
    \epsfig{file=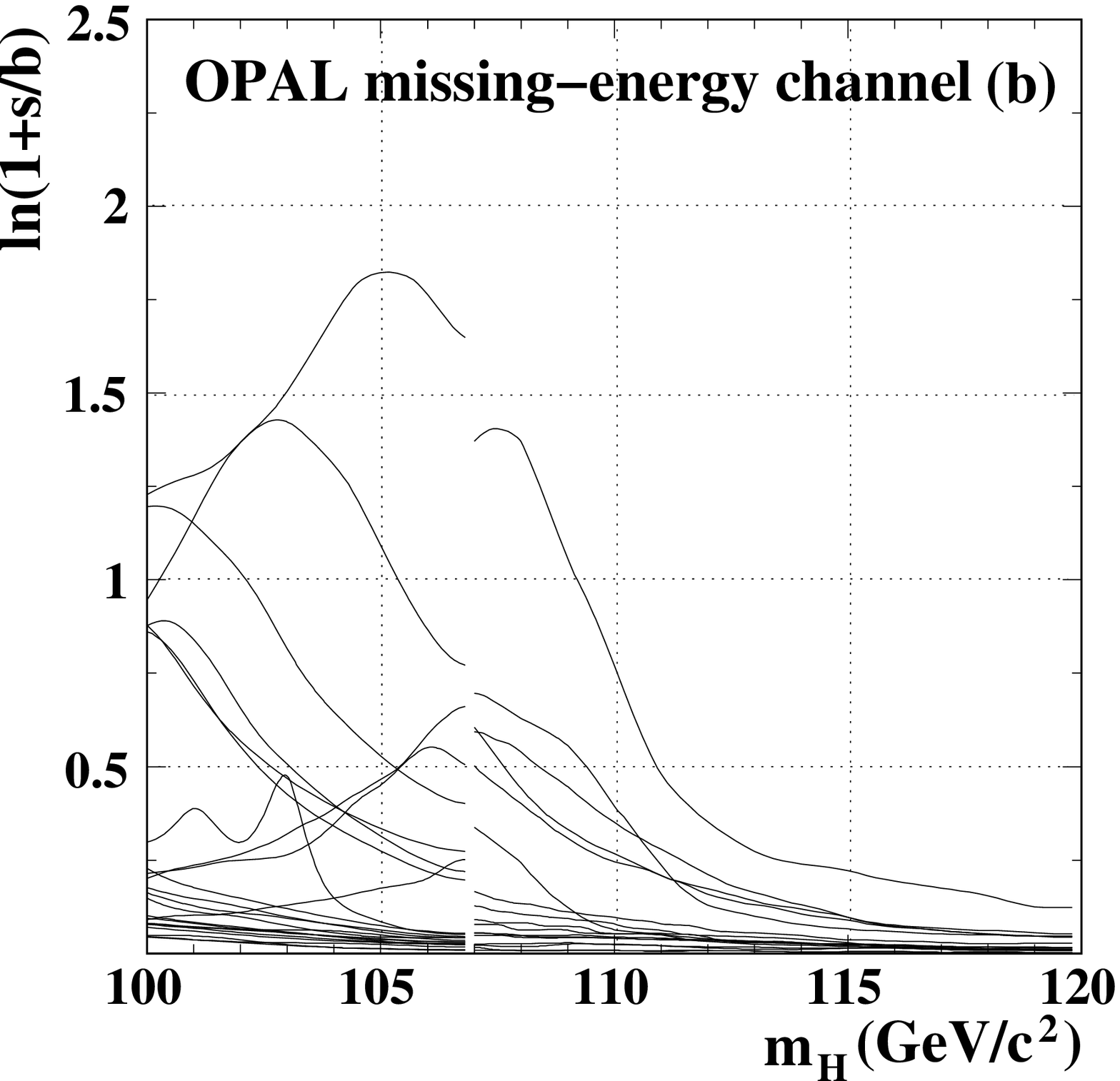,width=0.55\textwidth}}
  \centerline{
    \epsfig{file=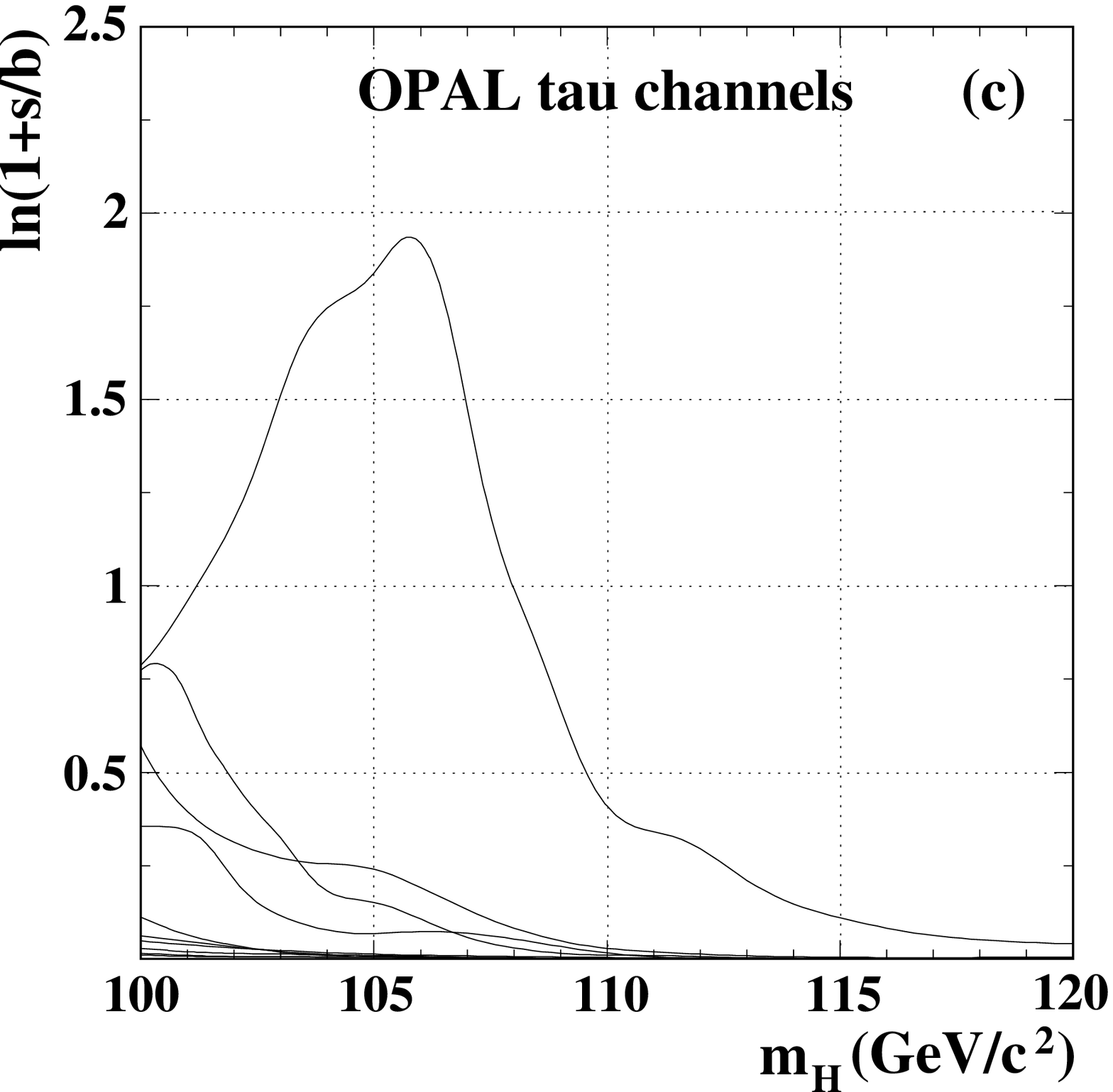,width=0.55\textwidth}
    \epsfig{file=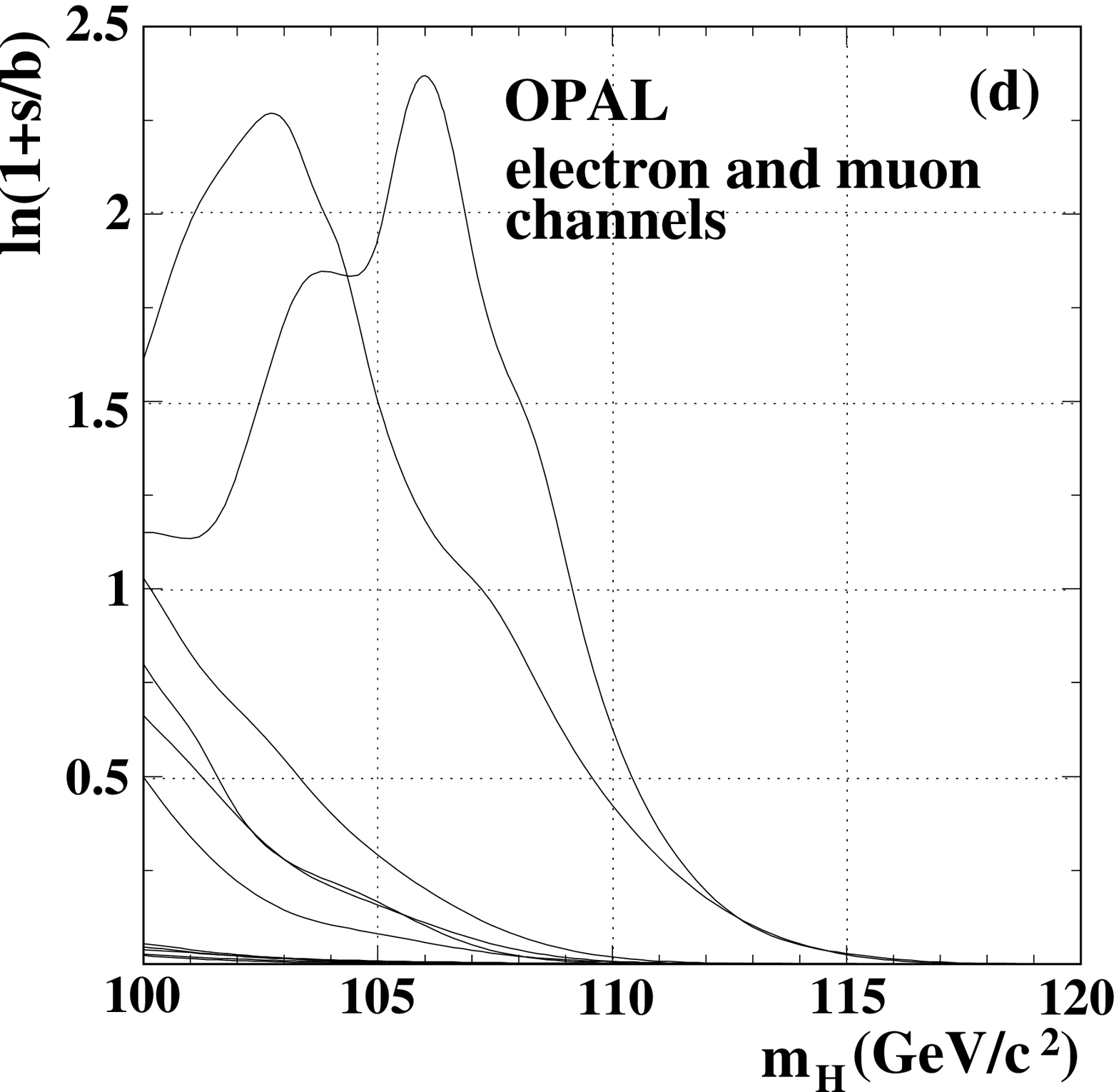,width=0.55\textwidth}}
  \vspace*{-0.1cm}
  \caption[]{\label{fig:evolution} \sl Distributions of event weights,
    $\ln(1+s/b)$, as a function of the Higgs boson test-mass, (a) the
    four-jet channel, (b) the missing-energy channel, (c) the tau
    channels, and (d) the electron and muon channels.  The
    discontinuities observed in the case of the missing-energy channel
    at $\mH=107$ \gevcs\ are due to the switching between the two ANN's,
    trained for low and high mass signals.}
\end{figure}

In this procedure an event-weight is assigned to each event.
These weights depend on the test-mass, and are shown for some of the
selected candidates in Figure~\ref{fig:evolution}(a)-(d).  The six
candidates with highest event weights ($w=\ln(1+s/b)>$0.1) at
\mH=115~\gevcs\ are listed in Table~\ref{tab:highestweights}.

To test the consistency of the data with the background hypothesis,
the confidence level \omclb\ is defined as
\begin{equation}
  \omclb = P(Q \geq Q_{\mathrm{obs}} | background),
\end{equation}
the fraction of experiments in a large ensemble of background-only
experiments which would produce results at least as background-like as the
observed data.

To test the consistency of the data with the signal+background
hypothesis, the confidence level \clsb\ is defined as
\begin{equation}
  \clsb = P(Q \leq  Q_{\mathrm{obs}} | signal+background),
\end{equation}
the fraction of experiments in a large ensemble of signal+background
experiments which would produce results less signal-like than the
observed data. By definition a signal+ background hypothesis is
excluded at the 95\% confidence level if $\clsb < 0.05$.  Statistical
downward fluctuations in the background can lead to deficits of
observed events which are inconsistent with the expected background
and this can cause the signal+background hypothesis to be excluded
even if the expected signal is so small that there is little or no
experimental sensitivity to it.
The confidence level \cls\ is defined to regulate this behaviour of
\clsb :
\begin{equation}
  \cls = \clsb / \clb .
\end{equation}
There is some loss of sensitivity by using \cls\ rather than \clsb,
but in no case is a limit more restrictive than the one obtained by
using \clsb.  We therefore consider a signal hypothesis to be excluded
at the 95$\%$~CL if $\cls < 0.05$.

Because all of the $s_i$, the $b_i$ (in general), and the candidates
in each bin depend on the test-mass, \clb, \clsb, and \cls\ all depend
on the test mass.  The limit on the Higgs boson mass is the smallest
test-mass \mH\ such that $\cls(\mH ) \geq 0.05$.  The sensitivity of
the analysis can be expressed by the median \cls\ in an ensemble of
background-only experiments.  It is used as the figure of merit to
optimise the analyses.  

Systematic uncertainties are incorporated into the confidence level
calculations by varying and reapplying the signal and background
estimations, taking correlations into account, assuming
Gaussian-distributed uncertainties. 

\section{Results}
\label{sect:results}
\begin{figure}[p]
  \centerline{
    \epsfig{file=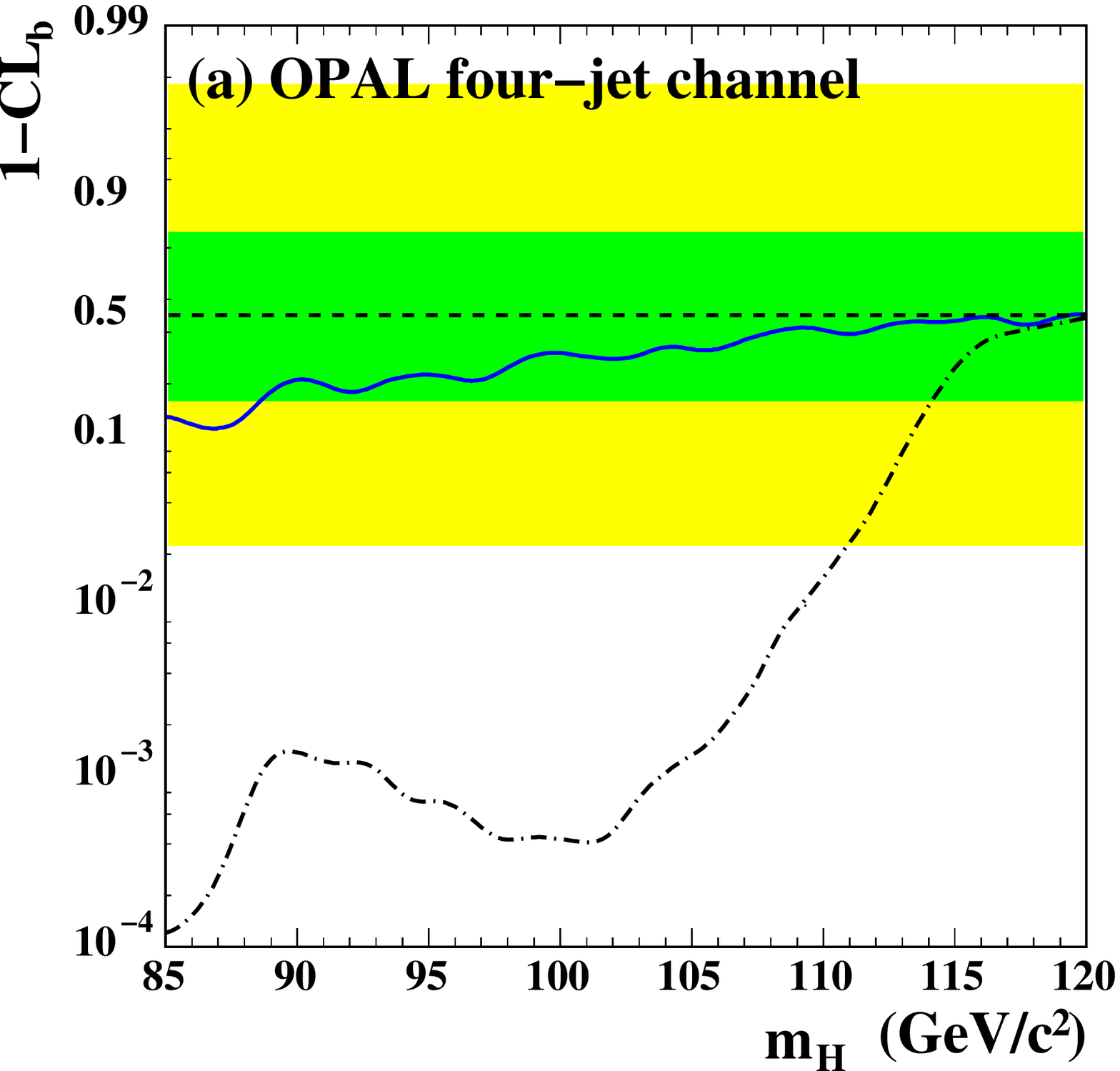,width=0.5\textwidth}
    \epsfig{file=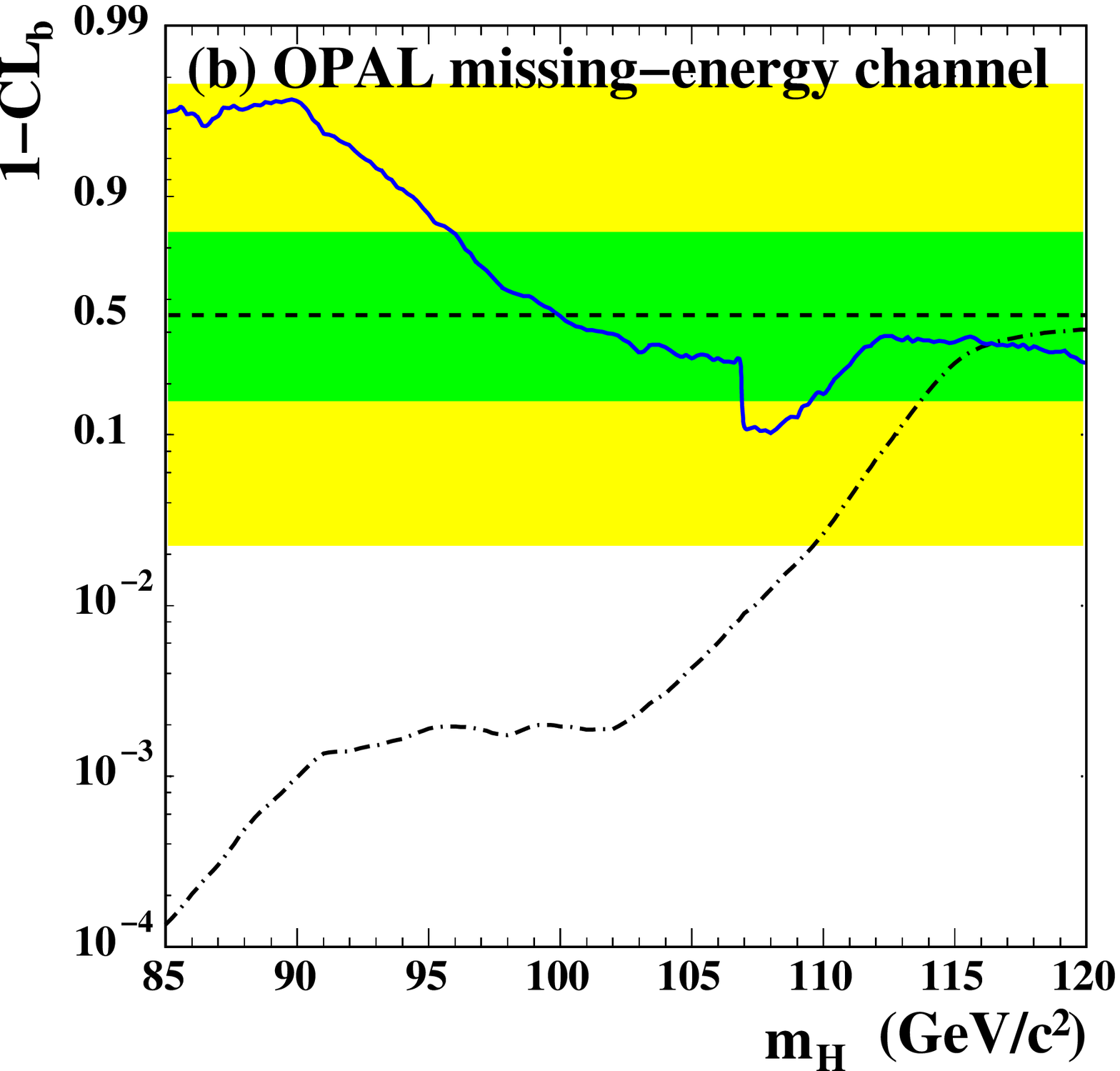,width=0.5\textwidth}}
  \centerline{
    \epsfig{file=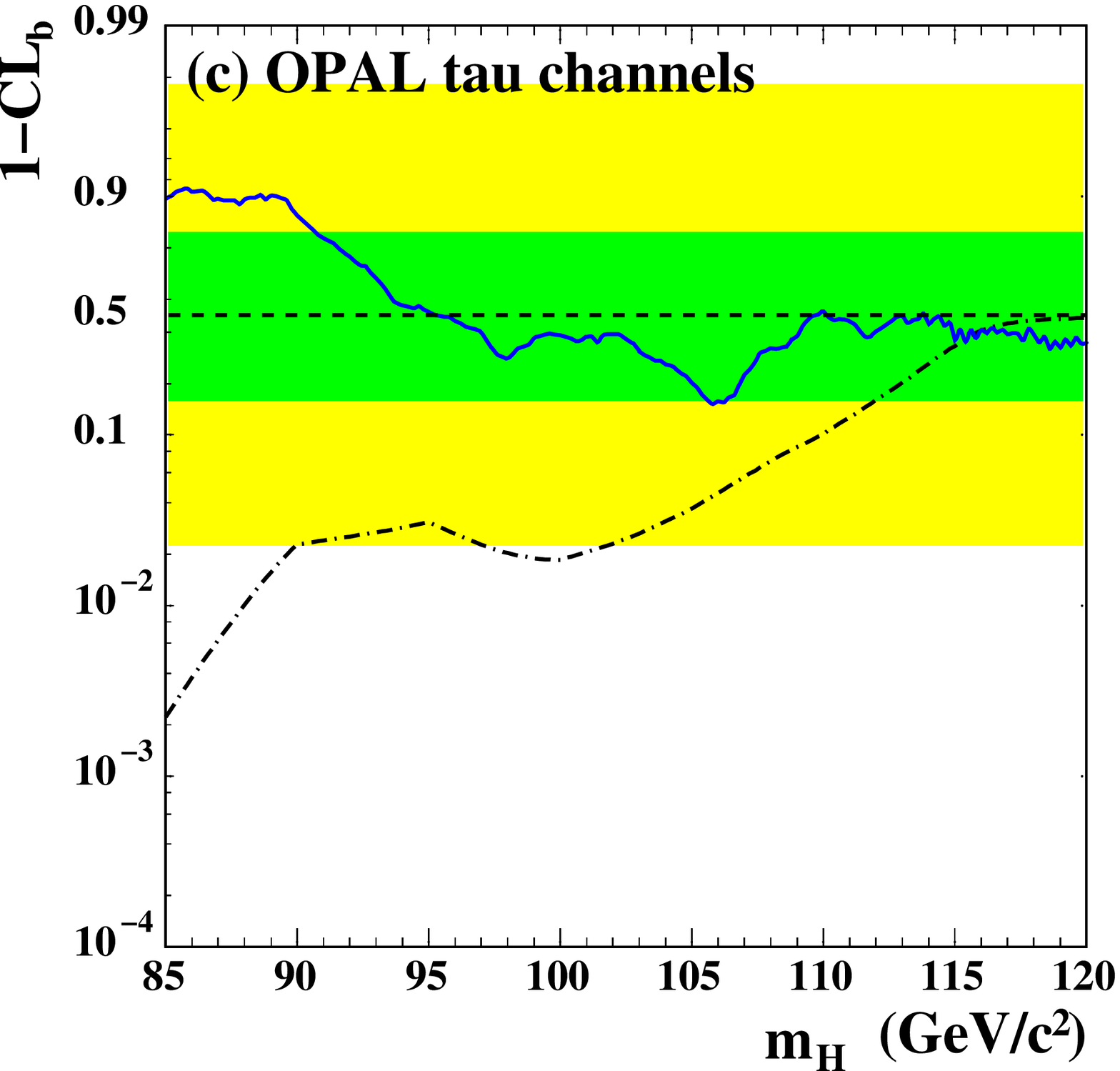,width=0.5\textwidth}
    \epsfig{file=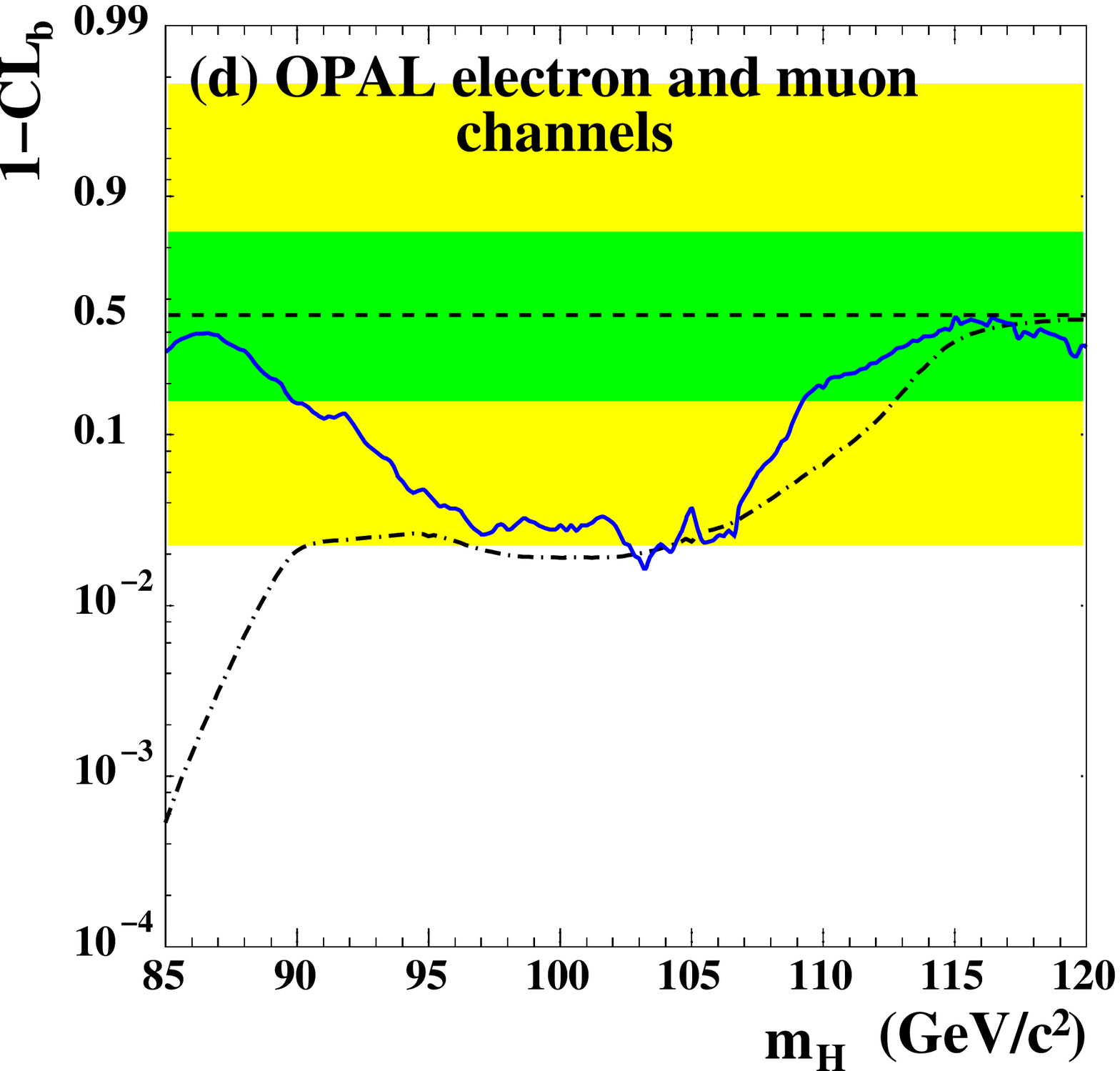,width=0.5\textwidth}}
  \caption[]{\label{fig:clbbychannel} \sl The confidence level
    $1-$\clb, as a function of the Higgs boson test-mass, separately
    for (a) the four-jet channel, (b) the missing-energy channel, (c)
    the tau channels and (d) the electron and muon channels.  The
    observations for the data are shown with solid lines.  The
    horizontal shaded bands indicate the 68\% and 95\% probability
    intervals centred on 0.5, the median expectation in the absence of
    a signal. The median expectation in the presence of a signal is
    presented with the dash-dotted line. }
\end{figure}

\begin{figure}[htbp]
  \centerline{
    \epsfig{file=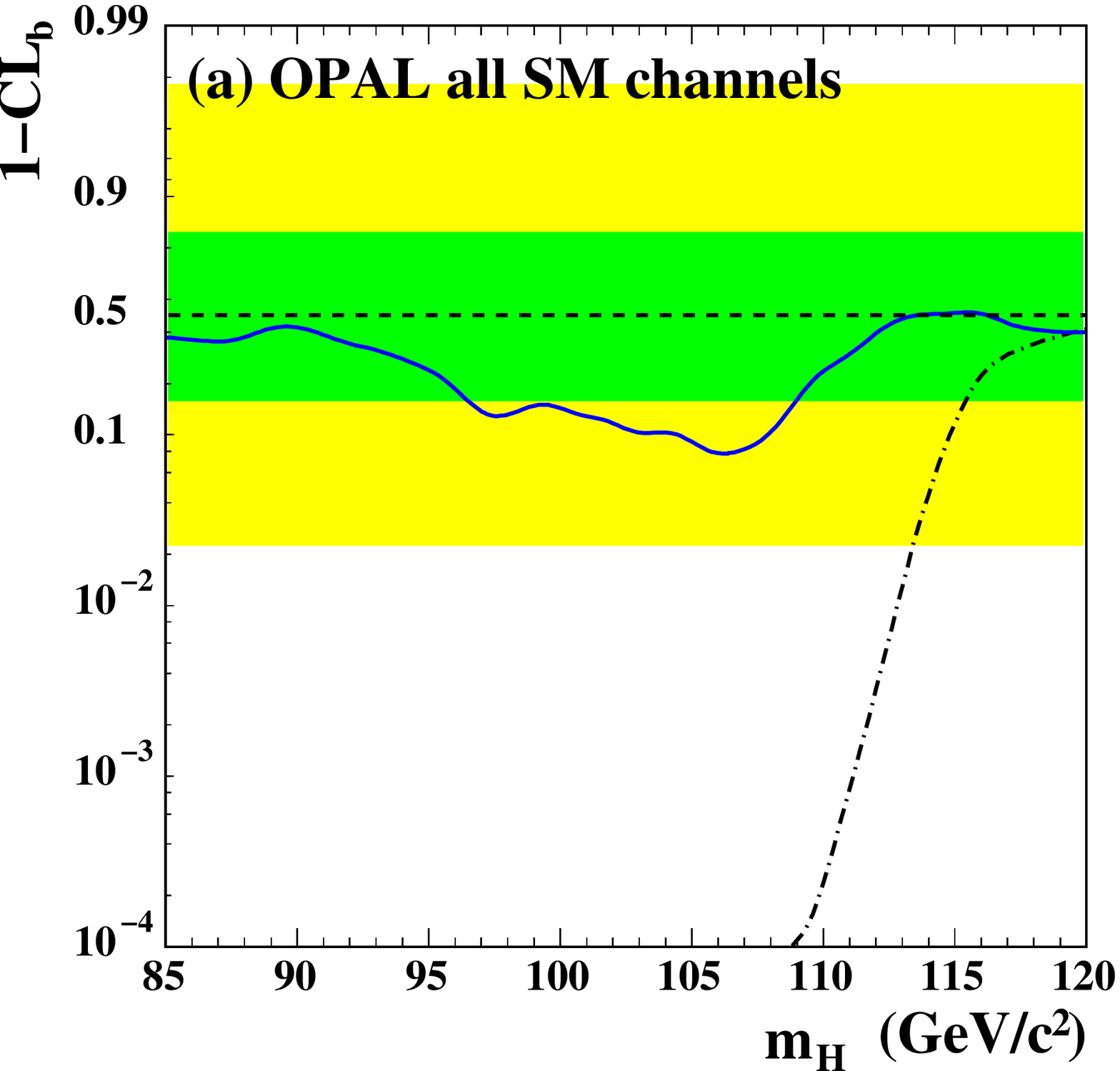,width=0.5\textwidth}\\
    \epsfig{file=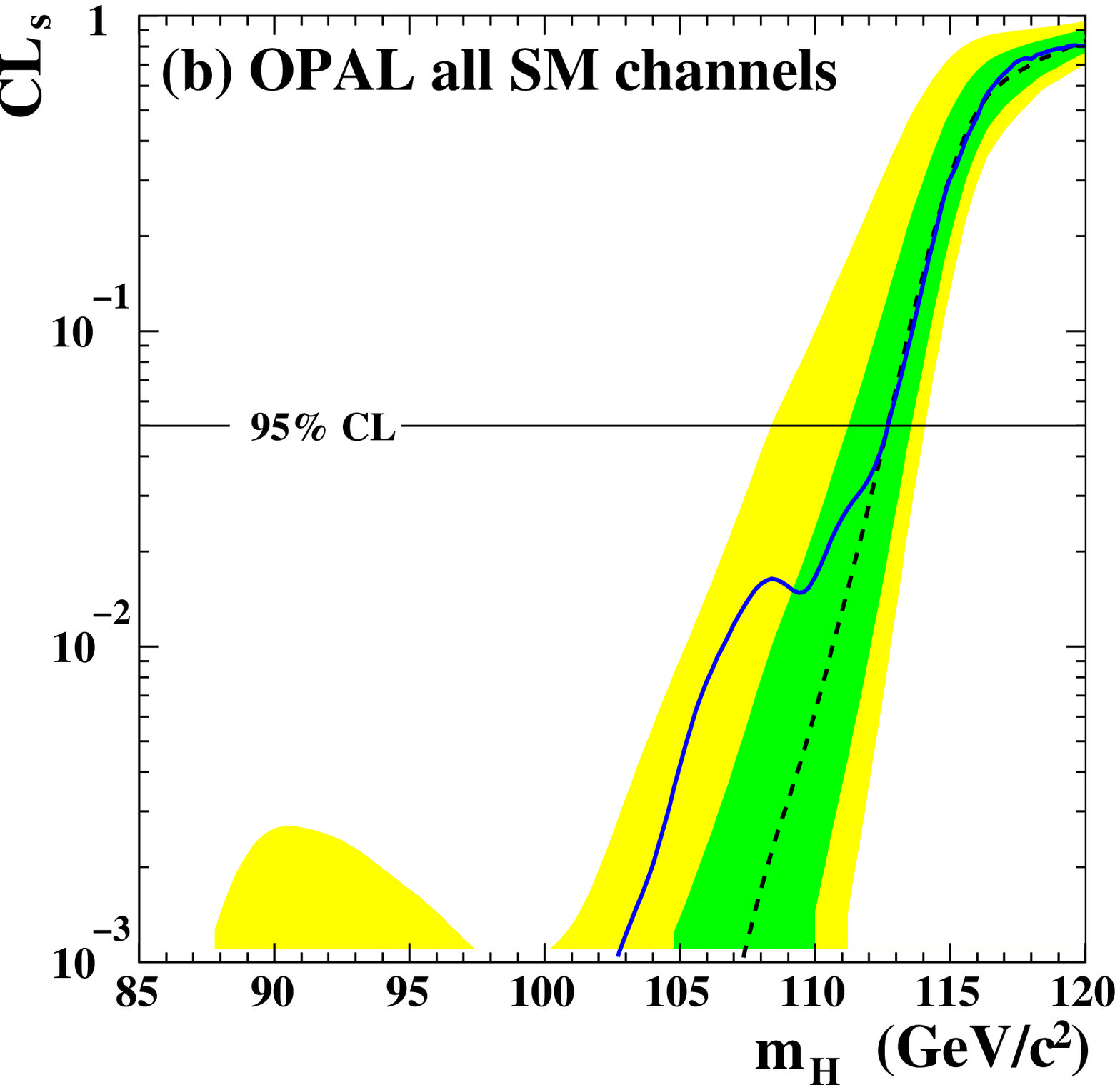,width=0.49\textwidth}}
  \caption[]{\label{fig:smclb}\sl For all search channels combined:
    (a) The confidence level $1-$\clb, as a function of the Higgs boson
    test mass.  The median expectation in the presence of a signal is
    presented with the dash-dotted line.  (b) The confidence level
    ratio \cls=\clsb/\clb\ versus the Higgs boson test-mass.  The
    observations for the data are shown with solid lines.  The shaded
    bands indicate the 68\% and 95\% probability intervals, with
    respect to the median expectation in the absence of a signal
    (dashed line). }
\end{figure}

In order to determine the compatibility of the observed data-set with
a background-only experiment, the confidence level $1-$\clb\ has been
computed as a function of the test-mass \mHtest. The results are shown
in Figure~\ref{fig:clbbychannel}, along with the distributions
expected in an ensemble of background-only experiments and
signal+background experiments. If the observed data agreed perfectly
with the prediction of the simulated background-only experiment, a
value of $1-$\clb=0.5 would be obtained. A lower (higher) value
would indicate an excess (deficit) of data. None of the channels show
evidence for a SM Higgs boson.  To increase the sensitivity of the
search, all channels are combined (Figure~\ref{fig:smclb}(a)).  

The largest deviation in $1-$\clb\ with respect to the expected SM
background is for a Higgs boson mass of 106~\gevcs\ with a
minimum $1-$\clb\ of about 0.08, which is a much smaller deviation
than expected for a Standard Model Higgs boson with a mass of
106~\gevcs.

Also shown in Figure~\ref{fig:smclb}(b) is the \cls\ confidence level,
for all search channels combined. The \cls\ shows the compatibility of
the data set with a signal+background hypothesis and also indicates
the sensitivity of the search. A signal hypothesis with a
\cls$\le$0.05 is considered to be excluded at the 95\% confidence
level. In this search a Higgs boson mass up to 112.7~\gevcs\ is
excluded at the 95\% CL, with the expected limit from background-only
hypotheses also being 112.7~\gevcs. As with other previous OPAL
publications~\cite{pr329,pr285} no indication of the signal has been
found. A comparison between the candidates' weights $w$ in the previous 
analysis and in the present paper can be found in 
Appendix~\ref{comparisonappendix}.
Figure~\ref{fig:smn95}(a) shows the upper limits on the signal
event rate at the 95\% CL.

The search results presented here are also used
to set 95\% CL upper bounds on the square of the HZZ coupling in
models which assume the same Higgs boson decay branching ratios as the
SM, but in which the HZZ coupling may be different.
Figure~\ref{fig:smn95}(b) shows the upper bound on
$\xi^2=(g_{\mathrm{HZZ}}/g_{\mathrm{HZZ}}^{\mathrm{SM}})^2$, the
square of the ratio of the coupling in such a model to the SM
coupling, as a function of the Higgs boson test-mass.  In the
evaluation of this ratio the \WW\ and \ZZ\ fusion processes are
assumed to scale also with $\xi$.  The new analyses in the missing
energy and four-jet channels are applied starting from \mH=80~\gevcs:
in Figure~\ref{fig:smn95}(b), a discontinuity can be observed 
corresponding to the transition point both in the expected and the
observed curves.

\begin{figure}[p]
\centerline{
  \epsfig{file=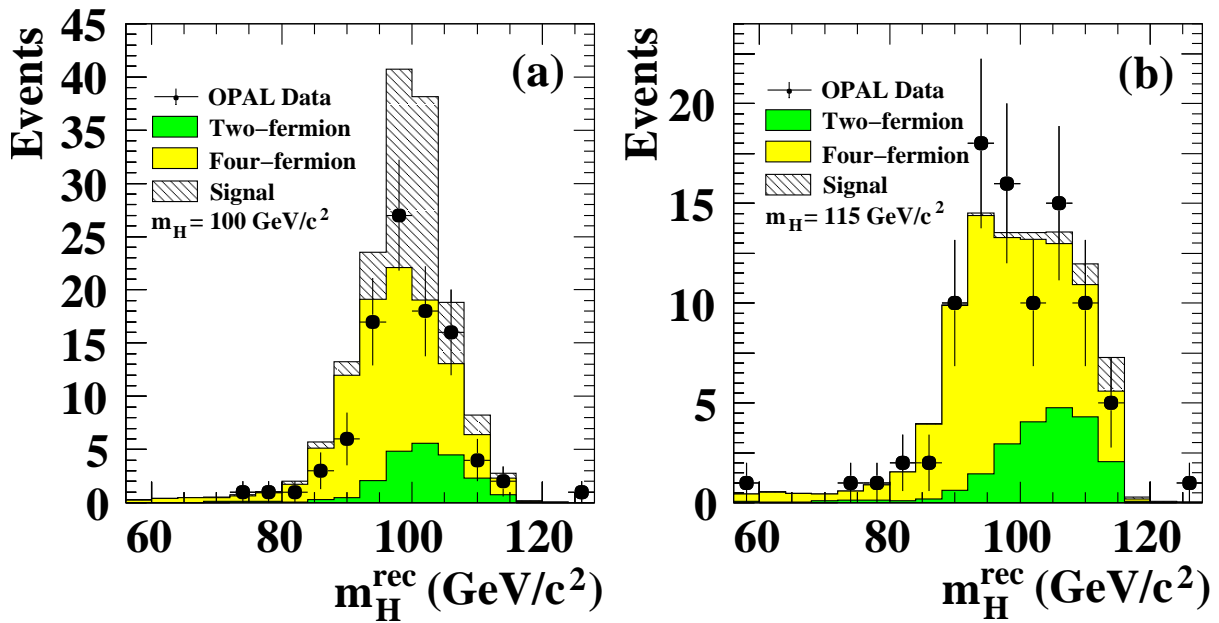,width=\textwidth}}
\vspace*{-0.3cm}
\caption[]{\label{fig:massplotcomb}\sl Distributions of the
  reconstructed mass for the selected events after a cut on the
  likelihood/ANN for all OPAL channels combined, with the expected
  contribution (a) from a \mH=100~\gevcs\ Higgs boson, and (b) from a
  \mH=115~\gevcs\ Higgs boson.  }
\end{figure}

The mass distributions for all channels combined, after a cut on the
likelihood/ANN value is shown in Figure~\ref{fig:massplotcomb}(a) and
(b) together with the contribution from a hypothetical SM Higgs boson
signal with \mH=100~\gevcs\ and \mH=115~\gevcs, respectively.

\section{Summary and Conclusion}
\label{sect:conclusion}
A search for the Standard Model Higgs boson has been performed with
the OPAL detector at LEP based on the full data sample collected at
$\sqrt{s}\approx$192--209~GeV in 1999 and 2000.  The largest deviation
with respect to the expected SM background in the confidence level for
the background hypothesis, $1-$\clb, is observed for a Higgs boson
mass of 106 GeV with a minimum $1-$\clb\ of about 0.08, much less
significant than that expected for a Standard Model
Higgs boson with a mass of 106~\gevcs.  A lower bound of 112.7~\gevcs\ on
the mass of the SM Higgs boson is obtained at the 95\% confidence
level for an expected limit from the background-only hypothesis of
112.7~\gevcs. The results do not confirm the excess at
\mH=115~\gevcs\ seen by ALEPH~\cite{alephfinal}, and are similar
to the results from L3~\cite{l3final} and DELPHI~\cite{delphifinal}.

\section*{Acknowledgements}
We particularly wish to thank the SL Division for the efficient operation
of the LEP accelerator at all energies
 and for their close cooperation with
our experimental group.  In addition to the support staff at our own
institutions we are pleased to acknowledge the  \\
Department of Energy, USA, \\
National Science Foundation, USA, \\
Particle Physics and Astronomy Research Council, UK, \\
Natural Sciences and Engineering Research Council, Canada, \\
Israel Science Foundation, administered by the Israel
Academy of Science and Humanities, \\
Benoziyo Center for High Energy Physics,\\
Japanese Ministry of Education, Culture, Sports, Science and
Technology (MEXT) and a grant under the MEXT International
Science Research Program,\\
Japanese Society for the Promotion of Science (JSPS),\\
German Israeli Bi-national Science Foundation (GIF), \\
Bundesministerium f\"ur Bildung und Forschung, Germany, \\
National Research Council of Canada, \\
Hungarian Foundation for Scientific Research, OTKA T-029328, 
and T-038240,\\
Fund for Scientific Research, Flanders, F.W.O.-Vlaanderen, Belgium.\\



\clearpage
\newpage

\begin{appendix}




  \section{Comparison to Previous Published Results}
  \label{comparisonappendix}
  In the four-jet and missing-energy channels new analyses have been
  developed to improve the sensitivity with respect to the analyses
  described in Ref.~\cite{pr329}.  In each of these channels a
  comparison between the candidates' weights $w=\ln(1+s/b)$ in the previous
  analysis and in the present paper has been performed.

  \subsection{Four-Jet Channel}
  The expected sensitivity of the new analysis has been improved over
  the whole mass range of hypothetical signal masses. For
  \mH=100~\gevcs, \mH=110~\gevcs\ and \mH=115~\gevcs\ the improvement
  of the sensitivity in \cls\ is about 200\%, 32\% and 12\%,
  respectively.  The comparison of the selected candidates in the four
  jet channel can be found in Table~\ref{tab:fourjcand}, which
  contains all events selected by the old analysis with their new
  weights. The main reasons for changes in the weights are
  reprocessing of the data, different assignment of the jets to bosons
  and the treatment of the highest b-tags.  The highest b-tags for
  candidate (16145:37028) changed from 0.595 and 0.251 to 0.344 and
  0.219 due to reprocessing. The likelihood value of
  ${\cal{L_{\mathrm{old}}}} =0.917$ calculated after reprocessing would
  not have met the old selection criteria of
  ${\cal{L_{\mathrm{old}}}}>0.96$.  Furthermore, the new jet pairing
  likelihood assigned a different jet pairing to the bosons and the
  presumed Higgs boson was reconstructed at 88.1~\gevcs. The value of
  the new discriminating variable for the 115~\gevcs\ selection of
  0.092 is not high enough to pass the cut value of 0.2.  In the case
  of the three candidates (13978:6299), (15353:24246) and (14847:5404)
  the jet pairing likelihood assigns the jets with two highest b-tags
  to different bosons due to kinematical constraints. The old analysis
  is sensitive only to the two highest b-tags in the event, regardless
  of the hypothetical boson the jets are assigned to. The new
  analysis is sensitive to the b-tags of the two jets assigned to the
  hypothetical Higgs boson. 

   \subsection{Missing-Energy Channel}
   The expected sensitivity of the new analysis has been improved over
   the whole mass range of hypothetical signal masses. For
   \mH=100~\gevcs, \mH=110~\gevcs\ and \mH=115~\gevcs\ the improvement
   of the sensitivity in \cls\ is about 4\%, 32\% and 8\%,
   respectively.  The comparison for the missing-energy channel can be
   found in Table~\ref{tab:misscand}, which contains events with
   $w\ge0.05$ for a \mH=115 GeV signal in either the new or the old
   analysis.  Only events with $\mathrm{ANN}\ge 0.5$ in the new
   analysis and likelihood~$\ge 0.2$ in the old analysis have been
   selected.  Among the remaining 14 candidates, 11 are selected by
   both analyses, and 7 out of these have weights larger than 0.05 in
   both cases.  One candidate (15587:18556) is selected by the new
   analysis only and two are selected by the old analysis only: these
   two candidates are lost by the new analysis due to the re-processing
   of the data.  Event (15886:54731) fails the preselection having a
   recoil mass $M_{\mathrm{miss}}$ = 46.3~\gevcs\ (the preselection cut
   is at 50~\gevcs) and event (15648:12732) has a low ANN value of 0.09
   since the b-tags of the two jets are low (0.40 and 0.21,
   respectively) and $M_{\mathrm{miss}}=57.9$~\gevcs.

\end{appendix}

\clearpage
\newpage


\begin{table}[h]
\begin{center}
\caption[]{\label{tab:lumi_st}\sl 
  Integrated luminosities of the data samples used for each search
  channel for the years 1999 and 2000. The differences between search
  channels are due to different requirements on detector functionality.\\}
\begin{tabular}{|l|r|r|}
\hline
\multicolumn{1}{|l|}{} & 
\multicolumn{2}{c|}{Integrated Luminosity (pb$^{-1}$)} \\\cline{2-3}
 & \multicolumn{1}{|c|}{Year 1999} & \multicolumn{1}{|c|}{Year 2000} \\ 
 Channel & 192-202 GeV & 200-209 GeV \\ \hline 
  $\Ho\Zo\rightarrow\bb\qq$ & 217.0 & 207.3 \\
$\Ho\Zo\rightarrow\bb\nunu$ & 212.7 & 207.2 \\
$\Ho\Zo\rightarrow\bb\tautau$/$\tautau\qq$
                            & 213.6 & 203.6 \\
$\Ho\Zo\rightarrow\qq\ee$   & 214.1 & 203.6 \\
$\Ho\Zo\rightarrow\qq\mm$   & 213.6 & 203.6 \\\hline

\end{tabular}
\end{center}
\end{table}

\begin{table}[htbp]
\begin{center}
\caption[]{\label{tab:smflow1999}\sl Cutflow of the selections,
  applied to the data taken in the year 1999.  Number of events after
  each cut, for the data and for the expected background, normalised
  to the data luminosity.  The two-photon background is not shown
  separately but is included in the total background.  The last column
  shows the luminosity-weighted average detection efficiencies for a
  Higgs boson mass of 100~\gevcs, considering \Ho\ra\bb\ in the
  four-jet channel, \Ho\ra all in the missing-energy, electron, and
  muon channels, and \Zo\Ho\ra\tautau(\Ho\ra all) or
  \Zo\Ho\ra\qq\tautau\ in the tau channel. The number
  of signal events expected after all cuts are given in parenthesis.\\}
\begin{tabular}{|c||r||r||r|r||c|}\hline
Cut&Data &Total & q\=q($\gamma$) & 4-fermi. & Efficiency (\%)\\
   &     &  bkg.&    bkg.        &   bkg.   & $\mH=100$~\gevcs \\\hline\hline
\multicolumn{6}{|c|}{Four-jet Channel ~~ 217.0 pb$^{-1}$}     \\ \hline
 (1) &  20848  &  20840.9  &  16232.0  &  4225.4  &   99.7 \\
 (2) &   7251  &   7217.8  &   4637.8  &  2556.1  &   94.1 \\
 (3) &   2430  &   2340.5  &    580.9  &  1754.2  &   87.0 \\
 (4) &   2413  &   2309.8  &    551.8  &  1752.6  &   86.9 \\
 (5) &   2167  &   2070.0  &    476.6  &  1590.8  &   82.6 \\
 (6) &   1984  &   1870.1  &    408.8  &  1459.8  &   79.5 \\\hline
${\cal D}(100)$  &  30 & 28.0 &  7.6 &  20.4 &   42.0(12.97$\pm$1.18) \\ \hline\hline
\multicolumn{6}{|c|}{Missing-energy Channel ~~ 212.7 pb$^{-1}$}  \\ \hline
 (1)  & 5821 &  5209.5  & 3290.4 & 956.7 & 83.5 \\
 (2)  & 3001 &  2944.2  & 2084.3 & 743.1 & 74.2 \\
 (3)  & 1534 &  1506.5  & 1079.1 & 414.7 & 71.9 \\
 (4)  &  625 &   585.7  &  258.9 & 321.6 & 70.8 \\
 (5)  &  371 &   358.4  &   70.1 & 285.7 & 65.0 \\
 (6)  &  205 &   191.8  &   61.7 & 130.1 & 63.1 \\
 (7)  &  152 &   156.8  &   34.6 & 122.2 & 61.0 \\\hline
${\cal L}^{\mathrm{HZ}}$& 10 & 13.9  & 2.8  & 11.1   & 46.9(5.10$\pm$0.21)   \\\hline\hline
\multicolumn{6}{|c|}{Tau Channel ~~ 213.6 pb$^{-1}$}  \\ \hline
(1)   & 5164   & 5328.3  &  2992.1  & 2335.9  & 78.0\\
(2)   &  816   &  886.1  &   102.0  &  784.2  & 62.4\\
(3)   &  207   &  214.6  &    55.1  &  159.6  & 50.4\\
(4)   &  170   &  186.0  &    53.8  &  132.1  & 50.0\\\hline
$\cal{L}^{\mathrm{HZ}}$ & 5  & 5.1  & 0.1  & 5.0 & 25.8(4.97$\pm$0.23)\\\hline\hline
\multicolumn{6}{|c|}{Electron Channel 214.1 pb$^{-1}$} \\\hline\hline
 (1) & 9560 & 9656.2 & 6587.4 & 3068.8 & 91.8 \\
 (2) &  189 &  158.1 &   73.9 &   84.2 & 75.0 \\
 (3) &  167 &  138.0 &   63.3 &   74.8 & 73.8 \\\hline
$\cal{L}^{\mathrm{HZ}}$ & 3 &  4.1 & 0.4 & 3.8 & 57.2(0.90$\pm$0.03)\\\hline\hline
\multicolumn{6}{|c|}{Muon Channel 213.6 pb$^{-1}$} \\\hline\hline
 (1) & 9526 & 9637.0 & 6574.3 & 3062.7 & 87.6 \\
 (2) &  120 &  113.2 &   88.0 &   25.2 & 77.1 \\
 (3) &   26 &   23.7 &   10.9 &   12.8 & 73.8 \\\hline
$\cal{L}^{\mathrm{HZ}}$ & 6 & 3.3 & 0.0 & 3.3 & 62.5(1.08$\pm$0.03)\\\hline
\end{tabular}

\end{center}
\end{table}


\begin{table}[htbp]
\vspace*{-1.0cm}
\begin{center}
\caption[]{\label{tab:smflow2000} \sl Cutflow of the selections,
similar to Table~\ref{tab:smflow1999}, but for the data taken during
the year 2000.  Efficiencies are given for a hypothetical Higgs boson
mass of 115~\gevcs\.\\}
\begin{tabular}{|c||r||r||r|r||c|}\hline
Cut & Data & Total & q\=q($\gamma$) & 4-fermi. & Efficiency (\%)\\
  &      &   bkg.&    bkg.        &   bkg.   & $\mH=115$~\gevcs \\\hline\hline
\multicolumn{6}{|c|}{Four-jet Channel ~~ 207.3 pb$^{-1}$}     \\ \hline
 (1) & 18242 & 17990.2 &  13697.3 &  4096.6 &    99.5  \\
 (2) &  6441 &  6430.7 &   3964.7 &  2456.1 &    93.2  \\
 (3) &  2215 &  2163.8 &    497.0 &  1664.2 &    85.9  \\
 (4) &  2186 &  2137.6 &    472.7 &  1662.5 &    85.1  \\
 (5) &  1982 &  1912.2 &    407.6 &  1503.4 &    81.3 \\ 
 (6) &  1786 &  1725.9 &    349.1 &  1376.0 &    78.5 \\ \hline
${\cal D}(115)$& 20 & 17.5 &  5.1 & 12.4 & 40.0(2.01$\pm$0.18)\\\hline\hline   
\multicolumn{6}{|c|}{Missing-energy Channel ~~ 207.2 pb$^{-1}$}  \\ \hline
(1) &  5417  &  5059.0 & 2979.8 & 1030.1 & 81.5 \\
(2) &  2791  &  2813.6 & 1869.8 & 788.0  & 73.1 \\
(3) &  1463  &  1515.0 & 1033.4 & 444.0  & 68.6 \\
(4) &   595  &   569.3 &  230.3 & 327.4  & 67.6 \\
(5) &   338  &   351.4 &   63.3 & 288.2  & 58.8 \\
(6) &   181  &   182.8 &   56.2 & 126.6  & 55.5 \\
(7) &   154  &   150.8 &   31.7 & 119.2  & 54.1 \\\hline
${\cal L}^{\mathrm{HZ}}$& 11 & 8.9 &2.6& 6.3&40.7(1.09$\pm$0.05)\\\hline\hline
\multicolumn{6}{|c|}{Tau Channel 203.6 pb$^{-1}$}  \\ \hline
(1)                 &   4783&4746.7         & 2476.9& 2269.8& 78.1\\
(2)                 &    777& 838.8         &   79.9&  758.9& 61.2\\
(3)                 &    190& 214.6         &   43.6&  171.0& 44.7\\
(4)                 &    166& 168.6         &   42.2&  126.4& 43.3\\\hline
${\cal L}^{\mathrm{HZ}}$ & 5&  4.5  & 0.2&    4.3 & 25.6(0.26$\pm$0.01)  \\\hline\hline
\multicolumn{6}{|c|}{Electron Channel 203.6 pb$^{-1}$} \\\hline
 (1) & 8593 & 8562.5 & 5609.3 & 2953.2 & 90.9 \\
 (2) &  169 &  156.7 &   69.0 &   87.7 & 75.1 \\
 (3) &  144 &  137.3 &   59.3 &   78.1 & 73.5 \\\hline
$\cal{L}^{\mathrm{HZ}}$ & 1 & 3.6 & 0.3 & 3.4 & 52.9(0.10$\pm$0.003)\\\hline\hline
\multicolumn{6}{|c|}{Muon Channel 203.6 pb$^{-1}$} \\\hline
 (1) &  8452 & 8441.0 &  5530.1 & 2910.9 & 88.4 \\
 (2) &  121  & 109.5  &  83.3   & 26.1   & 76.8 \\
 (3) &   29  &  23.3  &  10.2   & 13.1   & 71.8 \\\hline
$\cal{L}^{\mathrm{HZ}}$ &  4 & 3.4 & 0.0 & 3.4 & 59.2(0.14$\pm$0.003) \\\hline
\end{tabular}
\end{center}
\end{table}

\begin{table}[htbp]
\begin{center}
\caption[]{\label{tab:systable}\sl Relative systematic uncertainties
  (in \%) on the signal detection efficiencies and background
  estimations, broken down by source. Negligible sources not studied
  are denoted with a bar, and non-applicable sources with n/a.\\ }
\begin{tabular}{|l|c|c|c|c|c|c|}\cline{3-7}
\multicolumn{2}{}{} & \multicolumn{5}{|c|}{Channels} \\\hline
\multicolumn{2}{|l|}{Source} & \qq\Ho & \nunu\Ho & \tautau\qq & \ee\Ho & \mm\Ho  \\\hline
 Monte Carlo          & eff & 1.9  & 1.6 & 1.9 & 1.2 & 1.2 \\
 statistics           & bkg & 5.2  & 9.5 & 3.8 & 4.6 & 5.4 \\\hline
 $r\phi$-             & eff & 0.7  &-0.3 & 0.7 & 0.6 &-0.6 \\
 resolution           & bkg &-3.7  & 1.5 & 1.3 &-2.3 &-1.4 \\\hline
 $rz$-                & eff & 0.2  &-0.7 & 0.5 & 0.3 &-0.3 \\
 resolution           & bkg &-5.4  & 0.5 & 0.4 &-0.6 &-0.9 \\\hline
 microvertex $r\phi$ & eff &-1.0  & 0.1 &-0.2 & 0.7 &-0.4 \\
 hit efficiency       & bkg & 0.0  & 2.6 & 0.2 &-4.7 &-2.4 \\\hline
 microvertex $rz$    & eff &-2.5  &-0.1 &-0.3 & 0.3 &-0.6 \\
 hit efficiency       & bkg &-0.3  & 1.3 & 1.1 &-4.7 &-2.0 \\\hline
 b fragmentation      & eff & 3.9  & 0.9 &-0.5 & 1.7 & 1.6 \\
                      & bkg &-1.7  & 2.2 &-1.8 & 0.4 & 0.6 \\\hline
 B decay              & eff & 0.6  &-0.4 & 0.0 & 0.1 & 0.1 \\
 multiplicity         & bkg & 0.5  &-0.3 & 0.1 & 0.0 & 0.2 \\\hline
 c fragmentation      & eff & 0.2  & 0.1 &-0.2 & 0.2 & 0.1 \\
                      & bkg & 1.2  &-0.4 &-1.4 & 0.5 & 0.5 \\\hline
 c hadron             & eff & 0.7  & 0.2 &  -  &  -  &  -  \\
 fractions            & bkg & 2.6  & 0.4 &  -  &  -  &  -  \\\hline
 c hadron             & eff & 3.5  & 0.2 &  -  &  -  &  -  \\
 decay multiplicity   & bkg & 4.3  & 1.5 &  -  &  -  &  -  \\\hline
 SM MC                & eff & n/a  & n/a & n/a & n/a & n/a  \\
 comparison           & bkg & 0.4  & 0.5 & 4.2 &13.8 & 13.2 \\\hline
 Four-fermion         & eff & n/a  & n/a & n/a & n/a & n/a \\
 cross-section        & bkg & 1.4  & 1.5 & 1.9 & 1.9 & 2.0 \\\hline
Channel dependent     & eff &  6.6 & 3.6 &  6.1 & 1.2 & 0.5 \\
sources               & bkg & 11.4 & 5.1 & 15.0 & 8.5 & 3.5 \\ \hline
 \bf{Total}           & eff &  9.1 &  4.1 &  6.5 & 2.6 & 2.3 \\
                      & bkg & 15.2 & 11.7 & 16.4 &18.4 & 15.3 \\\hline
\end{tabular}

\end{center}
\end{table}

\begin{table}[h]
\begin{center}
\caption[]{\label{tab:highestweights}\sl Details for all candidates
  with an event weight $w=\ln(1+s/b)$ greater than 0.1 at
  \mHtest=115~\gevcs . In addition to the reconstructed mass, the
  values of the b-tags for the two jets of the Higgs boson candidate
  are listed.  For the tau channel the value of ${\cal B_{\tau}}$ is
  presented instead of ${\cal B}_1$.\\}
\begin{tabular}{|c|r|c|c|c|c|c|}\hline
Candidate & \mHrec\ &Channel &  ${\cal B}_1$
& ${\cal B}_2$ & $w_{115}$ & $\sqrt{s}$ \\\hline
1 &  111.2 & four-jet  & 0.94 & 0.47 & 0.4342 & 206.4 \\
2 &  108.2 & missing-energy & 0.75 & 0.24 & 0.1826 & 201.7 \\
3 &  107.2 & missing-energy & 0.99 & 0.11 & 0.1465 & 201.7 \\
4 &  112.6 & missing-energy & 0.31 & 0.24 & 0.1333 & 206.6 \\
5 &  105.0 & tau    & 0.99 & n/a  & 0.1315 & 205.2 \\ 
6 &  106.6 & four-jet  & 0.98 & 0.27 & 0.1295 & 206.4 \\ \hline
\end{tabular}
\end{center}
\end{table}

\begin{table}[htbp]
  \begin{center}
    \caption[]{\label{tab:fourjcand}\sl Four-jet channel: Comparison
      of $\ln(1+s/b)$-weights for candidates selected by the old
      analysis, and the weights attributed by the new analysis. A
      candidate which is not selected in one analysis, is denoted
      n/s. Events with a weight lower than $10^{-4}$ are listed with
      zero event-weight. Events indicated in bold are discussed in the
      text.\\ }
    \begin{tabular}{|c|c|c|c|c|c|c|c|c|}\cline{4-9}
       \multicolumn{3}{c|}{} &
       \multicolumn{2}{|c|}{\mHrec} &
       \multicolumn{2}{|c|}{$\ln(1+s/b)$-Weight} &
       \multicolumn{2}{|c|}{$\ln(1+s/b)$-Weight} \\
       \multicolumn{3}{c|}{} &
       \multicolumn{2}{|c|}{GeV} &
       \multicolumn{2}{|c|}{\mHtest=100~GeV} & 
       \multicolumn{2}{|c|}{\mHtest=115~GeV} \\\hline

       No. & \multicolumn{2}{|c|}{Run:Event} & 
       old & new & old & new & old & new \\\hline

    1 & 16242 &  8607         &  78.6 & 108.6 &  .1609 &  .3095 & .0424 & .0955\\ 
    2 & 16221 & 16299         &  90.9 &  91.1 &  .1742 &  .1473 & .0363 & .0302\\ 
    3 & 16167 & 26233         & 110.5 & 111.2 & 1.1286 & 1.4298 & .5324 & .4342\\   
  4 &{\bf 16145}&{\bf 37028}  & 112.0 & n/s   &  .2381 &   n/s  & .0796 &  n/s \\ 
    5 & 15935 & 12196         &  94.6 &  94.1 &  .0340 &  .0000 & .0118 & .0000\\
    6 & 15862 & 19366         & 107.2 & 106.8 &  .6312 &  .6101 & .1135 & .1295\\
    7 & 15821 & 29240         &  91.3 &  90.5 &  .0402 &  .0000 & .0116 & .0000\\
    8 & 15793 &  7877         &  99.6 &  99.4 &  .0368 &  .0835 & .0097 & .0297\\
    9 & 15737 & 30167         & 107.3 & 107.1 &  .3247 &  .3358 & .0584 & .0876\\
   10 & 15705 & 51066         & 105.0 &  50.3 &  .1344 &  .0000 & .0232 & .0000\\
   11 &{\bf 15353}&{\bf 24246}& 100.6 & 100.1 &  .4918 &  .1092 & .1491 & .0311\\
   12 & 15178 & 17218         &  90.0 &  90.7 &  .1341 &  .1379 & .0405 & .0296\\
   13 &{\bf 14847}&{\bf 5404} & 109.8 & 109.9 &  .9444 &  .1309 & .1396 & .0203\\
   14 & 14841 & 27410         &  91.0 &  91.4 &  .2381 &  .2825 & .0659 & .0477\\
   15 & 14827 & 32620         &  92.5 &  93.2 &  .1970 &  .1851 & .0424 & .0368\\
   16 & 14729 & 15555         &  97.8 &  97.5 &  .1627 &  .0423 & .0642 & .0138\\
   17 & 14560 & 24515         &  56.1 &  57.0 &  .1536 &  .0000 & .0626 & .0000\\
   18 & 14383 & 14394         &  98.8 &  99.2 &  .2069 &  .0507 & .0821 & .0186\\
   19 & 14268 & 24264         &  97.8 &  97.0 &  .0342 &  .0000 & .0097 & .0130\\
   20 & 14226 &  8270         & 100.7 & 102.1 &  .0845 &  .0610 & .0182 & .0130\\
   21 & 14134 & 74857         &   0.0 &  61.2 &  .1536 &  .0333 & .0626 & .0212\\
   22 & 14022 & 14470         &  94.1 &  94.7 &  .0269 &  .0000 & .0082 & .0000\\
   23 &{\bf 13978}&{\bf 6299} & 112.6 & 112.8 &  .7774 &  .2375 & .4007 & .0441\\
   24 & 13838 &  4447         & 100.5 & 102.1 &  .0758 &  .0230 & .0167 & .0054\\
   25 & 13275 &  9526         &  98.6 &  98.5 &  .1667 &  .2170 & .0434 & .0790\\
   26 & 13243 &  8535         &  89.7 &  89.2 &  .0572 &  .0641 & .0099 & .0069\\
   27 & 12972 & 10136         &  59.5 &  58.9 &  .1054 &  .1326 & .0088 & .0249\\\hline

     \end{tabular}

   \end{center}
 \end{table}

 \begin{table}[htbp]
   \begin{center}
     \caption[]{\label{tab:misscand}\sl Same as for
     Table~\ref{tab:fourjcand}, for the missing-energy channel.\\ }
     \vspace*{0.5cm}
    \begin{tabular}{|c|c|c|c|c|c|c|c|c|}\cline{4-9}
       \multicolumn{3}{c|}{} &
       \multicolumn{2}{|c|}{\mHrec} &
       \multicolumn{2}{|c|}{$\ln(1+s/b)$-Weight} &
       \multicolumn{2}{|c|}{$\ln(1+s/b)$-Weight} \\
       \multicolumn{3}{c|}{} &
       \multicolumn{2}{|c|}{GeV} &
       \multicolumn{2}{|c|}{\mHtest=100~GeV} &
       \multicolumn{2}{|c|}{\mHtest=115~GeV} \\\hline
       No. & \multicolumn{2}{|c|}{Run:Event} & 
        old & new & old & new & old & new \\\hline

   1 &{\bf 15886}&{\bf 54731}  & 112.1 & 111.9 & .4025 &  n/s  & .2044 &  n/s \\
   2 &  15761 &  28847         & 110.5 & 110.3 & .2986 & .1387 & .0707 &.0298 \\
   3 &{\bf 15648} &{\bf 12732} & 111.7 & 111.8 & .2234 &  n/s  & .0940 &  n/s \\
   4 &{\bf 15587} &{\bf 18556} & 100.7 & 109.0 &  n/s  & .2295 &  n/s  &.0579 \\
   5 &  15498 &  13971         & 112.8 & 100.4 & .1387 & .0820 & .0571 &.0283 \\
   6 &  15429 &  16226         &  99.2 & 112.6 & .1648 & .2173 & .1455 &.1333 \\
   7 &  15258 &  12168         &  99.1 &  99.0 & .1453 & .0998 & .0769 &.0350 \\
   8 &  15204 &  40272         & 104.0 &  99.3 & .3071 & .0528 & .1224 &.0233 \\
   9 &  15178 &  44741         & 110.1 & 105.1 & .6734 & .2672 & .2518 &.0908 \\
  10 &  15099 &  25825         & 101.8 & 109.9 & .2694 & .2173 & .0571 &.0579 \\
  11 &  14979 &  16964         & 108.2 & 101.6 & .3641 & .2419 & .1551 &.0968 \\
  12 &  14383 &  22908         & 108.2 & 108.1 & .5871 & .3535 & .1455 &.0968 \\
  13 &  12357 &  26618         & 106.9 & 108.2 & .5423 & .7580 & .0536 &.1826 \\
  14 &  12323 &  12944         & 107.2 & 107.2 & .5423 & .5290 & .1456 &.1465 \\\hline

 \end{tabular}

   \end{center}
 \end{table}

\clearpage
\newpage



\clearpage
\newpage


\clearpage
\newpage


\clearpage
\newpage


\clearpage
\newpage


\begin{figure}[htbp]
  \centerline{
    \epsfig{file=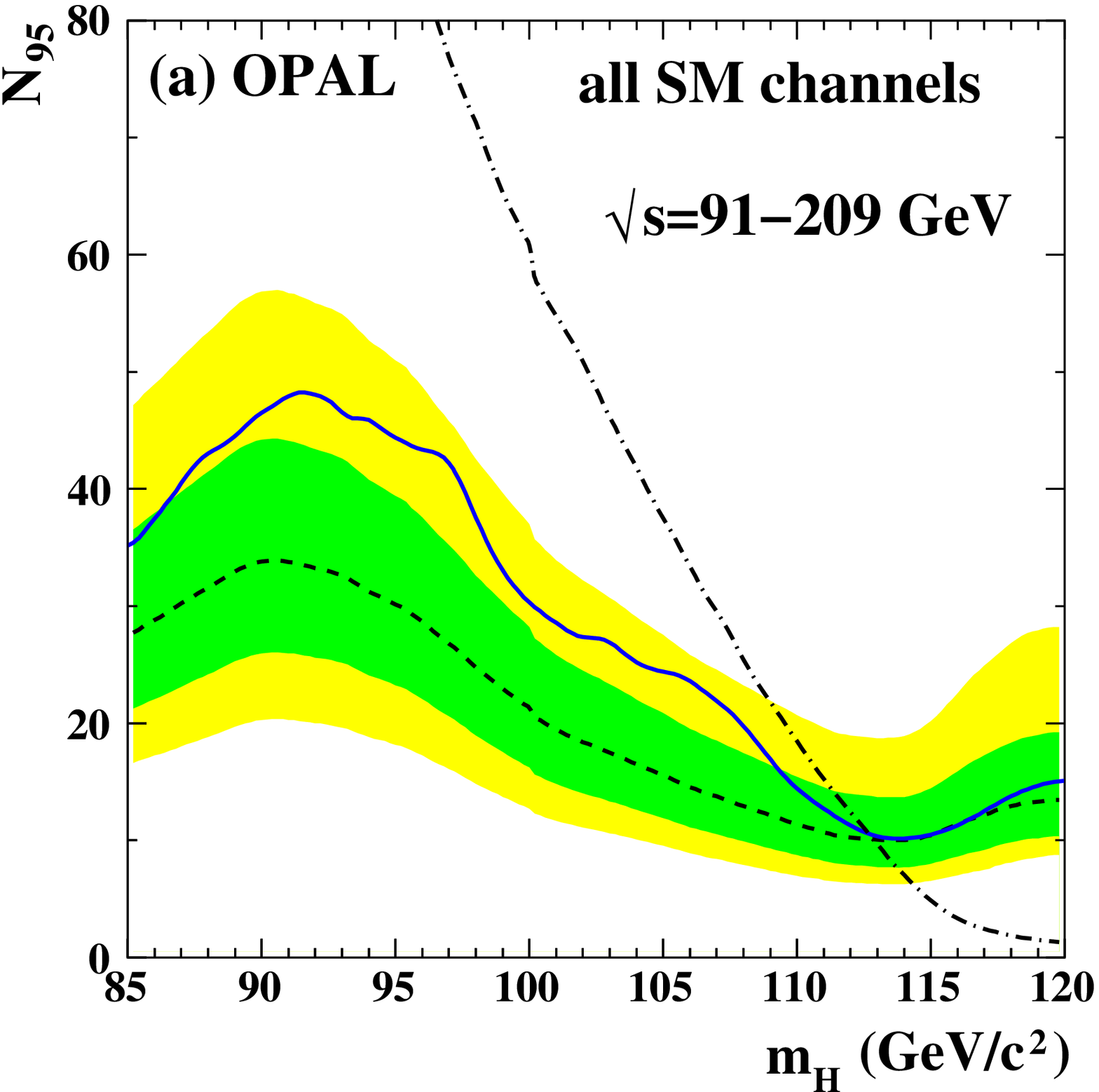,width=0.50\textwidth}\\
    \epsfig{file=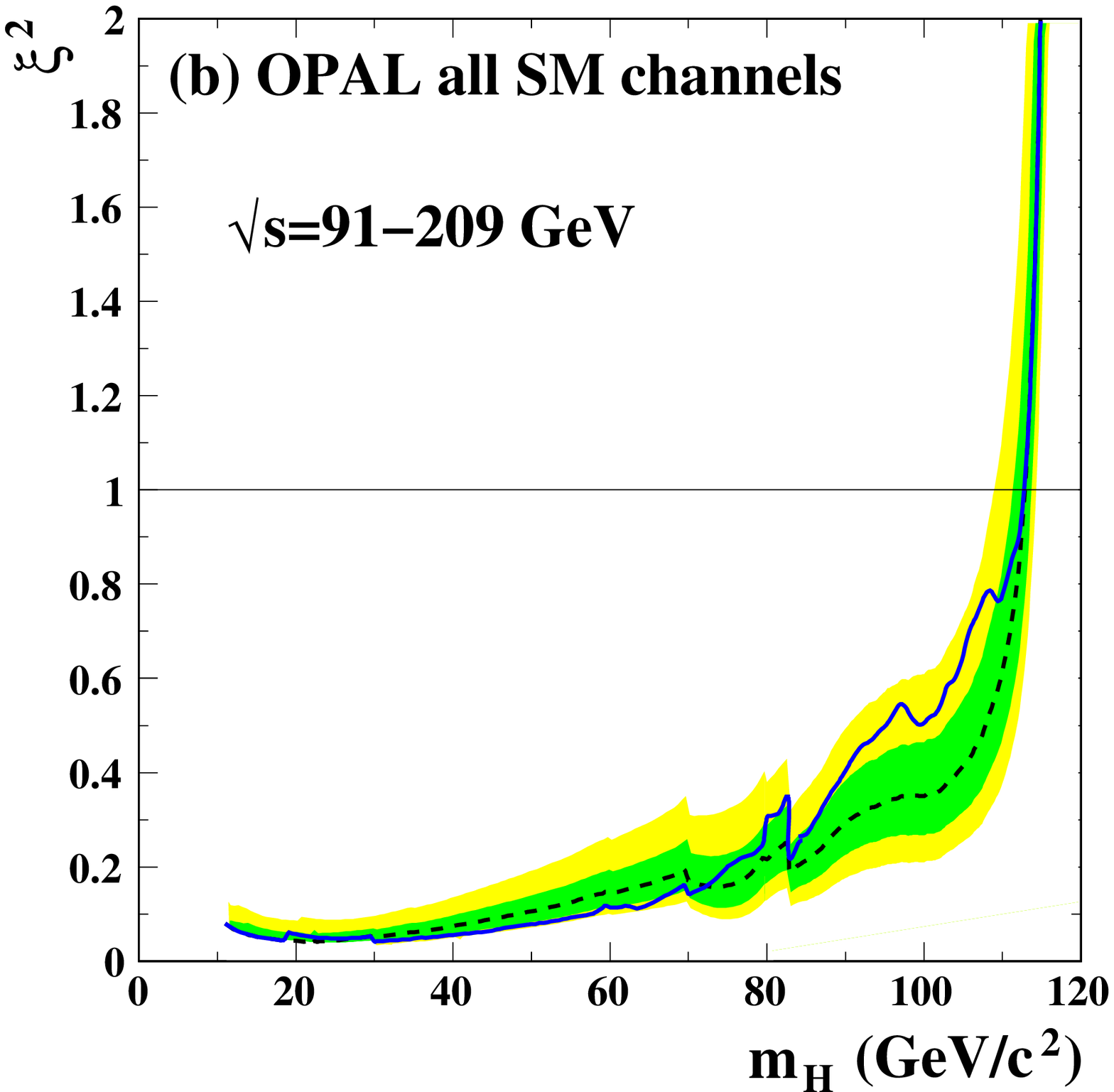,width=0.50\textwidth}}
  \caption[]{\label{fig:smn95}\sl  
    (a) Upper limits on the signal event rate at the 95\% confidence
    level, N$_{95}$, as observed (solid line) and the expected median
    (dashed line) for background-only hypotheses, as a function of the
    Higgs boson test mass.  The expected rate of the accepted signal
    counts for a Standard Model Higgs boson with a mass equal to the
    test-mass is shown with the dash-dotted line.  (b) The 95\% CL
    upper limit on $\xi^2$ is shown as a function of the Higgs boson
    mass \mH\ as the solid curve. The shaded bands indicate the 68\%
    and 95\% probability intervals, with respect to the median
    expectation in the absence of a signal. For the calculation of
    N$_{95}$ and $\xi^2$ all available OPAL data with 
    $\sqrt{s}=91-209$~\gevcs\ were used.}
\end{figure}

\end{document}